\documentclass[12pt, letterpaper, titlepage]{article}
\pdfoutput=1
\usepackage[utf8]{inputenc}
\usepackage{geometry}
\usepackage[english]{babel}
\usepackage{amsmath, amsfonts, graphicx}
\usepackage{amssymb}
\usepackage{mciteplus}
\usepackage{mathtools}
\usepackage{graphicx}
\usepackage{float}
\usepackage{cite}
\usepackage{bm}
\usepackage{pstricks}
\usepackage[labelfont=bf, font={small}]{caption}
\usepackage{footmisc}
\usepackage{multirow}
\usepackage{url}
\usepackage[breakable, theorems, skins]{tcolorbox}
\usepackage{enumerate}
\usepackage{bm}
\usepackage{epstopdf}
\usepackage{hyperref}

\newcommand{\Tr}{\operatorname{Tr}} 

\newlength{\abstractwidth}
\definecolor{mycolor}{rgb}{1,0,0}

\newcommand{\NS}{\text{NS}}
\newcommand{\R}{\text{R}}

\newcommand{\B}{\scriptscriptstyle \text{B}}
\newcommand{\T}{\scriptscriptstyle \text{T}}

\newcommand{\PB}{\scriptscriptstyle \text{PB}}
\newcommand{\+}{\scalebox{.4}{$+$}} 
\newcommand{\m}{\scalebox{.4}{$-$}} 

\edef\restoreparindent{\parindent=\the\parindent\relax}
\usepackage{parskip}
\restoreparindent

\numberwithin{equation}{section}
\allowdisplaybreaks

\begin{document}

\begin{titlepage}

\setcounter{page}{0} \baselineskip=15.5pt \thispagestyle{empty}



${}$
\vspace{0.4cm} 

\begin{center}

\def\thefootnote{\fnsymbol{footnote}}
\begin{changemargin}{0.05cm}{0.05cm}
\begin{center}
{\Large \bf From Quantum Groups to \\[10 pt] Liouville and Dilaton Quantum Gravity}
\end{center}
\end{changemargin}

~\\[1cm]
{Yale Fan${}^{\rm a}$\footnote{\href{mailto:yalefan@gmail.com}{\protect\path{yalefan@gmail.com}}} and Thomas G. Mertens${}^{\rm b}$\footnote{\href{mailto:thomas.mertens@ugent.be}{\protect\path{thomas.mertens@ugent.be}}}}
\\[0.3cm]
\vspace{0.7cm}
{\normalsize{\sl ${}^{\rm a}$Theory Group, Department of Physics,
\\[1mm]
University of Texas at Austin, Austin, TX 78712, USA}} \\[3mm]
{\normalsize{\sl ${}^{\rm b}$Department of Physics and Astronomy,
\\[1mm]
Ghent University, Krijgslaan, 281-S9, 9000 Gent, Belgium}} \\
\vspace{0.5cm}

\end{center}

\vspace{0.2cm}
\begin{changemargin}{1cm}{1cm}
{\small \noindent
\begin{center}
\textbf{Abstract}
\end{center}}
We investigate the underlying quantum group symmetry of 2d Liouville and dilaton gravity models, both consolidating known results and extending them to the cases with $\mathcal{N} = 1$ supersymmetry. We first calculate the mixed parabolic representation matrix element (or Whittaker function) of $\text{U}_q(\mathfrak{sl}(2, \mathbb{R}))$ and review its applications to Liouville gravity. We then derive the corresponding matrix element for $\text{U}_q(\mathfrak{osp}(1|2, \mathbb{R}))$ and apply it to explain structural features of $\mathcal{N} = 1$ Liouville supergravity. We show that this matrix element has the following properties: (1) its $q\to 1$ limit is the classical $\text{OSp}^+(1|2, \mathbb{R})$ Whittaker function, (2) it yields the Plancherel measure as the density of black hole states in $\mathcal{N} = 1$ Liouville supergravity, and (3) it leads to $3j$-symbols that match with the coupling of boundary vertex operators to the gravitational states as appropriate for $\mathcal{N} = 1$ Liouville supergravity. This object should likewise be of interest in the context of integrability of supersymmetric relativistic Toda chains. We furthermore relate Liouville (super)gravity to dilaton (super)gravity with a hyperbolic sine (pre)potential. We do so by showing that the quantization of the target space Poisson structure in the (graded) Poisson sigma model description leads directly to the quantum group $\text{U}_q(\mathfrak{sl}(2, \mathbb{R}))$ or the quantum supergroup $\text{U}_q(\mathfrak{osp}(1|2, \mathbb{R}))$.
\end{changemargin}
\vspace{0.3cm}

\vfil


\end{titlepage}

\newpage

\tableofcontents

\setcounter{tocdepth}{2}
\setcounter{footnote}{0}


\section{Introduction and Overview}
\label{s:intro}

Jackiw-Teitelboim (JT) gravity in two dimensions \cite{Jackiw:1984je, Teitelboim:1983ux, Almheiri:2014cka, Jensen:2016pah, Maldacena:2016upp, Engelsoy:2016xyb, Callebaut:2018nlq, Cotler:2016fpe, Bagrets:2016cdf, Stanford:2017thb, Kitaev:2018wpr, Mertens:2017mtv, Mertens:2018fds, Lam:2018pvp, Yang:2018gdb, Saad:2019lba, Blommaert:2019wfy, Saad:2019pqd, Blommaert:2020seb, Blommaert:2019hjr, Mertens:2019bvy, Blommaert:2020yeo, Lin:2019qwu, Johnson:2019eik, Stanford:2019vob, Almheiri:2020cfm} lies at the heart of the recent renaissance in the study of lower-dimensional gravitational models. Exploiting its first-order formulation in terms of $\mathfrak{sl}(2,\mathbb{R})$ BF theory \cite{Fukuyama:1985gg, Isler:1989hq, Chamseddine:1989yz, Jackiw_1992}, many amplitudes in this theory can be explicitly and exactly computed \cite{Blommaert:2018oro, Blommaert:2018iqz, Iliesiu:2019xuh}.

\begin{figure}[!htb]
\centering
\includegraphics[width=0.4\textwidth]{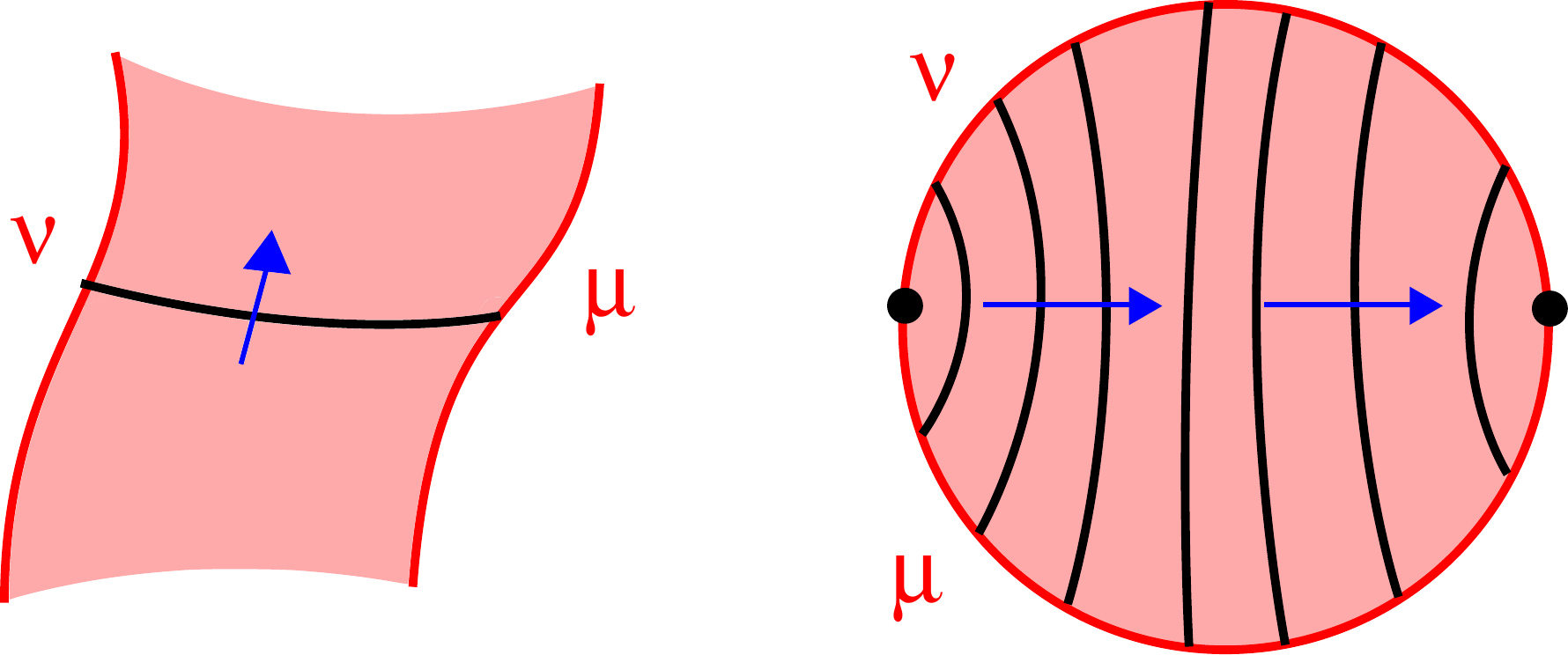}
\caption{Left: wavefunction $\psi_{R, \nu\mu}(g)$ in a mixed parabolic basis as a two-boundary state. Right: two-boundary slicing of the disk.}
\label{Whitslice}
\end{figure}

An important ingredient in the computation of gravitational amplitudes in BF language is the following representation matrix element of $\mathfrak{sl}(2,\mathbb{R})$:
\begin{equation}
\psi_{R,\nu\mu}(g) \equiv \left\langle R, \nu\right| g \left|R, \mu\right\rangle,
\end{equation}
where both indices $\mu$ and $\nu$ are fixed in a ``mixed parabolic basis.'' This fixing originates from the holographic boundary conditions, known from the work of Brown and Henneaux on 3d gravity \cite{Brown:1986nw} and applied to pure 3d gravity in \cite{Coussaert:1995zp, Henneaux:1999ib}. This basis is called ``mixed'' because the bra and ket of this equation are constrained by \emph{different} parabolic generators of $\mathfrak{sl}(2, \mathbb{R})$:
\begin{align}
\label{diagpar}
E^+ \left|R,\mu\right\rangle = -\mu \left|R,\mu\right\rangle, \qquad (E^-)^\dagger \left|R,\nu\right\rangle = \nu \left|R,\nu\right\rangle, \qquad \mu,\nu > 0,
\end{align}
where we write $\{H, E^+, E^-\}$ for the Cartan-Weyl basis of $\mathfrak{sl}(2, \mathbb{R})$. These matrix elements play the role of wavefunctions on Cauchy slices with both endpoints at holographic boundaries, as shown in Figure \ref{Whitslice} (left).

This description in terms of representation matrix elements can be derived directly in the BF or 2d Yang-Mills language \cite{Blommaert:2018iqz}. In particular, the fixing of the representation indices corresponds to considering a particular coset of the underlying SL$(2,\mathbb{R})$ structure.\footnote{In fact, one starts with a slightly different algebraic structure than a group: it was argued in \cite{Blommaert:2018oro, Blommaert:2018iqz, Fan:2021wsb} to be the positive subsemigroup SL$^+(2,\mathbb{R})$, whereas \cite{Iliesiu:2019xuh} considers a particular limit of the universal cover of SL$(2,\mathbb{R})$.} One can then utilize this slicing on different surfaces with boundaries to compute different amplitudes in these gravity models. An example is the disk amplitude shown in Figure \ref{Whitslice} (right).

In the mathematical literature, matrix elements with the particular parabolic constraints \eqref{diagpar} are called \emph{Whittaker functions} \cite{Jacquet, Schiffmann, Hashizume1, Hashizume2}. Due to their importance for gravitational calculations with holographic boundaries, we will sometimes also call them \emph{gravitational matrix elements}. These representation matrix elements are constructed from states that diagonalize the parabolic generators as in \eqref{diagpar}: these states are called \emph{Whittaker vectors}.

A surprising recent development is that a variety of different gravitational models also exhibit the same structure as in JT gravity. This observation applies in particular to Liouville gravity amplitudes in the fixed-length basis \cite{Mertens:2020hbs, Mertens:2020pfe}, where the gravitational matrix element in question is a Whittaker function of (the modular double of) the quantum group U$_q(\mathfrak{sl}(2,\mathbb{R}))$. Using this matrix element, one can for instance determine the disk boundary tachyon two-point function in Liouville gravity:
\begin{align}
\begin{tikzpicture}[baseline={([yshift=-.5ex]current bounding box.center)}, scale=0.7]
\draw[fill=mycolor,opacity=0.333] (0,0) ellipse (1.5 and 1.5);
\draw[thick,mycolor] (0,0) ellipse (1.5 and 1.5);
\draw[fill] (-1.5,0) circle (0.06);
\draw[fill] (1.5,0) circle (0.06); 
\node at (-2.2,0) {\small $\mathcal{B}_{\beta_M}$};
\node at (2.2,0) {\small $\mathcal{B}_{\beta_M}$};
\node at (0,-1.8) {\small $\ell_1$};
\node at (0,1.8) {\small $\ell_2$};
\end{tikzpicture}
= \left\langle \mathcal{B}_{\beta_M} \mathcal{B}_{\beta_M}\right\rangle_{\ell_1,\ell_2}.
\end{align}
The real parameter $\beta_M$ corresponds to the matter label of the primary operator in the matter CFT, with weight $\Delta_{\beta_M} = \beta_M(\mathfrak{q}+\beta_M)$ where $\mathfrak{q}=1/b - b$. The boundary tachyon vertex operators $\mathcal{B}_{\beta_M}$ are separated by length segments of length $\ell_1$ and $\ell_2$. The resulting Liouville gravity amplitude can be written explicitly as
\begin{align}
\left\langle \mathcal{B}_{\beta_M} \mathcal{B}_{\beta_M}\right\rangle_{\ell_1,\ell_2} = \int_{0}^{+\infty} ds_1\, ds_2\, &\rho(s_1) \rho(s_2) e^{-\ell_1 \frac{\cosh 2 \pi b s_1}{2\pi b^2 \sin \pi b^2}} e^{- \ell_2 \frac{\cosh 2 \pi b s_2}{2\pi b^2 \sin \pi b^2}} \nonumber \\
&\times \frac{S_{b}(\beta_M \pm i s_1 \pm i  s_2)}{S_{b}(2\beta_M)}, \label{final2}
\end{align}
where $\rho(s)= \sinh (2\pi b s) \sinh \left( \frac{2\pi s}{b}\right)$ is the density of states. More details on this notation, and how one obtains this amplitude from the non-critical string, are summarized in Appendix \ref{app:bosonic}.  Compared to \cite{Mertens:2020hbs}, these length segments are measured using a rescaled version of the Liouville metric that we will later want to identify with the physical boundary metric in a dilaton gravity theory with sinh potential. The relation between the boundary lengths measured in these different metrics is a simple rescaling:\footnote{In the JT limit $b\to 0$, this relation becomes $\ell_{\text{L}} = \frac{\ell_{\text{JT}}}{2\pi^2 b^4 \kappa}$, reproducing the scaling of \cite{Mertens:2020hbs} that matches with JT dilaton gravity amplitudes.}
\begin{equation}
\label{rescaleb}
\kappa \ell_{\text{L}} = \frac{\ell}{ 2 \pi b^2\sin \pi b^2}, \qquad \kappa= \sqrt{\frac{\mu}{\sin \pi b^2}}, 
\end{equation}
where $\ell_{\text{L}}$ is the length measured using the Liouville metric and $\ell$ is the physical boundary length of the dilaton gravity model, as we will explain in the main text; $\mu$ is the Liouville bulk cosmological constant. The object on the second line of \eqref{final2} is the coupling coefficient (or vertex function) of the boundary operators to the gravitational states, and is the main ingredient in this expression. The $\pm$ symbols on the second line indicate that one takes the product of four copies of the function, each with a different combination of signs.

The structure of the full equation \eqref{final2} is identical to that of the JT gravity boundary two-point function determined in \cite{Mertens:2017mtv}, and indeed, there exists a double-scaling limit in which $b\to 0$ that recovers precisely the JT gravity amplitudes. This connection to JT gravity was first pointed out for the disk partition function in the context of the minimal string in \cite{Saad:2019lba}, and then thoroughly investigated and extended in \cite{Mertens:2020hbs}. See also \cite{Johnson:2019eik, Okuyama:2019xbv, Gregori:2021tvs} for relevant recent work. From the group theory perspective, the quantum group U$_q(\mathfrak{sl}(2,\mathbb{R}))$ turns back into its classical Lie algebra $\mathfrak{sl}(2,\mathbb{R})$, as we will explain in depth. For JT gravity \cite{Blommaert:2018oro}, Liouville gravity \cite{Mertens:2020hbs}, and JT supergravity \cite{Fan:2021wsb}, it was shown that the object on the second line of \eqref{final2} has a group-theoretic interpretation as the square of a $3j$-symbol where one uses two mixed parabolic matrix elements (Whittaker functions) and one discrete operator insertion.

It has been proposed that $\mathcal{N}=1$ Liouville supergravity satisfies similar properties \cite{Mertens:2020pfe}. Since $\mathcal{N}=1$ JT supergravity and its amplitudes are described by a BF theory based on the $\mathfrak{osp}(1|2,\mathbb{R})$ superalgebra \cite{Montano:1990ru, Astorino:2002bj, Cardenas:2018krd, Fan:2021wsb},\footnote{See \cite{Stanford:2019vob, Johnson:2020heh, Johnson:2021owr, Okuyama:2020qpm} for various other perspectives on $\mathcal{N}=1$ JT supergravity.} the relevant group-theoretic structure for $\mathcal{N}=1$ Liouville supergravity would seem to be U$_q(\mathfrak{osp}(1|2,\mathbb{R}))$.\footnote{Interestingly, a different $q$-deformation is required when studying double-scaled supersymmetric SYK models \cite{Berkooz:2020xne}.} However, a lack of independent knowledge of the relevant representation matrix elements of U$_q(\mathfrak{osp}(1|2,\mathbb{R}))$ in the available literature prevented us from making the comparison more explicit.

In this work, we resolve this problem. In particular, we generalize the group-theoretic arguments of \cite{Kharchev:2001rs} to the $\mathcal{N}=1$ supersymmetric case to compute the mixed parabolic matrix element of (the modular double of) U$_q(\mathfrak{osp}(1|2,\mathbb{R}))$. Our result is:
\begin{align}
\psi^{\epsilon,\pm}_{s,g_\mu g_\nu}(x) = e^{-\pi i  s x} \int_{-\infty}^{+\infty} d\zeta\, & g_\mu^{i\zeta}g_\nu^{i\zeta +2 is} e^{-\pi i \frac{\epsilon}{2} (\zeta^2 + 2s \zeta)} e^{-\pi i \zeta x} \label{resdisp} \\
&\times\left[S_{\NS}(-i\zeta) S_{\R}(-2i s -i \zeta ) \pm S_{\R}(-i\zeta) S_{\NS}(-2i s -i \zeta )\right], \nonumber
\end{align}
where $s$ is the representation label of the continuous series irreps of U$_q(\mathfrak{osp}(1|2,\mathbb{R}))$, while $g_\mu^2$ and $g_\nu^2$ are the (suitably rescaled) eigenvalues of the parabolic generators \eqref{diagpar} in a sense that we will explain below.  The $\epsilon$ superscript parametrizes different deformations of the same underlying classical Whittaker function. \emph{A priori}, $\epsilon$ may be any real number, but we will focus on $\epsilon=\pm 1$ to make contact with the particular deformation relevant to Liouville supergravity. The $\pm$ superscript labels the pair of Whittaker functions that are present in a group with a nontrivial sCasimir operator in the scentre of the universal enveloping algebra, such as OSp$(1|2,\mathbb{R})$. Finally, the objects $S_{\NS}$ and $S_{\R}$ on the second line are suitable supersymmetric extensions of the well-known double sine function $S_b$ (see Appendix \ref{app:def}). Proving this formula is our first goal in this work.

As mentioned, our motivation for this calculation stems from the application to $\mathcal{N}=1$ Liouville supergravity. In particular, in \cite{Mertens:2020pfe}, the fixed-length boundary tachyon two-point function on the disk was found to be
\begin{align}
&\left\langle \mathcal{B}_{\beta_M} \mathcal{B}_{\beta_M}\right\rangle_{\ell_1,\ell_2} = \int_{0}^{+\infty} ds_1\, ds_2\, \rho(s_1) \rho(s_2) e^{-\ell_1 \frac{\sinh^2 \pi b s_1}{16 \sin^2 \frac{\pi b^2}{2}}} e^{- \ell_2 \frac{\sinh^2 \pi b s_2}{16 \sin^2 \frac{\pi b^2}{2}}} \times {} \label{final1} \\
&\left[\frac{S_{\R}(\beta_M \pm i( s_1 +  s_2))S_{\NS}(\beta_M \pm i( s_1 -  s_2))}{S_{\NS}(2\beta_M)} + \frac{S_{\NS}(\beta_M \pm i( s_1 +  s_2))S_{\R}(\beta_M \pm i(s_1 -  s_2))}{S_{\NS}(2\beta_M)}\right], \nonumber
\end{align}
where $\rho(s) = \cosh (\frac{\pi s}{b})\cosh (\pi b s)$. The right-hand side contains our new choice of length parameter, which is again rescaled compared to the Liouville length as\footnote{The $b\to 0$ limit of \eqref{rescale} is $\ell_{\text{L}} = \frac{\ell}{4\pi^2 b^4\kappa^2}$, and the length parameter $\ell$ is directly identified with the JT dilaton supergravity length scale $\ell_{\text{JT}}$ in this limit \cite{Mertens:2020pfe}.}
\begin{equation}
\label{rescale}
\kappa^2 \ell_{\text{L}} = \frac{\ell}{16 \sin^2 \frac{\pi b^2}{2}},\qquad \kappa= \sqrt{\frac{2\mu}{\cos \frac{\pi b^2}{2}}}, 
\end{equation}
where $\mu$ is the super-Liouville bulk cosmological constant. The left-hand side comes from transforming super-Liouville amplitudes \cite{Fukuda:2002bv} to the fixed-length basis. We refer to Appendix \ref{app:bosonic} for some of the details of this procedure.\footnote{This corresponds to a choice $\eta = +1$ of local fermionic boundary condition. We refer to \cite{Fukuda:2002bv, Douglas:2003up, Seiberg:2003nm} for more details on the super-Liouville SCFT ingredients of this result.} The matter label $\beta_M$ corresponds to the weight $\Delta_{\beta_M} = \frac{1}{2}\beta_M(\mathfrak{q}+\beta_M)$ of a primary operator in the matter SCFT.

In this work, as an application of our newly determined Whittaker function \eqref{resdisp} of U$_q(\mathfrak{osp}(1|2,\mathbb{R}))$, we will show that the vertex function in $\mathcal{N}=1$ Liouville supergravity has a similar interpretation as that in JT (super)gravity and Liouville gravity. In particular, the quantity on the second line of \eqref{final1} is equal to
\begin{equation}
\int_{-\infty}^{+\infty} dx\, \psi^{\epsilon}_{s_1} (x)  \psi^{\epsilon}_{s_2} (x)^\ast e^{- \beta_M \pi x},
\end{equation}
up to unimportant prefactors. Moreover, the density of states $\rho(s)$ in \eqref{final1} is computed to be the Plancherel measure of this same Whittaker function:
\begin{equation}
\int_{-\infty}^{+\infty} dx\, \psi^{\epsilon}_{s_1} (x)  \psi^{\epsilon}_{s_2} (x)^\ast = \frac{\delta(s_1-s_2)}{\rho(s_1)}.
\end{equation}
Finally, the energy variable in the exponentials multiplying the lengths $\ell_i$ is precisely the Casimir operator of these same representations of U$_q(\mathfrak{osp}(1|2,\mathbb{R}))$.

Having understood the relation between the representation-theoretic objects appearing in amplitudes such as \eqref{final1}, it remains to explain \emph{why} they appear in the first place. Doing so requires understanding the Liouville (super)gravity theory directly from a Lagrangian perspective, and thereby identifying the relevant quantum (super)group as a symmetry. A natural language for achieving such an understanding is that of the (graded) Poisson sigma model description of dilaton gravity, where the quantization of the model (in a physical sense) entails passing to the quantized version of the Poisson algebra. For the particular case of a hyperbolic sine dilaton (pre)potential, this procedure results in either the U$_q(\mathfrak{sl}(2,\mathbb{R}))$ or the U$_q(\mathfrak{osp}(1|2,\mathbb{R}))$ quantum algebra as the quantized charge algebra.\footnote{Notice that the words ``quantum'' and ``quantized'' in this sentence correspond to different procedures: by ``quantum algebra,'' we mean the $q$-deformed algebra, whereas by ``quantized charge algebra,'' we mean the algebra of symmetry charges of the physically quantized (in $\hbar$) dynamical system.} Understanding this approach is the second goal of this work.

The remainder of this work is structured as follows.

In \textbf{Section \ref{s:two}}, we compute the mixed parabolic matrix element for the bosonic quantum group U$_q(\mathfrak{sl}(2,\mathbb{R}))$, reproducing the results of \cite{Kharchev:2001rs} from a perspective more amenable to supersymmetrization. In \textbf{Section \ref{s:three}}, which comprises the main part of this work, we calculate the mixed parabolic matrix element (or Whittaker function) of U$_q(\mathfrak{osp}(1|2,\mathbb{R}))$. The result \eqref{resdisp} was already stated above, and will be checked and matched to supergravity results.

To better understand the origin of this quantum group symmetry, we give arguments in \textbf{Section \ref{s:dilgrav}} as to how Liouville gravity and $\mathcal{N}=1$ Liouville supergravity relate to dilaton (super)gravity with a sinh (pre)potential, which can in turn be written as a (graded) Poisson sigma model, finally unveiling the U$_q(\mathfrak{sl}(2,\mathbb{R}))$ or U$_q(\mathfrak{osp}(1|2,\mathbb{R}))$ quantum group structure. This section can be read independently of the somewhat more technical preceding sections.

We conclude in \textbf{Section \ref{s:concl}} with some open problems and speculations on the bigger picture.

As mentioned above, \textbf{Appendix \ref{app:bosonic}} provides some background material on Liouville gravity and supergravity. \textbf{Appendices \ref{app:spec}}, \textbf{\ref{alternative}}, \textbf{\ref{app:besseli}}, \textbf{\ref{app:NS}}, and \textbf{\ref{app:nonlinear}} contain some additional complementary material that is not required to understand the main story.

\section{\texorpdfstring{U$_q(\mathfrak{sl}(2, \mathbb{R}))$}{Uq(sl(2, R))} Gravitational Matrix Element}
\label{s:two}

In this section and the next, we present the calculation of the Whittaker function from a group-theoretic perspective. This section deals with the bosonic case, first discussed in \cite{Kharchev:2001rs}. The next section follows with the supersymmetric generalization. By their nature, both this section and the next are rather technical.

The Whittaker vectors and resulting Whittaker function of U$_q(\mathfrak{sl}(2,\mathbb{R}))$ were determined by Kharchev, Lebedev, and Semenov-Tian-Shansky \cite{Kharchev:2001rs} using a carrier space for the representations of the quantum group that is slightly different than the usual one of Ponsot and Teschner \cite{Ponsot:1999uf, Ponsot:2000mt, Hadasz:2013bwa}. We first demonstrate how one can translate the results of \cite{Kharchev:2001rs} into a more convenient carrier space, akin to the one used in \cite{Ponsot:1999uf, Ponsot:2000mt, Hadasz:2013bwa}, which leads to slightly more elegant expressions for the Whittaker vectors. The resulting Whittaker function is the same, since it depends solely on the representation labels and not on the precise construction underlying it. The main benefit of reformulating the calculation in this way is that it facilitates the supersymmetric generalization in Section \ref{s:three}.

\subsection{\texorpdfstring{$q$}{q}-Deformed Algebra: Definition and Classical Limits}

We first define the quantum group U$_q(\mathfrak{sl}(2,\mathbb{R}))$ and its self-dual continuous series representations. In particular, we will compare different realizations of these in the literature, and take the classical limit ($q\to 1$) whenever possible to develop intuition for these objects.

The $q$-deformed SL$(2,\mathbb{R})$ algebra consists of three generators $K, E^+, E^-$ satisfying the following commutator relations:
\begin{align}
\label{qsl}
KE^\pm = q^{\pm 1}E^{\pm}K , \qquad [E^+,E^-] = \frac{K^2-K^{-2}}{q-q^{-1}} = \frac{\sin 2 \pi b^2H}{\sin \pi b^2},
\end{align}
where $K= q^H$ and $q=e^{\pi ib^2}$ is the deformation parameter. We require $b^2 \in \mathbb{R}$, such that $q$ is a phase factor, for compatibility with the $*$-relation that we specify further on.

The Casimir operator $\mathcal{C}$ that commutes with all generators is given by
\begin{align}
\label{cas}
\left(\frac{2\pi ib^2}{q-q^{-1}}\right)\mathcal{C} &= \frac{(q+q^{-1})(K^2+K^{-2})}{2(q-q^{-1})^2} + \frac{1}{2}E^- E^+ +  \frac{1}{2}E^+ E^- \\
&= E^- E^+ + \frac{qK^2+q^{-1}K^{-2}}{(q-q^{-1})^2}.
\end{align}
The prefactor can be chosen arbitrarily at this point, and we have made a specific choice that will turn out to be natural from the dilaton gravity perspective.

The continuous series representations of this quantum algebra can be constructed on the space of entire functions, restricted to the real line $\mathbb{R}$, with suitable asymptotic restrictions specified in \cite{Kharchev:2001rs} that allow one to drop ``pieces at infinity'' when doing contour deformations. On this space, we define the action of the shift operator:
\begin{equation}
T_\Delta f(t) \equiv f(t+\Delta), \quad t\in\mathbb{R}.
\end{equation}
Kharchev et al.\ \cite{Kharchev:2001rs} then define the following set of generators:\footnote{We have set $\lambda_{\text{there}} = -2 \alpha_{\text{here}}$ in order to have $\alpha = +Q/2 + is$ with $s\in\mathbb{R}$ as the representation label. We also set $K_{\text{there}} = - \mathcal{K}^{-2}$, $E_{\text{there}} = \mathcal{E}^-$, $F_{\text{there}} = -\mathcal{E}^+$.}
\begin{align}
\mathcal{K} &= -ie^{\pi i b \alpha}T_{ib/2}, \nonumber \\
\mathcal{E}^+ &= -\frac{q e^{2\pi b t}}{q-q^{-1}} \left(e^{-2\pi i b \alpha} - e^{2\pi i b \alpha}T_{ib}\right), \label{KLT} \\
\mathcal{E}^- &= \frac{e^{-2\pi b t}}{q-q^{-1}}\left( 1- T_{-ib}\right), \nonumber
\end{align}
which can be readily shown to satisfy \eqref{qsl}. We have defined the quantity $\alpha = Q/2 + is$, where $Q \equiv b+ 1/b$ and $s\in \mathbb{R}$ denotes the representation label. With this value of $\alpha$, and with respect to the inner product
\begin{equation}
\label{asymmetric}
(f,g) \equiv \int_\mathbb{R} dt \, e^{2\pi Q t} \overline{f(t)} g(t),
\end{equation}
the above operators satisfy the following hermiticity conditions:
\begin{equation}
\label{herm}
\mathcal{K}^\dagger = \mathcal{K}, \qquad (\mathcal{E}^+)^\dagger = - \mathcal{E}^+, \qquad (\mathcal{E}^-)^\dagger = - \mathcal{E}^-.
\end{equation}
Identifying $\mathcal{K} = q^{\mathcal{H}}$, one has $\mathcal{H}^\dagger = -\mathcal{H}$.

The representations thus constructed are precisely the continuous series representations of U$_q(\mathfrak{sl}(2,\mathbb{R}))$. These irreps are distinguished as being the unique representations for which the carrier space is simultaneously the carrier space of the continuous series representations of the dual quantum algebra, which is found by setting $b\to 1/b$ in all of the relations above. (Following common convention, we will always denote dual generators with tildes, e.g., $\tilde{K}$, $\tilde{E}$, etc.) For this reason, they are called \emph{self-dual}. This means that one can think of them as nontrivial representations of the modular double \cite{Faddeev:1999fe, Ponsot:1999uf, Ponsot:2000mt, Bytsko:2002br, Bytsko:2006ut, Ip}
\begin{equation}
\label{md}
\text{U}_q(\mathfrak{sl}(2,\mathbb{R})) \otimes \text{U}_{\tilde{q}}(\mathfrak{sl}(2,\mathbb{R})),
\end{equation}
where $\tilde{q} = e^{\pi i /b^2}$, and where the generators $(\mathcal{K}^2, \mathcal{E}^+, \mathcal{E}^-)$ commute with the dual generators $(\tilde{\mathcal{K}}^2, \tilde{\mathcal{E}}^+, \tilde{\mathcal{E}}^-)$. This ``modular-doubled'' quantum group is a natural object to consider and has been studied in many different contexts in the literature. One illuminating property is that operators that commute with both quantum algebras must be scalar. This is the $q$-analogue of Schur's lemma, which shows that combining the quantum group with its dual into the modular double seems to give the most natural $q$-analogue of classical group theory.

The carrier space construction \eqref{KLT} has the benefit of having a well-known classical limit $b\to 0$, which we now explain. We relate the carrier space coordinate $t$ to a new half-space coordinate $x$ by $x=e^{2\pi b t}$. Setting $\alpha = 1/2b - bj$, we obtain the Borel-Weil realization of $\mathfrak{sl}(2,\mathbb{R})$:
\begin{align}
\hat{\mathcal{H}} &= x\partial_x - j, \nonumber \\
\hat{\mathcal{E}}^+ &= -x^2 \partial_x + 2 j x, \label{BW} \\
\hat{\mathcal{E}}^- &= \partial_x, \nonumber 
\end{align}
satisfying
\begin{equation}
\label{clalgsl}
[H, E^\pm] = \pm E^\pm, \qquad [E^+, E^-] = 2H,
\end{equation}
where $j = -1/2 + ik$ with $k \in \mathbb{R}$. The quadratic Casimir is\footnote{Note that the $b\to 0$ limit of the operator \eqref{cas} is $\frac{1}{2}E^+E^- + \frac{1}{2}E^-E^+ + H^2 + \frac{1}{4}$, where the constant term is conventionally dropped for the Casimir of $\mathfrak{sl}(2,\mathbb{R})$ as written.}
\begin{equation}
\mathcal{C} \equiv \frac{1}{2}E^+E^- + \frac{1}{2}E^-E^+ + H^2 = j(j+1) = -\frac{1}{4} - k^2.
\end{equation}
We emphasize that the coordinate $x$ lives on $\mathbb{R}^+$ due to the exponential mapping relating $t$ and $x$. We hence immediately land on the positive subsemigroup SL$^+(2,\mathbb{R})$, for which the carrier space coordinate is positive: $x > 0$. This is the most direct way of appreciating the link between the $q$-deformed modular double quantum group and the positive subsemigroup in the classical limit.

The realization \eqref{BW} of $\mathfrak{sl}(2,\mathbb{R})$ exponentiates to
\begin{equation}
\label{pssl}
(g\circ f)(x) = (bx + d)^{2j}f\left(\frac{ax + c}{bx + d}\right),
\end{equation}
which defines the principal series representations of SL$^+(2,\mathbb{R})$ ($ad - bc = 1$, $a, b, c, d > 0$).

However, we will not adhere to this particular choice of carrier space for the continuous series irreps in this work. It is instead convenient to write $\alpha = 1/2b + \lambda$ with $\lambda = b/2 + is$. Now we apply the following isomorphism that preserves the algebra \eqref{qsl}:
\begin{align}
\mathcal{K} &= e^{-2\pi \lambda t}K e^{2\pi \lambda t}, \nonumber \\
\mathcal{E}^{+} &=  q e^{-2\pi \lambda t} E^+ e^{2\pi \lambda t} K, \label{iso} \\
\mathcal{E}^{-} &=  e^{-2\pi \lambda t} E^- e^{2\pi \lambda t} K^{-1}, \nonumber
\end{align}
thus obtaining new generators $K, E^+, E^-$. This maps the carrier space with inner product \eqref{asymmetric} to the more ``symmetric'' carrier space $L^2(\mathbb{R})$. Explicitly, the generators become:
\begin{align}
K &= T_{ib/2}, \nonumber \\
E^+ &= -e^{2\pi b t} \frac{e^{\pi ib \lambda}T_{ib/2} - e^{-\pi ib \lambda}T_{-ib/2}}{q-q^{-1}}, \label{PT} \\
E^- &= e^{-2\pi b t} \frac{e^{-\pi ib \lambda}T_{ib/2} - e^{\pi ib \lambda}T_{-ib/2}}{q-q^{-1}}. \nonumber
\end{align}
This is almost the same set of generators used in \cite{Ponsot:1999uf, Ponsot:2000mt, Hadasz:2013bwa}, the only differences being that $\lambda = b/2 + is$ and the hermiticity conditions. Indeed, with the new inner product
\begin{equation}
\label{newinner}
(f,g) \equiv \int_{\mathbb{R}} dt \, \overline{f(t)} g(t),
\end{equation}
these new operators satisfy the same hermiticity conditions as in \eqref{herm}:\footnote{The generators used in \cite{Ponsot:1999uf, Ponsot:2000mt, Hadasz:2013bwa} with $\lambda = Q/2 + is$, on the other hand, satisfy
\begin{equation}
K^\dagger = K, \qquad (E^+)^\dagger = E^+, \qquad (E^-)^\dagger = E^-,
\end{equation}
with the same inner product \eqref{newinner}.}
\begin{equation}
K^\dagger = K, \qquad (E^+)^\dagger = -E^+, \qquad (E^-)^\dagger = -E^-,
\end{equation}
provided that $\lambda = b/2 + is$ with $s \in \mathbb{R}$.

Setting $K = q^H$, the operator $H$ is explicitly:
\begin{equation}
\label{defH}
H = \frac{1}{2\pi b}\partial_t.
\end{equation}
The Casimir operator \eqref{cas} in this representation is given by
\begin{equation}
\label{casirrepbos}
\mathcal{C} = -\frac{\cos 2\pi b (\lambda-b/2)}{2\pi b^2\sin \pi b^2} = -\frac{\cosh 2\pi b s}{2\pi b^2\sin \pi b^2}.
\end{equation}
As $b\to 0$, we have $\mathcal{C} \to -k^2$ where $s=-bk$ (again dropping the constant). The expression \eqref{casirrepbos} appears in the arguments of the exponential functions in \eqref{final2}.

Upon setting $x=e^{2\pi b t}$ and $\lambda = -bj$, we find that in the classical $b\to 0$ limit, \eqref{PT} reduces to:
\begin{align}
\hat{H} &= x \partial_x, \nonumber \\
\hat{E}^+ &= - x^2\partial_x +jx, \label{clmi} \\
\hat{E}^- &= \partial_x +\frac{j}{x}, \nonumber
\end{align}
satisfying the $\mathfrak{sl}(2,\mathbb{R})$ algebra \eqref{clalgsl} with Casimir $\mathcal{C} = j(j+1)$. These operators are antihermitian on $\mathbb{R}^+$ with respect to the measure $d\mu(x) = dx/x$:
\begin{align}
\int_0^{+\infty} \frac{dx}{x} \left( \overline{f(x)} \mathcal{O} g(x) \right) = \int_0^{+\infty} \frac{dx}{x}\left( \overline{-\mathcal{O} f(x)} g(x)\right),
\end{align}
provided that $j=-1/2+ik$ with $k\in \mathbb{R}$.

Exponentiating the realization \eqref{clmi} of the $\mathfrak{sl}(2,\mathbb{R})$ algebra leads to the following group action on functions on $\mathbb{R}^+$:
\begin{equation}
\label{pssl2}
(g\circ f)(x) = \frac{(ax + c)^{j}(bx + d)^{j}}{x^j}f\left(\frac{ax + c}{bx + d}\right).
\end{equation}
One can indeed prove that this group representation is unitary for $j=-1/2+ik$ in the sense that\footnote{Up to subtleties involving the boundaries of integration.}
\begin{align}
\int_0^{+\infty} \frac{dx}{x}(g \circ f_1)^*(x) (g \circ f_2)(x) = \int_0^{+\infty} \frac{dx}{x}f_1^*(x) f_2(x),
\end{align}
where the transformation of the $1/x$ in the measure precisely compensates for the new pieces in \eqref{pssl2} compared to \eqref{pssl}.

The isomorphism \eqref{iso} between the different realizations reduces in the classical $b\to 0$ limit to the following equivalence of differential operators preserving the $\mathfrak{sl}(2,\mathbb{R})$ algebra:
\begin{align}
\hat{\mathcal{H}} &= x^{j} \hat{H} x^{-j}, \nonumber \\
\hat{\mathcal{E}}^{+} &=  x^{j} \hat{E}^+ x^{-j}, \\
\hat{\mathcal{E}}^{-} &=  x^{j} \hat{E}^- x^{-j}. \nonumber
\end{align}
This gives a convenient way to relate \eqref{pssl} and \eqref{pssl2}.

\subsection{Whittaker Vectors and Whittaker Function}
\label{s:whitbos}

Let us now determine the Whittaker vector by diagonalizing $E^+$ and demanding it to be a simultaneous eigenvector of the dual ($b\to 1/b$) quantum group.

For a rank-one Lie algebra, as stated in \eqref{diagpar}, the Whittaker vector is defined as diagonalizing the generator $E^+$ associated with the positive root: 
\begin{equation}
E^+ \phi(t) = -\mu \phi(t).
\end{equation}
When generalizing to the $q$-deformed algebra, there is a one-parameter family of extensions given by diagonalizing the same generator up to a possible action by the Cartan generator:
\begin{equation}
E^+ \phi_\alpha(t) = -\mu q^{2 \alpha H} \phi_\alpha(t), \qquad \alpha \in \mathbb{R}.
\end{equation}
The resulting solution to this equation is not unique. However, in the case of the modular double, one needs to combine this relation with its dual ($b\to 1/b$), which turns out to be far more restrictive and leads to a more-or-less unique solution. We will exemplify this phenomenon further on. A simultaneous eigenvector of both $E^+$ and $\tilde{E}^+$ can be found for a suitable relation between the eigenvalues as follows:
\begin{align}
\label{eigwhit}
E^+ \phi_\alpha^{\+}(t) &= -i\frac{g^b}{q-q^{-1}}q^{2 \alpha H} \phi_\alpha^{\+}(t), \\
\tilde{E}^+ \phi_\alpha^{\+}(t) &= -i\frac{g^{1/b}}{\tilde{q}- \tilde{q}^{-1}} \tilde{q}^{2 \alpha \tilde{H}} \phi_\alpha^{\+}(t),
\end{align}
where $g$ is a parameter that we call the eigenvalue of the modular double quantum group.

Inserting the explicit expressions \eqref{PT}, we obtain the finite difference equations
\begin{align}
\label{fde}
e^{\pi i b \lambda} \phi_{\alpha}^{\+}\left(t+\frac{ib}{2}\right) - e^{-\pi i b \lambda} \phi_{\alpha}^{\+}\left(t-\frac{ib}{2}\right) &= ig^b e^{-2\pi b t} \phi_{\alpha}^{\+}(t+ib \alpha), \\
e^{\pi i \lambda/b} \phi_{\alpha}^{\+}\left(t+\frac{i}{2b}\right) - e^{-\pi i \lambda/b} \phi_{\alpha}^{\+}\left(t-\frac{i}{2b}\right) &= ig^{1/b}e^{-2\pi t/b} \phi_{\alpha}^{\+}\left(t+ \frac{i \alpha}{b}\right).
\end{align}
The additional quantum parameter $\alpha$ represents a freedom akin to ordering ambiguities in quantization. To find the solution, we can Fourier-transform these difference equations by writing $\phi_{\alpha}^{\+}(t)$ in terms of $F(-i\zeta)$ in a 1:1 fashion as
\begin{align}
\phi_{\alpha}^{\+}(t) = e^{-2\pi \lambda t}\int_{\mathcal{C}} d\zeta\, g^{i\zeta} F(-i\zeta) e^{\pi i \alpha \zeta^2} e^{-2\pi i \zeta (t - s \alpha)}.
\end{align}
The difference equations then reduce to:\footnote{The first functional equation is the $q$-deformation of the classical recursion relation $\Gamma(z+1)=z\Gamma(z)$ for the gamma function, which, with an additional log-convexity assumption, is enough to prove uniqueness in that case. In our case, adding the second functional equation will be sufficient. This equation ``scales away'' and hence becomes vacuous in the $b \to 0$ limit.}
\begin{align}
\label{shiftfde}
2\sin(\pi b \zeta) F(\zeta) &= F(\zeta +b), \\
\label{shiftfde2}
2\sin\left(\frac{\pi \zeta}{b}\right) F(\zeta) &= F\left(\zeta +\frac{1}{b}\right),
\end{align}
which form the pair of functional equations satisfied by the double sine function $S_b(\zeta)$. We get a unique solution, up to overall normalization, if we make the following assumptions:
\begin{itemize}
\item $F$ is continuous.
\item $b$ and $1/b$ are incommensurate.
\end{itemize}
We can appreciate these conditions as follows. If one imposes only the first relation \eqref{shiftfde}, then one is free to multiply any solution $F$ by any periodic function in $\zeta$ with period $b$, such as $e^{k \sin(2\pi m\zeta/b)}$. This possibility is excluded when accounting for the second relation, provided that $b$ is sufficiently generic. If, e.g., $b=1$, then both equations are the same and there is clearly no unique solution. But if $b$ and $1/b$ are incommensurate and hence $b^2 \neq p'/p$ for coprime $p',p \in \mathbb{N}$, then one can use the above relations successively to provide a dense covering of $\zeta \in \mathbb{R}$. Combining this with the continuity of $F$ determines $F$ up to a single value, e.g., $F(b)$, or the overall normalization factor. We choose this normalization at this point to match with the Whittaker vectors of SL$^+(2,\mathbb{R})$ in the $b\to 0$ limit \cite{Blommaert:2018oro, Blommaert:2018iqz}.

Starting with a finite value of $F(b)$,\footnote{If $F(b)$ is zero, then the zeros at $\zeta = Q + nb + m/b$ for $n,m \in \mathbb{Z}_{\geq 0}$ are double zeros. But then ``doubling back'' shows that one has (at least simple) zeros at $\zeta = Q + nb + m/b$ for $n,m \in \mathbb{Z}$, which is again dense in $\mathbb{R}$. If $F(b)$ is infinity, then the poles at $\zeta = -nb-m/b$ for $n,m \in \mathbb{Z}_{\geq 0}$ are double poles. Again reversing direction shows that all points $\zeta = -nb-m/b$ for $n,m \in \mathbb{Z}$ are (at least simple) poles, making the function infinite almost everywhere.} we can apply \eqref{shiftfde} successively to find that the function $F(\zeta)$ has zeros at $\zeta = Q + nb + m/b$ for $n,m \in \mathbb{Z}_{\geq 0}$ and simple poles at $\zeta = -nb-m/b$ for $n,m \in \mathbb{Z}_{\geq 0}$.\footnote{It is straightforward to show that these are the only zeros and poles of the solution. Suppose that to accuracy $\epsilon$, we can write a given real number $x$ as $x=n_\epsilon b+m_\epsilon/b$ for $n_\epsilon, m_\epsilon\in \mathbb{Z}$: $\left|x-n_\epsilon b-m_\epsilon /b\right| < \epsilon$. We drop the $\epsilon$ subscript from here on. If $nm>0$, then one lands on one of the known cases: $x$ is either a simple pole or a simple zero of $F$, or $x=b$ or $x=1/b$, which yield finite values for $F$. When $nm<0$, one is in the generic case. One can get to this case by first finding $x=nb$, for which $F$ is nonzero by consecutive applications of \eqref{shiftfde}, followed by $m$ applications of \eqref{shiftfde2} in the reverse direction, which again gives finite multiplicative factors at each step.} It can be smoothly extended to a meromorphic function on the complex plane, the double sine function $S_b(\zeta)$ defined in Appendix \ref{app:def}.

Notice that if $b^2 \in \mathbb{Q}$, then the above is still \emph{a} solution to the system of equations \eqref{shiftfde}, \eqref{shiftfde2}; it is just not automatically unique. In Liouville gravity language, this case corresponds to the minimal string where the matter sector is a $(p', p)$ minimal model.

\begin{figure}[!htb]
\centering
\includegraphics[width=0.3\textwidth]{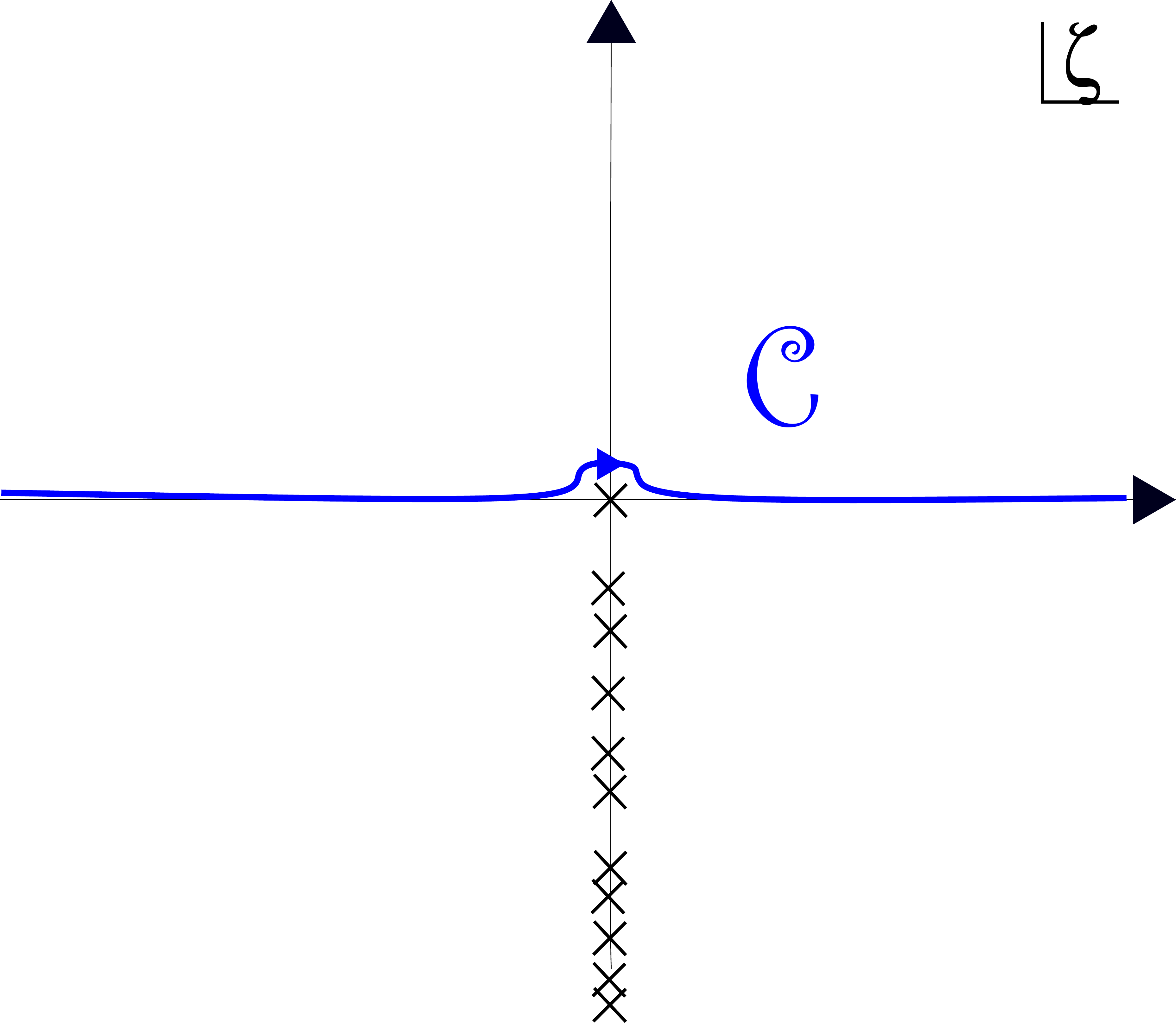}
\caption{The contour $\mathcal{C}$ used to define $\phi_{\alpha}^{\protect\+}(t)$.}
\label{contourWhit}
\end{figure}

The solution can hence be written as a contour integral in terms of $\lambda = b/2 + is$:
\begin{align}
\label{rwv}
\boxed{\phi_{\alpha}^{\+}(t) = e^{-2\pi \lambda t}\int_{\mathcal{C}} d\zeta\, g^{i\zeta} S_b(-i\zeta) e^{\pi i \alpha \zeta^2} e^{-2\pi i \zeta (t - s \alpha)}.}
\end{align}
The contour $\mathcal{C}$ follows the real axis above the poles of the function $S_b(-i\zeta)$ (Figure \ref{contourWhit}).\footnote{One can check that this is the correct contour by plugging \eqref{rwv} back into \eqref{fde} and checking that it is a solution. After some manipulation and substitution in the integral, this requires on the left-hand side a deformation of the contour downward from $ib$ to the real axis without any pole-crossing to match with the right-hand side. This in turn requires the contour to pass above all of the poles of the double sine function.}

At this point, we can already take the classical limit $b\to 0$:
\begin{align}
\phi_{\alpha}^{\+}(t(x)) \,\, \to \,\, \frac{1}{2\pi} e^{2\pi b t j}\int_{\mathcal{C}} du\, \Gamma(-iu) g^{ibu}(2\pi b^2)^{-iu}e^{-2\pi i b t u} = x^{j} e^{- \frac{\mu}{x}},
\end{align}
where $j=-1/2+ ik$ and we have set $g = (2\pi b^2 \mu)^{1/b}$. This limiting function indeed satisfies $\hat{E}^{+} \phi_{\alpha}^{\+}(x) = -\mu \phi_{\alpha}^{\+}(x)$ with $\hat{E}^{+}$ given in \eqref{clmi}, which can also be seen by taken the $b\to 0$ limit of the eigenvalue problem \eqref{eigwhit}.

Likewise, the Whittaker vector diagonalizing $(E^-)^\dagger$ is:
\begin{align}
\label{lwv}
\phi_\alpha^{\m}(t) = e^{+2\pi \lambda t}\int_{\mathcal{C}} d\zeta\, g^{i\zeta} S_b(-i\zeta) e^{\pi i \alpha \zeta^2} e^{2\pi i \zeta (t + s\alpha)}.
\end{align}
Classically, this becomes
\begin{align}
\phi_\alpha^{\m}(t(x)) \,\, \to \,\,  x^{-j} e^{- \nu x}, \qquad g=(2\pi b^2 \nu)^{1/b},
\end{align}
which satisfies $(\hat{E}^{-})^{\dagger} \phi_\alpha^{\m}(x) = \nu \phi_\alpha^{\m}(x)$ with $\hat{E}^{-}$ given in \eqref{clmi}.

The Whittaker function is found by computing the following inner product, where the Cartan element is the only nontrivial entry:
\begin{align}
\label{defWhit}
\psi^{\epsilon}_s(x) \equiv e^{\pi b x} \int_\mathbb{R} dt \, \overline{\phi_{\alpha_1}^{\m}(t)} e^{2\phi H}\phi_{\alpha_2}^{\+}(t) = e^{\pi b x} \int_\mathbb{R} dt \, \overline{\phi_{\alpha_1}^{\m}(t)} \phi_{\alpha_2}^{\+}(t+x).
\end{align}
Here, $H$ is the translation operator in \eqref{PT} given by \eqref{defH}, and we have set $x = \phi/\pi b$. This expression depends only on the difference $\epsilon \equiv \alpha_1-\alpha_2$, as will become clear. The prefactor $e^{\pi b x}$ is a choice that will allow us to work with a flat measure on $x$-space.

Again, let us first examine the classical limit, for which we get (now with the half-space coordinate $x = e^{2\pi b t}$, not to be confused with the $x$ in the previous paragraph):
\begin{align}
\label{clWhit}
\frac{1}{2\pi b} e^{-2ik \phi} \int_{\mathbb{R}^+} \frac{dx}{x} x^{-2ik} e^{-\nu x} e^{-e^{-2 \phi}\mu/x} = \frac{1}{\pi b} \left(\frac{\nu}{\mu}\right)^{ik} K_{2ik}(2\sqrt{\nu\mu}e^{-\phi}).
\end{align}
Up to the prefactor of $\frac{1}{2\pi b}$, this is the known Whittaker function for SL$^+(2,\mathbb{R})$, where the $\phi$-coordinate parametrizes the single Cartan direction on the coset manifold, with a flat measure $d\mu(\phi) = d\phi$.

Now to the $q$-deformed case. Inserting \eqref{rwv} and \eqref{lwv} into \eqref{defWhit}, the $t$-integral boils down to
\begin{equation}
\int dt\, e^{-2\pi i t (\zeta + \tilde{\zeta} + 2s)} = \delta( \zeta + \tilde{\zeta} + 2s),
\end{equation}
leading to the Whittaker function
\begin{align}
&e^{-2\pi i s x}\int_{\mathcal{C}} d\zeta\, g_\mu^{i\zeta} g_\nu^{i\zeta + 2is}S_b(-i\zeta) S_b(-i\zeta-2is)e^{-\pi i \alpha_1 (\zeta+2s)^2} e^{\pi i \alpha_2 \zeta^2}e^{2\pi i \alpha_1 s (\zeta +2s)} e^{2\pi i \alpha_2 s\zeta} e^{-2\pi i \zeta x}
\end{align}
or:\footnote{Using the limiting values
\begin{equation}
\label{limits}
S_b(bx) \to \frac{1}{\sqrt{2\pi}}(2\pi b^2)^{x-1/2}\Gamma(x), \qquad S_b\left(\frac{1}{2b}+bx\right) \to 2^{x-1/2}
\end{equation}
and the definition
\begin{equation}
\label{mbk}
K_\nu(z) = \frac{1}{4\pi i }\left(\frac{z}{2}\right)^{\nu} \int_{-i\infty}^{+i\infty} dt\, \Gamma(t) \Gamma(t-\nu) \left(\frac{z}{2}\right)^{-2t},
\end{equation}
we match onto the classical Whittaker function \eqref{clWhit}:
\begin{equation}
\lim_{b\to 0}\psi^{\epsilon}_s \left(\frac{\phi}{\pi b}\right) = \frac{1}{\pi b} \left(\frac{\nu}{\mu}\right)^{ik} K_{2 i s/b}\left(2 \sqrt{\nu\mu}e^{-\phi}\right).
\end{equation}}
\begin{align}
\label{qmelbb}
\boxed{\psi^{\epsilon}_s(x) =  e^{-2\pi i s x} \int_{\mathcal{C}} d\zeta\, g_\mu^{i\zeta} g_\nu^{i\zeta + 2is}S_b(-i\zeta) S_b(-i\zeta-2is)e^{-\pi i \epsilon(\zeta^2 + 2 s \zeta)} e^{-2\pi i \zeta x}.}
\end{align}
We indeed see that the final result depends only on the difference $\epsilon = \alpha_1-\alpha_2$. Importantly, the Whittaker vectors themselves depend on the precise way in which we realize the representation (they depend on the auxiliary coordinate $t$), whereas the Whittaker function is independent of this choice and depends only on the underlying algebra and representation. Indeed, comparing our results to those of \cite{Kharchev:2001rs}, the Whittaker vectors \eqref{rwv} and \eqref{lwv} are different, but the Whittaker function \eqref{qmelbb} is the same, up to redefining $x\to -x$.\footnote{Curiously, the realization used in \cite{Kharchev:2001rs} and written above in \eqref{KLT} leads to the Whittaker function in the so-called Gauss-Euler integral representation, while we obtain it directly in the Mellin-Barnes integral representation, which is of more interest when comparing to the relativistic Toda chain expressions of \cite{Kharchev:2001rs}.}

The Whittaker function \eqref{qmelbb} is a solution to the Casimir eigenvalue equation, which reads in this case as follows (the translation operators below act on $x$ rather than $t$):
\begin{align}
\left(T_{ib} + T_{-ib} + g_\mu^b g_\nu^b  q^{\epsilon} e^{-2\pi b x} T_{-i b \epsilon}\right)f(x) = 2 \cosh (2\pi b s) f(x).
\end{align}
In the $b\to 0$ limit, this reduces to a Schr\"odinger equation for a particle in an exponential potential, with solution \eqref{clWhit}. Of course, this second-order ODE has two independent solutions, the second one being a modified Bessel function of the first kind. It is interesting to derive the analogue of this solution in the $q$-deformed setting. We do so in Appendix \ref{app:besseli}.

\subsection{\texorpdfstring{$3j$}{3j}-Symbols or Vertex Functions}

We now compute the $3j$-symbols relevant for gravitational calculations. For this purpose, next to the above self-dual continuous series representation matrix elements, we also need a discrete representation matrix element with eigenvalue zero for the parabolic generators ($E^\pm=0$). It is actually quite easy to write down a solution to the finite difference equations \eqref{fde} that meets these demands:
\begin{equation}
\phi^{\+}(t) = e^{-2\pi \lambda t}, \qquad \phi^{\m}(t) = e^{+2\pi \lambda t}.
\end{equation}
Applying the Cartan generator to this state and dismissing a constant factor, we get:
\begin{equation}
\label{qdis}
\psi^{\text{discrete}}(x) = e^{-2\pi \lambda x}.
\end{equation}
Now taking two continuous irrep Whittaker functions \eqref{qmelbb} and one discrete series Whittaker function \eqref{qdis} (with representation label $\lambda = \beta_M$), the resulting mixed parabolic $3j$-symbol was calculated in \cite{Mertens:2020hbs} as the group (coset) integral
\begin{equation}
\label{idwhitsb}
\int_{-\infty}^{+\infty} dx \, \psi^{\epsilon}_{s_1, g_\mu g_\nu} (x) \psi^{\epsilon}_{s_2, g_\mu g_\nu} (x)^\ast e^{-2 \pi \beta_M x} = \left(\frac{g_\nu}{g_\mu}\right)^{is_1-is_2}\frac{1}{(g_\nu g_\mu)^{\beta_{M}}}\frac{S_b(\beta_M \pm is_1 \pm is_2)}{S_b(2\beta_M)},
\end{equation}
for $\epsilon = \pm 1$. To match this expression with the coupling coefficients in bosonic Liouville gravity (the second line of \eqref{final2}) that describe the coupling of an operator insertion to an in- and an out-state, we set $g_\mu = g_\nu$ equal to an arbitrary constant.

Taking the $b\to 0$ limit of both sides of \eqref{idwhitsb}, we obtain the equality
\begin{align}
\left(\frac{1}{2\sqrt{\nu\mu}}\right)^{2h}\frac{1}{(\pi b)^3} \left(\frac{\nu}{\mu}\right)^{ik_1-ik_2} &\int_\mathbb{R} dx\, K_{2ik_1}(e^{x}) K_{2ik_2}(e^{x}) e^{2hx} \label{3jcl} \\
&= \left(\frac{\nu}{\mu}\right)^{ik_1-ik_2} \frac{1}{(\nu\mu)^{h}(2\pi b)^3} \frac{\Gamma(h\pm i k_1 \pm ik_2)}{\Gamma(2h)}, \nonumber
\end{align}
which leads to the JT gravity vertex functions if we set $\mu = \nu = 1$ by convention. This constraint corresponds to the Brown-Henneaux gravitational coset constraints imposed at the holographic boundary. It would be interesting to obtain a similar understanding of the analogous $g_\mu = g_\nu$ constraint in Liouville gravity. Presumably, one route would be through its formulation as a deformed dilaton gravity model. We come back to this reformulation of Liouville (super)gravity in terms of dilaton (super)gravity in Section \ref{s:dilgrav}.

\subsection{Plancherel Measure}
\label{s:planchbos}

Finally, we can take the limit where $\beta_M \to 0$ in the $3j$-symbol squared expressions \eqref{idwhitsb}. We set $\beta_M = \epsilon$ as a regulator. Since $S_{b}(\epsilon) \to \frac{1}{2\pi \epsilon}$ and diverges, the resulting expression \eqref{idwhitsb} vanishes. An exception occurs when $s_1 =\pm s_2$. Considering the integral over $s_1$ and applying the $q$-deformed first Barnes lemma \eqref{qbarnebos}, we find
\begin{equation}
\int_{-\infty}^{+\infty} ds_1 \frac{S_b(\epsilon \pm i s_1 \pm i s_2)}{S_b(2\epsilon)} = \frac{S_b(2\epsilon) S_b(2i s_2) S_b(-2is_2) S_b(2\epsilon)}{S_b(2\epsilon) S_b(4\epsilon)} = 2 S_b(2i s_2) S_b(-2is_2).
\end{equation}
From this, we get the identity
\begin{equation}
\lim_{\epsilon \to 0}\frac{S_b(\epsilon \pm i s_1 \pm i s_2)}{S_b(2\epsilon)} = \frac{\delta(s_1 - s_2)}{4 \sinh (2\pi b s_2) \sinh \left(\frac{2\pi s_2}{b}\right)}, \qquad s_1, s_2 > 0,
\end{equation}
or
\begin{equation}
\boxed{\int_{-\infty}^{+\infty} dx \, \psi^{\epsilon}_{s_1}(x) \psi^{\epsilon}_{s_2}(x)^* = \frac{\delta(s_1 - s_2)}{4 \sinh (2\pi b s_2) \sinh \left(\frac{2\pi s_2}{b}\right)},}
\end{equation}
from which we read off the Plancherel measure:
\begin{equation}
\rho(s) \sim \sinh (2\pi b s) \sinh \left(\frac{2\pi s}{b}\right).
\end{equation}
This is indeed the measure in the result \eqref{final2}, thereby completing our identification of the ingredients of \eqref{final2} with suitable group-theoretic objects.

Our next goal will be to perform the analogous computations for U$_q(\mathfrak{osp}(1|2,\mathbb{R}))$.

\section{\texorpdfstring{U$_q(\mathfrak{osp}(1|2, \mathbb{R}))$}{Uq(osp(1|2, R))} Gravitational Matrix Element}
\label{s:three}

In this section, we generalize the previous arguments to the supergroup U$_q(\mathfrak{osp}(1|2,\mathbb{R}))$. Our main results are the Whittaker vectors \eqref{qwhitv} and the Whittaker function \eqref{qmelb}.

\subsection{\texorpdfstring{$q$}{q}-Deformed Algebra: Definition and Classical Limits}

The U$_q(\mathfrak{osp}(1|2, \mathbb{R}))$ quantum algebra consists of three generators $K, F^+, F^-$, the latter two of which are fermionic, satisfying the relations:\footnote{In the math literature, one typically uses instead
\begin{align}
\label{salgebm}
kf^\pm = q^{\pm 1} f^{\pm}k , \qquad \{f^+,f^-\} = \frac{k-k^{-1}}{q-q^{-1}},
\end{align}
which is related to our conventions by
\begin{align}
k = K^2, \qquad f^\pm = \pm 2\sqrt{2}(q^{1/2}+q^{-1/2})^{-1/2} F^\pm.
\end{align}
Our conventions are related to those in \cite{Hadasz:2013bwa}, up to rescalings of $F^\pm$ and with $\mathcal{Q}$ and $\mathcal{C}$ chosen such that the $b\to 0$ limit agrees with the classical $\mathfrak{osp}(1|2, \mathbb{R})$ algebra, as we elaborate on below.}
\begin{align}
\label{salgeb}
KF^\pm = q^{\pm\frac{1}{2}}F^{\pm}K , \qquad \{F^+, F^-\} = -\frac{K^2-K^{-2}}{8(q^{1/2}-q^{-1/2})} = -\frac{\sin 2\pi b^2 H}{8\sin \frac{\pi b^2}{2}}.
\end{align}
Again, $K = q^H$ and $q = e^{\pi ib^2}$. This algebra has a sCasimir element $\mathcal{Q}$ that commutes with $K$ and anticommutes with $F^\pm$, given by the expression \cite{Arnaudon:1996qe}
\begin{equation}
\label{Scas}
\left(\frac{1}{q^{1/2}+q^{-1/2}}\right) \mathcal{Q} = [F^-, F^+] + \frac{K^2+K^{-2}}{8(q^{1/2}+q^{-1/2})} = \frac{q^{1/2}K^2 - q^{-1/2}K^{-2}}{4(q-q^{-1})} + 2 F^- F^+.
\end{equation}
It squares to an element commuting with $K$ and $F^\pm$, the Casimir of the algebra:
\begin{equation}
\label{qsusycas}
\left(\frac{1}{q^{1/2}+q^{-1/2}}\right)^2 \mathcal{C} = \frac{qK^4+q^{-1}K^{-4}-2}{16(q-q^{-1})^2} - \frac{(qK^2+q^{-1}K^{-2})F^-F^+}{2(q^{1/2}+q^{-1/2})} - 4(F^{-})^{2}(F^{+})^{2},
\end{equation}
with relation
\begin{equation}
\label{relion}
\mathcal{Q}^2 = \mathcal{C}.
\end{equation}
In the classical limit $b\to 0$, this algebra becomes the $\mathfrak{osp}(1|2, \mathbb{R})$ Lie superalgebra:\footnote{In \cite{Fan:2021wsb}, we called this the \emph{opposite} superalgebra, since all anticommutators have an extra minus sign compared to the usual $\mathfrak{osp}(1|2, \mathbb{R})$ algebra. The principal series representations are defined by exponentiating this opposite superalgebra. The discrepancy comes from the fact that for finite irreps, the generators are supermatrices with bosonic entries, whereas for the principal series representations, one uses Grassmann-valued differential operators.}
\begin{align}
\label{salgebcl}
[H, F^\pm] = \pm\frac{1}{2}F^\pm, \qquad \{F^+, F^-\} = -\frac{1}{2}H,
\end{align}
where we include the bosonic generators $E^\pm$ via the definitions $\{F^\pm, F^\pm\} \equiv \mp \frac{1}{2} E^{\pm}$. The (s)Casimir operators then reduce to those of $\mathfrak{osp}(1|2, \mathbb{R})$:
\begin{align}
\label{Scascl}
\frac{1}{2}\mathcal{Q} \quad &\to \quad F^-F^+ - F^+F^- + \frac{1}{8} = \frac{1}{2}H + 2 F^- F^+ + \frac{1}{8}, \\
\label{cascl}
\mathcal{C} - \frac{1}{16} \quad &\to \quad H^2 + \frac{1}{2}(E^+E^- + E^-E^+) + (F^+F^- - F^-F^+),
\end{align}
adjusting for constant factors and terms.

The continuous series representations can be found by acting on the graded Hilbert space $L^2(\mathbb{R}^{1|1})$, where we write the functions in the representation space as:
\begin{equation}
\label{funcspace}
f(t,\vartheta) \in L^2(\mathbb{R}^{1|1}), \qquad  f(t,\vartheta) \equiv f_{\B}(t) + \vartheta f_{\T}(t) \equiv \left(\begin{array}{c}
f_{\B}(t) \\
\hline
f_{\T}(t)
\end{array}\right).
\end{equation}
The generators are then defined in terms of $t$-translation operators as follows:
\begin{align}
K &= T_{ib/2} \left(\begin{array}{c|c}
1 &0 \\
\hline
0 &1 
\end{array}\right), \\
F^+ &= \frac{1}{2\sqrt{2}} e^{\pi b t}\left(  \begin{array}{c|c}
0 & \frac{e^{\frac{i\pi b\lambda}{2}}T_{ib/2} - e^{-\frac{i\pi b\lambda}{2}}T_{-ib/2}}{q^{1/2}-q^{-1/2}} \\
\hline
\frac{e^{\frac{i\pi b\lambda}{2}}T_{ib/2} + e^{-\frac{i\pi b\lambda}{2}}T_{-ib/2}}{q^{1/2}+q^{-1/2}} & 0 
\end{array}\right), \\
F^- &= -\frac{1}{2\sqrt{2}} e^{-\pi b t} \left( \begin{array}{c|c}
0 & \frac{e^{-\frac{i\pi b\lambda}{2}}T_{ib/2} - e^{\frac{i\pi b\lambda}{2}}T_{-ib/2}}{q^{1/2}-q^{-1/2} }\\
\hline
\frac{e^{-\frac{i\pi b\lambda}{2}}T_{ib/2} + e^{\frac{i\pi b\lambda}{2}}T_{-ib/2}}{q^{1/2}+q^{-1/2}} & 0 
\end{array}\right).
\end{align}
One readily checks that these satisfy \eqref{salgeb}.

These generators were first written down in \cite{Hadasz:2013bwa, Pawelkiewicz:2013wga} and shown to generate the continuous self-dual representations of U$_q(\mathfrak{osp}(1|2,\mathbb{R}))$. These are again simultaneous representations of the quantum group and its dual $b\to 1/b$ (see also \cite{Ip:2013qba}), forming the modular-doubled quantum group:
\begin{equation}
\label{mdp}
\text{U}_q(\mathfrak{osp}(1|2,\mathbb{R})) \otimes \text{U}_{\tilde{q}}(\mathfrak{osp}(1|2,\mathbb{R})).
\end{equation}
As in the case of $\text{U}_q(\mathfrak{sl}(2,\mathbb{R}))$ (Section \ref{s:two}), we set $\lambda = b/2 + is$ as the representation label and impose suitable hermiticity conditions with respect to the inner product
\begin{equation}
(f,g) \equiv \int_\mathbb{R} dt\, d\vartheta\, \overline{f(t,\vartheta)} g(t,\vartheta).
\end{equation}
The generator $K$ satisfies $K = K^\dagger$, and the two fermionic generators satisfy a somewhat modified hermiticity constraint. If the functions $f_{\B}(t)$ and $f_{\T}(t)$ in the function space \eqref{funcspace} are even, then we require the following hermiticity constraint for $F^+$:\footnote{There are related statements when the parity of $f_{\B}$ or $f_{\T}$ is different, but for the purposes of computing the Whittaker vectors further on, we will only require this particular case.}
\begin{equation}
\label{dagger}
(F^+)^\dagger = \frac{1}{2\sqrt{2}} e^{\pi b t}\left(  \begin{array}{c|c}
0 & -\frac{e^{\frac{i\pi b\lambda}{2}}T_{ib/2} - e^{-\frac{i\pi b\lambda}{2}}T_{-ib/2}}{q^{1/2}-q^{-1/2}} \\
\hline
\frac{e^{\frac{i\pi b\lambda}{2}}T_{ib/2} + e^{-\frac{i\pi b\lambda}{2}}T_{-ib/2}}{q^{1/2}+q^{-1/2}} & 0 
\end{array}\right),
\end{equation}
where the top right entry picks up a minus sign but the bottom left entry does not. There is an analogous relation for $F^-$. These are the same hermiticity relations as those found in \cite{Fan:2021wsb} for $\mathfrak{osp}(1|2,\mathbb{R})$, and we will indeed see that in the $b\to 0$ limit here as well.

For completeness, the bosonic elements $E^{\pm} \equiv \mp 4(F^{\pm})^2$ can be explicitly computed as:
\begin{align}
E^{+} &= \frac{-e^{2\pi b t}}{2(q-q^{-1})}\left(  \begin{array}{c|c}
\begin{gathered} q^{1/2}(e^{i\pi b\lambda}T_{ib} + 1) \\ {} - q^{-1/2}(e^{-i\pi b\lambda}T_{-ib} + 1) \end{gathered} & 0 \\
\hline
0 & \begin{gathered} q^{1/2}(e^{i\pi b\lambda}T_{ib} - 1) \\ {} - q^{-1/2}(e^{-i\pi b\lambda}T_{-ib} - 1) \end{gathered}
\end{array}\right), \\
E^{-} &= \frac{e^{-2\pi b t}}{2(q-q^{-1})}\left(  \begin{array}{c|c}
\begin{gathered} q^{-1/2}(e^{-i\pi b\lambda}T_{ib} + 1) \\ {} - q^{1/2}(e^{i\pi b\lambda}T_{-ib} + 1) \end{gathered} & 0 \\
\hline
0 & \begin{gathered} q^{-1/2}(e^{-i\pi b\lambda}T_{ib} - 1) \\ {} - q^{1/2}(e^{i\pi b\lambda}T_{-ib} - 1) \end{gathered}
\end{array}\right).
\end{align}
These bosonic elements are antihermitian: $(E^+)^\dagger = -E^+$ and $(E^-)^\dagger = -E^-$, just as for U$_q(\mathfrak{sl}(2,\mathbb{R}))$ (as discussed in Section \ref{s:two}).

The (s)Casimir elements \eqref{Scas} and \eqref{qsusycas} are readily evaluated to be
\begin{align}
\label{Scasexpl}
\mathcal{Q} &= \frac{\sin \pi b (\lambda-b/2)}{4\sin \frac{\pi b^2}{2}}\left(\begin{array}{c|c}
1 &0 \\
\hline
0 &-1 
\end{array}\right) = \frac{i\sinh \pi b s}{4\sin \frac{\pi b^2}{2}}\left(\begin{array}{c|c}
1 &0 \\
\hline
0 &-1 
\end{array}\right), \\
\label{Casexpl}
\mathcal{C} &= \frac{\sin^2 \pi b (\lambda-b/2)}{16\sin^2 \frac{\pi b^2}{2}}\left(\begin{array}{c|c}
1 &0 \\
\hline
0 &1 
\end{array}\right) = -\frac{\sinh^2 \pi b s}{16\sin^2 \frac{\pi b^2}{2}}\left(\begin{array}{c|c}
1 &0 \\
\hline
0 &1 
\end{array}\right),
\end{align}
which satisfy the relation \eqref{relion}. This expression for $\mathcal{C}$ matches the energy variables in the exponentials in \eqref{final1}, up to a conventional minus sign in the definition of $\mathcal{C}$.

To match with the classical $b \to 0$ limit, we identify $\lambda = -2bj$ and $x=e^{2\pi b t}$. In this limit, the quantum algebra leads to the following ``symmetric'' realization of the principal series representations of $\mathfrak{osp}(1|2,\mathbb{R})$:
\begin{align}
\hat{H} &= \left(\begin{array}{c|c}
x\partial_x &0 \\
\hline
0 &x\partial_x 
\end{array}\right), \\
\label{hatFp}
\hat{F}^+ &= \frac{1}{2\sqrt{2}}\left(  \begin{array}{c|c}
0 & 2x^{3/2}\partial_x -2j x^{1/2} \\
\hline
x^{1/2} & 0 
\end{array}\right), \\
\hat{F}^- &= -\frac{1}{2\sqrt{2}}\left(  \begin{array}{c|c}
0 & 2x^{1/2}\partial_x +2j x^{-1/2} \\
\hline
x^{-1/2} & 0 
\end{array}\right), \\
\hat{E}^+ &= \left(\begin{array}{c|c}
-x^2\partial_x + (j - \frac{1}{2})x &0 \\
\hline
0 & -x^2\partial_x  + jx
\end{array}\right), \\
\hat{E}^- &= \left(\begin{array}{c|c}
\partial_x  + (j - \frac{1}{2})x^{-1} &0 \\
\hline
0 &\partial_x + jx^{-1}
\end{array}\right), 
\end{align}
which satisfy the Lie superalgebra
\begin{alignat}{2}
[H, E^\pm] &= \pm E^\pm, \quad & [E^+, E^-] &= 2H, \nonumber \\
[H, F^\pm] &= \pm\frac{1}{2}F^\pm, \quad & [E^\pm, F^\mp] &= -F^\pm, \label{osplinerev} \\
\{F^+, F^-\} &= -\frac{1}{2}H, \quad & \{F^\pm, F^\pm\} &= \mp \frac{1}{2}E^\pm, \nonumber
\end{alignat}
with sCasimir
\begin{equation}
\frac{1}{2}\mathcal{Q} = F^-F^+ - F^+ F^- +\frac{1}{8} = -\left(\frac{j}{2}+\frac{1}{8} \right)\left(\begin{array}{c|c}
1 &0 \\
\hline
0 &-1 
\end{array}\right).
\end{equation}
The generators $\hat{H}$, $\hat{E}^+$, $\hat{E}^-$ are antihermitian on $\mathbb{R}^+$ provided that $j = -1/4 + ik/2$ with $k \in \mathbb{R}$, so we indeed recover the principal series representations of OSp$^+(1|2,\mathbb{R})$. The odd generators $\hat{F}^\pm$ satisfy the same modified version of hermiticity when acting on functions on superspace that have bosonic components, just like \eqref{dagger} above. We refer to Footnote 27 of \cite{Fan:2021wsb} for the corresponding statement regarding the Borel-Weil realization of $\mathfrak{osp}(1|2,\mathbb{R})$.

It is possible to write down a different realization of these $q$-deformed representations that limits to the standard Borel-Weil realization of $\mathfrak{osp}(1|2,\mathbb{R})$. We present this in Appendix \ref{alternative}.

\subsection{Whittaker Vectors and Whittaker Function}

To find the Whittaker vector, we will ``diagonalize'' the fermionic generator $F^+$. The bosonic generator $E^+=-4(F^+)^2$ is then automatically diagonalized as well. Writing the superspace Whittaker vector as
\begin{align}
\Phi_\alpha(t,\vartheta) = \phi_{\alpha,\B}(t) + \vartheta  \phi_{\alpha,\T}(t) = \left(\begin{array}{c} \phi_{\alpha,\B}(t) \\  \phi_{\alpha,\T}(t) \end{array}\right),
\end{align}
we define the Whittaker vector of the modular double of U$_q(\mathfrak{osp}(1|2,\mathbb{R}))$ as the solution to the following eigenvalue problem:
\begin{align}
F^+ \phi^{\+}_{\alpha,\T}(t) &= \frac{i}{2\sqrt{2}}\frac{\epsilon g^b}{q^{1/2}-q^{-1/2}} q^{2\alpha H} \phi^{\+}_{\alpha,\B}(t), \quad F^+ \phi^{\+}_{\alpha,\B}(t) = \frac{1}{2\sqrt{2}}\frac{\epsilon g^b}{q^{1/2}+q^{-1/2}} q^{2\alpha H} \phi^{\+}_{\alpha,\T}(t), \nonumber \\
\tilde{F}^+ \phi^{\+}_{\alpha,\T}(t) &= \frac{i}{2\sqrt{2}}\frac{\epsilon g^{1/b}}{\tilde{q}^{1/2}-\tilde{q}^{-1/2}} \tilde{q}^{2\alpha \tilde{H}} \phi^{\+}_{\alpha,\B}(t), \quad \tilde{F}^+ \phi^{\+}_{\alpha,\B}(t) = \frac{1}{2\sqrt{2}}\frac{\epsilon g^{1/b}}{\tilde{q}^{1/2}+\tilde{q}^{-1/2}} \tilde{q}^{2\alpha \tilde{H}} \phi^{\+}_{\alpha,\T}(t), \label{Whitsusydef}
\end{align}
where we have the freedom to choose a sign $\epsilon = \pm 1$ while leaving invariant the eigenvalue of $E^+ = -4(F^+)^2$. Indeed, applying \eqref{Whitsusydef} twice, we get
\begin{align}
E^+ \Phi^{\+}_{\alpha}(t,\vartheta) &=  -\frac{i}{2}\frac{g^{2b}}{q-q^{-1}} q^{\alpha}q^{4\alpha H} \Phi^{\+}_{\alpha}(t,\vartheta), \\
\tilde{E}^+ \Phi^{\+}_{\alpha}(t,\vartheta) &=  -\frac{i}{2}\frac{g^{2b}}{\tilde{q}-\tilde{q}^{-1}} \tilde{q}^{\alpha}\tilde{q}^{4\alpha \tilde{H}} \Phi^{\+}_{\alpha}(t,\vartheta),
\end{align}
which form the analogues of the bosonic relations \eqref{eigwhit}. Strictly speaking, \eqref{Whitsusydef} is not a diagonalization. We simply demand that the transformation of the top component be proportional to the bottom component, and vice versa. The proportionality factor is not the same in both cases, but upon squaring to get the bosonic generator, one does get a genuine diagonalization. For such a system, it is always possible to rescale the eigenvectors to obtain a genuine diagonalization problem for either the quantum group \emph{or} its dual, but not for both. This is why we prefer to write \eqref{Whitsusydef} as above for the general case. When comparing with the classical $b\to 0$ results, the dual quantum group scales out, and it is convenient to rescale the eigenvectors as follows:
\begin{align}
\phi^{\prime\+}_{\alpha,\T}(t) \equiv (8\pi b^2)^{1/4} \left( \frac{q^{1/2}-q^{-1/2}}{i(q^{1/2}+q^{-1/2})}\right)^{1/4} \phi^{\+}_{\alpha,\T}(t), \nonumber \\
\phi^{\prime\+}_{\alpha,\B}(t) \equiv (8\pi b^2)^{1/4} \left( \frac{i(q^{1/2}+q^{-1/2})}{q^{1/2}-q^{-1/2}}\right)^{1/4} \phi^{\+}_{\alpha,\B}(t). \label{whitresc}
\end{align}
This maps the first eigenvalue system in \eqref{Whitsusydef} into a genuine diagonalization problem:
\begin{align}
F^+ \phi^{\prime\+}_{\alpha,\T}(t) &=  \frac{\epsilon g^b}{2\sqrt{2}}\sqrt{\frac{i}{q - q^{-1}}} q^{2\alpha H} \phi^{\prime\+}_{\alpha,\B}(t), \qquad F^+ \phi^{\prime\+}_{\alpha,\B}(t) =  \frac{\epsilon g^b}{2\sqrt{2}}\sqrt{\frac{i}{q - q^{-1}}} q^{2\alpha H} \phi^{\prime\+}_{\alpha,\T}(t),
\end{align}
with the same proportionality factor in both equations.

The full eigenvalue problem \eqref{Whitsusydef} can be written explicitly as the following set of coupled difference equations:
\begin{align}
e^{\frac{\pi i b \lambda}{2}} \phi^{\+}_{\alpha,\T}\left(t+\frac{ib}{2}\right) - e^{-\frac{\pi i b \lambda}{2}} \phi^{\+}_{\alpha,\T}\left(t-\frac{ib}{2}\right) &= i \epsilon g^b  e^{-\pi b t} \phi^{\+}_{\alpha,\B}(t+ib \alpha), \nonumber \\
e^{\frac{\pi i b \lambda}{2}} \phi^{\+}_{\alpha,\B}\left(t+\frac{ib}{2}\right) + e^{-\frac{\pi i b \lambda}{2}} \phi^{\+}_{\alpha,\B}\left(t-\frac{ib}{2}\right) &= \epsilon g^b  e^{-\pi b t} \phi^{\+}_{\alpha,\T}(t+ib \alpha),  \nonumber \\
e^{\frac{\pi i \lambda}{2 b}} \phi^{\+}_{\alpha,\T}\left(t+\frac{i}{2b}\right) - e^{-\frac{\pi i \lambda}{2 b}} \phi^{\+}_{\alpha,\T}\left(t-\frac{i}{2b}\right) &= i \epsilon g^{\frac{1}{b}} e^{-\frac{\pi t}{b}} \phi^{\+}_{\alpha,\B}\left(t+\frac{i \alpha}{b}\right),  \nonumber \\
e^{\frac{\pi i \lambda}{2 b}} \phi^{\+}_{\alpha,\B}\left(t+\frac{i}{2b}\right) + e^{-\frac{\pi i \lambda}{2 b}} \phi^{\+}_{\alpha,\B}\left(t-\frac{i}{2b}\right) &= \epsilon g^{\frac{1}{b}}  e^{-\frac{\pi t}{b}} \phi^{\+}_{\alpha,\T}\left(t+\frac{i \alpha}{b}\right), \label{susyfde}
\end{align}
where the last two equations are the $b \to 1/b$ counterparts of the first two. The solution to the system \eqref{susyfde} is:
\begin{align}
\label{qwhitv}
\phi^{\+}_{\alpha,\B}(t) &= e^{-\pi \lambda t}\int_{\mathcal{C}} d\zeta\, g^{i\zeta} S_{\R}(-i\zeta) e^{\frac{1}{2}\pi i \alpha \zeta^2} e^{-\pi i \zeta (t - s \alpha)}, \\
\label{qwhitv2}
\phi^{\+}_{\alpha,\T}(t) &= \epsilon_{\+} e^{-\pi \lambda t}\int_{\mathcal{C}} d\zeta\, g^{i\zeta} S_{\NS}(-i\zeta) e^{\frac{1}{2}\pi i \alpha \zeta^2} e^{-\pi i \zeta (t - s \alpha)},
\end{align}
where the contour $\mathcal{C}$ is the same as in the bosonic case in Section \ref{s:whitbos}. The supersymmetric double sine functions are defined by
\begin{align}
\label{Sindef}
S_{\NS}(x) = S_b\left(\frac{x}{2}\right) S_b\left(\frac{x}{2} + \frac{Q}{2}\right), \qquad S_{\R}(x) = S_b\left(\frac{x}{2} + \frac{b}{2}\right) S_b\left(\frac{x}{2} + \frac{1}{2b}\right),
\end{align}
and they satisfy the fundamental shift properties
\begin{alignat}{2}
S_{\NS}(x+b) &= 2 \cos \left(\frac{\pi b x}{2} \right) S_{\R}(x), \qquad & S_{\NS}\left(x+\frac{1}{b}\right) &= 2 \cos \left(\frac{\pi x}{2b} \right) S_{\R}(x), \nonumber \\
S_{\R}(x+b) &= 2 \sin \left(\frac{\pi b x}{2} \right) S_{\NS}(x), \qquad & S_{\R}\left(x+\frac{1}{b}\right) &= 2 \sin \left(\frac{\pi x}{2b} \right) S_{\NS}(x). \label{susyshift}
\end{alignat}
One can prove uniqueness of this solution in the same way as in the bosonic case of Section \ref{s:whitbos}. Assuming continuity and incommensurability of $2b$ and $1/2b$, one proves directly that the Fourier transform of the system \eqref{susyfde} leads to the shift relations \eqref{susyshift} of the supersymmetric double sine functions, which have a unique solution up to a single overall normalization.

In the classical $b \to 0$ limit, we set $g_\mu = (4\pi b^2 \mu)^{\frac{1}{2b}}$ and $g_\nu =  (4\pi b^2 \nu)^{\frac{1}{2b}}$, and we obtain for the Whittaker vectors \eqref{whitresc} upon inserting the solutions \eqref{qwhitv} and \eqref{qwhitv2} that
\begin{align}
\phi^{\prime\+}_{\alpha,\B}(t) &\to \sqrt{2\mu}x^{j-1/2}e^{-\frac{\mu}{x}}, \\
\phi^{\prime\+}_{\alpha,\T}(t) &\to x^{j}e^{-\frac{\mu}{x}},
\end{align}
which match with a direct diagonalization of the differential operator \eqref{hatFp}: $\hat{F}^+ \Phi^{\prime\+}(t) = \frac{\sqrt{\mu}}{2} \Phi^{\prime\+}(t)$.

Analogously, we get when diagonalizing $(F^-)^{\dagger}$ that
\begin{align}
\phi^{\m}_{\alpha,\B}(t)& = e^{+\pi \lambda t}\int_{\mathcal{C}} d\zeta\, g^{i\zeta} S_{\R}(-i\zeta) e^{\frac{1}{2}\pi i \alpha \zeta^2} e^{\pi i \zeta (t + s \alpha)}, \\
\phi^{\m}_{\alpha,\T}(t)& = \epsilon_{\m} e^{+\pi \lambda t}\int_{\mathcal{C}} d\zeta\, g^{i\zeta} S_{\NS}(-i\zeta) e^{\frac{1}{2}\pi i \alpha \zeta^2} e^{\pi i \zeta (t + s \alpha)}.
\end{align}
As mentioned above, these solutions automatically diagonalize $E^+$ and $(E^-)^\dagger$.

Next, we define the Whittaker function by evaluating the integral
\begin{align}
\psi^{\epsilon,\pm}_{s,g_\mu g_\nu} (x) &\equiv e^{\frac{\pi b x}{2}} \int dt\, d\vartheta \, \overline{\Phi^{\m}_{\alpha_1}(t,\vartheta)} e^{2\pi b x H} \Phi^{\+}_{\alpha_2}(t,\vartheta) \nonumber \\
&= e^{\frac{\pi b x}{2}} \int dt\, d\vartheta \, \overline{\Phi^{\m}_{\alpha_1}(t,\vartheta)} \Phi^{\+}_{\alpha_2}(t+x,\vartheta) \label{whitintegral} \\
&= e^{\frac{\pi b x}{2}} \int dt \left(\epsilon_{\+} \overline{\phi^{\m}_{\alpha_1,\B}(t)} \phi^{\+}_{\alpha_2,\T}(t+x) + \epsilon_{\m}\overline{\phi^{\m}_{\alpha_1,\T}(t)} \phi^{\+}_{\alpha_2,\B}(t+x) \right), \nonumber
\end{align}
where we have used the definition \eqref{defH} of $H$. This leads to two independent Whittaker functions, depending on whether $\epsilon_{\+}$ and $\epsilon_{\m}$ have the same or opposite signs. These two cases are denoted by the $\pm$ superscript on the left-hand side. We choose to set $\epsilon_{\+}=1$ and match $\epsilon_{\m}$ with the sign on the left.

One can immediately obtain a system of difference equations satisfied by the Whittaker function by inserting the sCasimir $\mathcal{Q}$:
\begin{equation}
e^{\frac{\pi b x}{2}} \int dt\, d\vartheta\, \overline{\Phi^{\m}(t,\vartheta)} e^{2\pi b x H}\mathcal{Q} \Phi^{\+}(t,\vartheta) = -\frac{i\sinh \pi b s}{4\sin \frac{\pi b^2}{2}}\psi^{\epsilon,\mp}_s (x),
\end{equation}
where we used the expression \eqref{Scasexpl}. Inserting instead \eqref{Scas}, and utilizing that the Whittaker vectors diagonalize $F^+$ for the ket and $(F^-)^\dagger$ for the bra, we obtain the equality
\begin{equation}
\label{fdewhit}
\left(\frac{e^{ib\partial_x} - e^{-ib\partial_x}}{q^{1/2} - q^{-1/2}} \mp \frac{i g_\mu^{b}g_\nu^b}{q^{1/2} - q^{-1/2}} e^{-\pi b x} \right) \psi^{\epsilon=0,\pm}_s (x) = -\frac{i\sinh \pi b s}{\sin \frac{\pi b^2}{2}}\psi^{\epsilon=0,\mp}_s (x).
\end{equation}
For simplicity, we have set $\epsilon=0$ here. By explicitly computing the Whittaker function, we will check that it is a solution to this system of difference equations.

Evaluating the integral \eqref{whitintegral} requires a calculation identical to the one presented above for the bosonic case. The result is our proposal for the Whittaker function of U$_q(\mathfrak{osp}(1|2,\mathbb{R}))$:
\begin{equation}
\begin{aligned}
\psi^{\epsilon,\pm}_{s,g_\mu g_\nu}(x) = e^{-\pi i  s x} \int_{-\infty}^{+\infty} & d\zeta\, g_\mu^{i\zeta}g_\nu^{i\zeta +2 is} e^{-\pi i \frac{\epsilon}{2} (\zeta^2 + 2s \zeta)} e^{-\pi i \zeta x} \\
&\times\left[S_{\NS}(-i\zeta) S_{\R}(-2i s -i \zeta ) \pm S_{\R}(-i\zeta) S_{\NS}(-2i s -i \zeta )\right].
\end{aligned}
\label{qmelb}
\end{equation}
This function satisfies a system of difference equations:\footnote{For convenience, we note that
\begin{align}
2\sinh (\pi b (\zeta +s)) &= 4 \cosh \frac{\pi b}{2}(\zeta +2s) \sinh \frac{\pi b \zeta}{2} +2 \sinh \pi b s \\
&= 4 \sinh \frac{\pi b}{2}(\zeta +2s) \cosh \frac{\pi b \zeta}{2} - 2 \sinh \pi b s.
\end{align}}
\begin{align}
\label{susyfde2}
\left( T_{ib} - T_{-ib}- ig_\mu^b g_\nu^b e^{-\pi b x} q^{\epsilon/2} T_{-i b\epsilon} \right) \psi^{\epsilon,+}_{s,g_\mu g_\nu}(x) &= 2 \sinh \pi b s \, \psi^{\epsilon,-}_{s,g_\mu g_\nu}(x), \nonumber \\
\left( T_{ib} - T_{-ib} + ig_\mu^b g_\nu^b e^{-\pi b x} q^{\epsilon/2} T_{-i b\epsilon} \right) \psi^{\epsilon,-}_{s,g_\mu g_\nu}(x) &= 2 \sinh \pi b s \, \psi^{\epsilon,+}_{s,g_\mu g_\nu}(x),
\end{align}
which match with \eqref{fdewhit}, as announced before. Combining them, we obtain the decoupled second-order difference equation(s)
\begin{align}
\big( T_{2ib} +T_{-2ib} - 2 &+ g_\mu^{2b} g_\nu^{2b} e^{-2\pi b x} q^{2\epsilon} T_{-2i b\epsilon} \\
&\pm ig_\mu^b g_\nu^b e^{-\pi b x}q^{\epsilon/2}(q-1) T_{-ib(\epsilon+1)} \nonumber \\
&\pm ig_\mu^b g_\nu^b e^{-\pi b x}q^{\epsilon/2}(1-q^{-1}) T_{-ib(\epsilon-1)} \big) \psi^{\epsilon,\pm}_{s,g_\mu g_\nu}(x) = 4 \sinh^2 \pi b s \, \psi^{\epsilon,\pm}_{s,g_\mu g_\nu}(x). \nonumber
\end{align}
This equation can be found by applying the sCasimir $\mathcal{Q}$ again to the difference equation \eqref{fdewhit}, and is interpretable as the Casimir eigenvalue equation obtained by applying the expression \eqref{qsusycas} for $\mathcal{C}$ directly. In the $b \to 0$ limit, all of these equations become differential equations, and in particular, the Casimir equation reduces to a Schr\"odinger problem for a particle in a Morse potential \cite{Fan:2021wsb}.
 
The expression \eqref{qmelb} has the correct classical $b\to 0$ limit, as we already pointed out in \cite{Fan:2021wsb}. We define the new variables
\begin{equation}
\zeta = 2ibt, \qquad x = \phi/\pi b, \qquad s = -bk.
\end{equation}
The double sine functions \eqref{Sindef} have the following small-$b$ limits:
\begin{align}
S_{\NS}(bx) \, &\to \, \frac{1}{\sqrt{2\pi}} 2^{\frac{x}{2}} (2\pi b^2)^{\frac{x}{2}-\frac{1}{2}} \Gamma\left(\frac{x}{2} \right), \\
S_{\R}(bx) \, &\to \, \frac{1}{\sqrt{2\pi}} 2^{\frac{x}{2} - \frac{1}{2}} (2\pi b^2)^{\frac{x}{2}} \Gamma\left(\frac{x}{2} + \frac{1}{2} \right).
\end{align}
Using the integrals
\begin{align}
e^{ik\phi} \int_{i\mathbb{R}} dt\, \Gamma(t) \Gamma(t+ik+1/2) e^{2\phi t} &= 4\pi ie^{-\phi/2} K_{-ik-1/2}(2 e^{-\phi}), \\
e^{ik\phi} \int_{i\mathbb{R}} dt\, \Gamma(t+1/2) \Gamma(t+ik) e^{2\phi t} &= 4\pi ie^{-\phi/2} K_{-ik+1/2}(2 e^{-\phi}),
\end{align}
we reproduce the OSp$^+(1|2,\mathbb{R})$ Whittaker functions:
\begin{equation}
\psi^{\epsilon, \pm}_{s,g_\mu g_\nu}\left(x\right) \to e^{-\phi/2} (\mu\nu)^{1/4}\left(\frac{\mu}{\nu}\right)^{ik/2}\left(K_{ik+1/2}(2 \sqrt{\mu\nu}e^{-\phi}) \pm K_{-ik+1/2}(2 \sqrt{\mu\nu}e^{-\phi}) \right),
\end{equation}
as determined before in Section 4.3 of \cite{Fan:2021wsb}.

\subsection{\texorpdfstring{$3j$}{3j}-Symbols or Vertex Functions}

Let us now evaluate 
\begin{equation}
\int_{-\infty}^{+\infty} dx\, \psi^{\epsilon}_{s_1,g_\mu g_\nu} (x)  \psi^{\epsilon}_{s_2,g_\mu g_\nu} (x)^* e^{-\beta_M \pi x}.
\end{equation}
Just as in the bosonic case, we focus on the particular deformation where $\epsilon = \pm 1$. Only these values seem to reproduce the Liouville (super)gravity amplitudes; the deeper reason for this eludes us. The result of this calculation should yield the vertex function of $\mathcal{N}=1$ Liouville supergravity boundary correlators. Mimicking the argument in the bosonic case, we start with
\begin{equation}
\int_{-\infty}^{+\infty} dx\, e^{\pi i x (-s_1 + s_2 -  \zeta_1 +  \zeta_2) - \beta_M \pi x} = 2\delta(-\zeta_1+\zeta_2 - s_1 + s_2 + i \beta_M).
\end{equation}
We get four terms of the type
\begin{align}
&e^{-\pi i \frac{\epsilon}{2}(\beta_M^2-s_1^2+s_2^2+2i s_1 \beta_M)}\int_{-\infty}^{+\infty} d\zeta_1\, e^{\pi \epsilon \beta_M \zeta_1} \\
&\times S_{\NS}(-i\zeta_1) S_{\R}(-i \zeta_1-2is_1) S_{\NS}(i\zeta_1+is_1 -is_2 + \beta_M)S_{\R}(i\zeta_1+is_1 + is_2 + \beta_M), \nonumber
\end{align}
where the other terms have other combinations of $S_{\NS}$ and $S_{\R}$ on the second line. In order to evaluate this integral, we make use of a $q$-deformed supersymmetric version of the Barnes identity, which we provide in Appendix \ref{app:qbarnes}. In particular, using \eqref{qbarnes}, we can express the integral as a sum of two terms:
\begin{align}
\label{idwhitssb}
&\int_{-\infty}^{+\infty} dx \hspace{0.1cm} \psi^{\epsilon}_{s_1,g_\mu g_\nu} (x) \psi^{\epsilon}_{s_2,g_\mu g_\nu} (x)^* e^{-2 \beta_M \pi x} =  \left(\frac{g_\mu}{g_\nu}\right)^{is_2-is_1}\frac{4}{(g_\mu g_\nu)^{\beta_{M}}} \times {} \\
&\left[\frac{S_{\R}(\beta_M \pm i( s_1 +  s_2))S_{\NS}(\beta_M \pm i( s_1 -  s_2))}{S_{\NS}(2\beta_M)} + \frac{S_{\NS}(\beta_M \pm i( s_1 +  s_2))S_{\R}(\beta_M \pm i(s_1 -  s_2))}{S_{\NS}(2\beta_M)}\right]. \nonumber
\end{align}
Upon setting $g_\mu = g_\nu$, this indeed coincides with the vertex function obtained in $\mathcal{N}=1$ Liouville supergravity \cite{Mertens:2020pfe} and written in equation \eqref{final1}.

\subsection{Plancherel Measure}
\label{s:planch}

We can also take the limit where $\beta_M \to 0$. Since $S_{\NS}(\epsilon) \to \frac{1}{\pi \epsilon}$, the above vertex function becomes zero except when $s_1 =\pm s_2$. Setting $\beta_M = \epsilon$ and then taking the integral over the $s_1$-variable, we obtain:
\begin{align}
\int_{-\infty}^{+\infty} ds_1 & \left[\frac{S_{\R}(\epsilon \pm i( s_1 +  s_2))S_{\NS}(\epsilon \pm i( s_1 -  s_2))}{S_{\NS}(2\epsilon)} + \frac{S_{\NS}(\epsilon \pm i( s_1 +  s_2))S_{\R}(\epsilon \pm i(s_1 -  s_2))}{S_{\NS}(2\epsilon)}\right] \nonumber \\
&= \frac{2S_{\NS}(2\epsilon )S_{\R}( 2i  s_2) S_{\R}( -2i  s_2)S_{\NS}(2\epsilon )}{S_{\NS}(4\epsilon)S_{\NS}(2\epsilon)} = 4 S_{\R}( 2i  s_2) S_{\R}( -2i  s_2),
\end{align}
where we again used the $q$-Barnes superlemma \eqref{qbarnes}. Hence the quantity in brackets has $\delta$-function support:
\begin{align}
\lim_{\epsilon \to 0} & \left[\frac{S_{\R}(\epsilon \pm i( s_1 +  s_2))S_{\NS}(\epsilon \pm i( s_1 -  s_2))}{S_{\NS}(2\epsilon)} + \frac{S_{\NS}(\epsilon \pm i( s_1 +  s_2))S_{\R}(\epsilon \pm i(s_1 -  s_2))}{S_{\NS}(2\epsilon)}\right] \nonumber \\
&= \frac{\delta(s_1-s_2)}{2\cosh \frac{\pi s_2}{b} \cosh \pi b s_2}, \qquad s_1, s_2 > 0.
\end{align}
This confirms that no insertion is present in this limit, and that the Whittaker functions defined above give the Plancherel measure in the sense that
\begin{align}
\boxed{\int_{-\infty}^{+\infty} dx \, \psi^{\epsilon}_{s_1,g_\mu g_\nu} (x) \psi^{\epsilon}_{s_2,g_\mu g_\nu} (x)^* = \frac{2\delta(s_1-s_2)}{\cosh \frac{\pi s_2}{b} \cosh \pi b s_2}.}
\end{align}
This Plancherel measure defines the density of black hole states in the gravitational interpretation of this model:
\begin{equation}
\rho(s) \sim \cosh \left(\frac{\pi s}{b} \right)\cosh (\pi b s),
\end{equation}
as found in equation \eqref{final1}.

In the study of $\mathcal{N}=1$ Liouville supergravity, there is a second sector of amplitudes where the local fermionic boundary condition at the boundary of the disk is $\psi = \eta \bar{\psi}$ with $\eta=-1$ and $\psi$ is the spin-$1/2$ field accompanying the Liouville field. We refer the reader to \cite{Mertens:2020pfe} for the precise amplitudes in this second sector where $\eta=-1$. We have chosen $\eta=+1$ everywhere for our story up to this point, since only this sector makes contact with $\mathfrak{osp}(1|2,\mathbb{R})$ objects in the $b\to 0$ limit. For the $\eta=-1$ sector, one obtains instead a linear combination of $\mathfrak{sl}(2,\mathbb{R})$ objects in the $b\to 0$ limit. This suggests that if these amplitudes have a quantum group interpretation in terms of representation theory, then it must be one without a classical $b\to 0$ counterpart. We have not reached a satisfying understanding of this situation. Nonetheless, we can guess a ``Whittaker function'' that does the job in producing the correct super-Liouville amplitudes. We present it in Appendix \ref{app:NS}.

\section{Liouville Gravity as 2d Dilaton Gravity}
\label{s:dilgrav}

In the previous two sections, we uncovered the underlying U$_q(\mathfrak{sl}(2,\mathbb{R}))$ and U$_q(\mathfrak{osp}(1|2,\mathbb{R}))$ structure of bosonic and $\mathcal{N}=1$ supersymmetric Liouville gravity amplitudes, respectively. In this section, we attempt to explain this structure by reinterpreting the Liouville gravity model directly in terms of dilaton (super)gravity with a modified (relative to JT) dilaton potential, after which we perform the quantum analysis of this theory in the Poisson sigma model framework and uncover the same $q$-deformed algebra as a symmetry.

\subsection{From Liouville Gravity to Dilaton Gravity}

There is a relatively direct way to relate Liouville gravity to dilaton gravity models \cite{StanfordSeiberg, Mertens:2020hbs} (see also \cite{Goel:2020yxl, Suzuki:2021zbe}). In the bosonic case, the argument was presented in Appendix F of \cite{Mertens:2020hbs}, and we summarize it here.

The starting point is the Lagrangian description of Liouville gravity, for which we provide some review in Appendix \ref{app:bosonic}. In this language, the argument proceeds by writing the matter sector in terms of a timelike Liouville field.\footnote{This might seem like a restriction; in particular, the minimal models are at first sight not contained in this description. However, recent work has shown how, in the bosonic case, the minimal models can be described using timelike Liouville CFT \cite{Kapec:2020xaj}.} The Liouville and matter sectors then have a bulk action
\begin{equation}
S = \frac{1}{4\pi} \int_{\Sigma} d^2 x\left[ (\hat{\nabla} \phi)^2 + 4 \pi \mu e^{2 b \phi} \right] + \frac{1}{4\pi} \int_{\Sigma} d^2 x\left[ -(\hat{\nabla} \chi)^2 + 4 \pi \mu_M e^{2 b \chi} \right],
\end{equation} 
where we have chosen a flat reference metric $\hat{g}$ for simplicity. Classically, the relation between sinh dilaton gravity and Liouville gravity is \cite{Mertens:2020hbs}
\begin{align}
\label{bosredef}
\phi = \rho/b - \pi b\Phi, \qquad \chi = \rho/b + \pi b\Phi,
\end{align}
mapping the Liouville field $\phi$ and the timelike Liouville field $\chi$ into the conformal factor of the dilaton gravity metric $ds^2 = e^{2\rho} dz\, d\bar{z}$ and the dilaton field $\Phi$. The bulk action then becomes, with $\mu_M = -\mu$:\footnote{The $bc$ ghost CFT that accompanies the non-critical string is also present in this dilaton gravity language, arising from gauge-fixing the dilaton gravity metric to conformal gauge.}
\begin{align}
S &= - \int d^2 x\, \partial_\mu \Phi \partial^\mu \rho  - 2\mu \int d^2 x \, e^{2 \rho} \sinh 2\pi b^2 \Phi \\
\label{actgen} 
&= -\frac{1}{2}\int d^2 x\sqrt{g} \left[\Phi R + W(\Phi)\right] - \oint dx \sqrt{h} \, \Phi K,
\end{align}
where $W(\Phi) = \frac{\sinh 2\pi b^2 \Phi}{\sin \pi b^2}$ provided that we identify the bulk Liouville cosmological constant as
\begin{equation}
\label{mu}
\mu = \frac{1}{4\sin \pi b^2} \quad \implies \quad \kappa \equiv \sqrt{\frac{\mu}{\sin \pi b^2}} = \frac{1}{2 \sin \pi b^2}.
\end{equation}

For the $\mathcal{N}=1$ supersymmetric case, if we work directly in superspace, then we can proceed in an almost identical fashion. Describing the $\hat{c}_m<1$ matter sector of the $\mathcal{N}=1$ Liouville supergravity model with a timelike version of the super-Liouville CFT, we write the matter + Liouville sector in superspace as:
\begin{equation}
S = \frac{1}{4\pi}\int d^2 x\, d^2\theta \left[D \Phi_L \bar{D} \Phi_L + 8\pi  \mu e^{b\Phi_L}\right] - \frac{1}{4\pi}\int d^2 x\, d^2\theta \left[D \chi \bar{D} \chi + 8\pi \mu e^{b\chi}\right].
\end{equation}
Defining the field combinations
\begin{equation}
\label{fredef}
\Phi_L = \Sigma/b - 2\pi b\Phi, \qquad \chi = \Sigma/b + 2\pi b\Phi,
\end{equation}
we can rewrite the action in a suggestive way as:
\begin{equation}
S = -2\int d^2 x\, d^2\theta \left[D \Sigma \bar{D} \Phi + 2 \mu e^{\Sigma}\sinh 2\pi b^2 \Phi \right].
\end{equation}
The supercurvature $R_{+-}$ is defined in terms of the superconformal parameter $\Sigma$ as
\begin{equation}
R_{+-} \equiv 2e^{-\Sigma} D \bar{D} \Sigma,
\end{equation}
which, after a partial integration, allows us to finally write:
\begin{equation}
\label{superact}
S = -\int d^2 x\, d^2\theta\, E \left[\Phi R_{+-} + 4\mu \sinh 2\pi b^2 \Phi\right] - 2\oint dx\, d\theta \, \Phi K,
\end{equation}
where $E=e^{+\Sigma}$ is the superdeterminant of the superzweibein. Since (the bulk piece of) a generic 2d dilaton supergravity model can be written as
\begin{equation}
\label{superactgen}
S = -\int d^2 x\, d^2\theta\, E \left[\Phi R_{+-} + u(\Phi)\right],
\end{equation}
we are led to claim that $\mathcal{N}=1$ Liouville supergravity is a dilaton supergravity model with a hyperbolic sine prepotential $u(\Phi) = \sinh (2\pi b^2 \Phi)/(4 \sin \frac{\pi b^2}{2})$ if we identify the Liouville cosmological constant as follows:
\begin{equation}
\mu = \frac{1}{16 \sin \frac{\pi b^2}{2}} \quad \implies \quad \kappa \equiv \sqrt{\frac{2\mu}{\cos \frac{\pi b^2}{2}}} = \frac{1}{2 \sqrt{ \sin \pi b^2}}.
\end{equation}
We can view this choice of $\mu$ as a choice of scale in Liouville gravity that allows for a direct comparison to the underlying quantum group, and that has a clean $b \to 0$ limit to $\mathcal{N} = 1$ JT supergravity.

Just as in the bosonic gravity model, the above argument remains to be clarified further, which we postpone to future work.  Instead, we will present indirect evidence that this specific dilaton (super)gravity theory indeed makes contact with statements about Liouville (super)gravity. To achieve this goal, we next review how generic dilaton gravity models have a useful interpretation in terms of nonlinear gauge theory.

\subsection{From Dilaton Gravity to Poisson Sigma Models}

A theory of 2d dilaton gravity, in turn, admits a group-theoretic description as a Poisson sigma model \cite{Ikeda:1993aj, Ikeda:1993fh, Schaller:1994es, Strobl_1999, Ertl_2001}. See also \cite{Grumiller:2020elf, Grumiller:2021cwg} for some recent generalizations.

In the bosonic case, we start with the second-order formulation of dilaton gravity and rewrite it in first-order variables, introducing the zweibein $e$ and the spin connection $\omega$:
\begin{align}
\label{secondorder}
S &= \frac{1}{2}\int d^2 x\sqrt{-g}\, (\Phi R + W(\Phi)) \\
&= \int \left[\Phi\, d\omega + \frac{1}{4}W(\Phi)\epsilon^{ab} e_a \wedge e_b + X^a (de_a + \epsilon_a{}^b \omega \wedge e_b)\right],
\end{align}
where $g_{\mu\nu} = \eta^{ab}e_{a\mu} e_{b\nu}$ and $\omega$ is torsion-free. In this subsection, we work in Lorentzian signature (as will be convenient for our later discussion of quantization), with $\epsilon^{01} = +1$. This theory can be identified with a topological Poisson sigma model with three-dimensional target space, of the type
\begin{equation}
\label{PSMbos}
S = \int \left( A_i \wedge dX^i + \frac{1}{2} P^{ij}(X) A_i \wedge A_j \right),
\end{equation}
where $A_i = (e_0, e_1, \omega)$ and $X^i = (X^0, X^1, \Phi)$. We read off the Poisson algebra
\begin{equation}
\label{bospoi}
\left\{X^0, X^1\right\}_{\PB} = \frac{W(X^2)}{2}, \qquad \left\{X^a, X^2\right\}_{\PB} = \epsilon^a{}_b X^b.
\end{equation}
For $W(\Phi)=2\Phi$, this becomes the $\mathfrak{so}(2,1)$ Lie algebra. Defining the ``lightcone generators'' $E^\pm \equiv -X^0\pm X^1$ and setting $H \equiv X^2$, we get:
\begin{align}
\label{bosalg}
\left\{H,E^\pm\right\}_{\PB} = \pm E^\pm, \qquad \left\{E^+,E^-\right\}_{\PB} = 2 V(H),
\end{align}
where 
\begin{equation}
V(H)\equiv \frac{1}{2}W(H)
\end{equation}
is a rescaled version of the dilaton potential $W$. For $V(H) = H$, this becomes the $\mathfrak{sl}(2,\mathbb{R})$ Lie algebra.

Next, we write down the equations for the case of supergravity. The component form of the action \eqref{superactgen} is
\begin{align}
L = \Phi\, d\omega + X^a (de_a + \epsilon_a{}^b \omega e_b &+ 2 i \bar{\psi}\gamma_a\psi) - \left(2uu'-\frac{iu''}{16}\bar{\chi}\chi\right)\epsilon^{ab}e_a e_b \nonumber \\
&+ 4 iu \bar{\psi} \gamma^3 \psi + i u' \bar{\chi} e_a \gamma^a \psi + i \bar{\chi}\left(d\psi + \frac{1}{2}\omega \gamma^3 \psi\right).
\end{align}
This is to be compared with the generic form of a graded Poisson sigma model:
\begin{align}
S = \int \left( A_i \wedge dX^i - \frac{1}{2} A_i \wedge A_j P^{ji}(X) \right),
\end{align}
where the fields $A_i$ and $X^i$ and, by extension, $P^{ji}(X)$ are graded fields. Namely, some of the components are even variables while some are odd. We have written this action with a specific ordering of the factors \cite{Strobl_1999, Ertl_2001}. It is equivalent to \eqref{PSMbos} only for even fields, but differs from that action when the fields are graded. Identifying the $3|2$-dimensional target space coordinates as
\begin{equation}
X^i \equiv (X^a, \chi^\alpha, \Phi), \qquad A_i \equiv (e_a, i \bar{\psi}_\alpha, \omega),
\end{equation}
we can read off the relevant graded Poisson tensor.\footnote{We use the following realization of the 2d Dirac algebra:
\begin{equation}
(\gamma^0)_{\alpha}{}^{\beta} = \left(\begin{array}{cc}
0 &-1 \\
1 & 0 
\end{array}\right) = -i\sigma^2, \quad (\gamma^1)_{\alpha}{}^{\beta} = \left(\begin{array}{cc}
0 &1 \\
1 &0 
\end{array}\right)=\sigma^1, \quad (\gamma^3)_{\alpha}{}^\beta \equiv \gamma^1\gamma^0 = \left(\begin{array}{cc}
1 &0 \\
0 &-1
\end{array}\right) = \sigma_3
\end{equation}
for a 2d space where $a = 0, 1$ and $\eta_{ab} = \operatorname{diag}(-, +)$, as well as the raised versions:
\begin{equation}
(\gamma^0)^{\alpha\beta} = \left(\begin{array}{cc}
1 & 0 \\
0 & 1 
\end{array}\right), \quad (\gamma^1)^{\alpha\beta} = \left(\begin{array}{cc}
1 & 0 \\
0 & -1 
\end{array}\right), \quad (\gamma^3)^{\alpha\beta} = \left(\begin{array}{cc}
0 & -1 \\
-1 & 0
\end{array}\right),
\end{equation}
where indices are raised and lowered with $\epsilon_{\alpha\beta} = \epsilon^{\alpha\beta} = \left(\begin{array}{cc}
0 & 1 \\
-1 & 0
\end{array}\right)$.}
The nonlinear Poisson algebra describing $\mathcal{N}=1$ dilaton supergravity has five generators $X^0, X^1, \chi^0, \chi^1, \Phi$, with the following graded bracket relations \cite{Strobl_1999, Ertl_2001}:
\begin{gather}
\big\{\chi^\alpha,\chi^\beta\big\}_{\PB} = P^{\alpha \beta} = -8iu(\gamma^3)^{\alpha\beta} - 4 i X_a (\gamma^a)^{\alpha\beta}, \vphantom{\frac{}{2}} \nonumber \\
\big\{X^a, \Phi\big\}_{\PB} = P^{a 2} =\epsilon^a{}_{b}X^b, \qquad \big\{\chi^\alpha,\Phi\big\}_{\PB} = P^{\alpha 2} = - \frac{1}{2}(\gamma^3 \chi)^\alpha, \\
\big\{X^a,\chi^\alpha\big\}_{\PB} = P^{a\alpha} = u' (\gamma^a \chi)^\alpha, \qquad \big\{X^a,X^b\big\}_{\PB} = P^{ab} = - \epsilon^{ab}\left(4uu'+\frac{1}{8i}u'' \bar{\chi}_\alpha \chi^\alpha\right), \nonumber
\end{gather} 
in terms of a single superpotential function $u(\Phi)$. Defining the lightcone variables
\begin{align}
\chi^\alpha = \left(\begin{array}{c}
\chi^+ \\
\chi^- 
\end{array}\right), \qquad X^\pm = \pm X^0 - X^1,
\end{align}
one can rewrite this algebra into a more suggestive form by defining
\begin{equation}
F^\pm \equiv \chi^{\pm}/(4\sqrt{i}), \qquad E^\pm \equiv X^\pm/2, \qquad H \equiv \Phi,
\end{equation}
which yields:
\begin{gather}
\big\{F^+,F^+\big\}_{\PB} = \frac{1}{2} E^+, \qquad \big\{F^-,F^-\big\}_{\PB} = -\frac{1}{2} E^-, \qquad \big\{F^+,F^-\big\}_{\PB} = \frac{1}{2} u(H), \nonumber \\
\big\{H, F^\pm\big\}_{\PB} = \pm \frac{1}{2} F^\pm, \qquad \big\{H, E^\pm\big\}_{\PB} = \pm E^\pm, \nonumber \\
\big\{E^\pm,F^\mp\big\}_{\PB} = - u'(H)F^\pm,  \qquad \big\{E^\pm,F^\pm\big\}_{\PB} = 0, \vphantom{\frac{1}{2}} \nonumber \\
\big\{E^+,E^-\big\}_{\PB} = 2(u(H)u'(H) + u''(H) F^- F^+). \vphantom{\frac{1}{}} \label{Poisalg}
\end{gather}
We will later on recognize this Poisson superalgebra as the classical limit of the $q$-deformed algebra of $\mathfrak{osp}(1|2,\mathbb{R})$ for a specific choice of $u(H)$. This will require a quantization of the Poisson sigma model. Before going there, however, we present one more classical argument that sheds light on the asymptotic gravitational boundary conditions we are using.

\subsection{Semiclassical Black Hole First Law}

Using the above results, we can directly compare the semiclassical $\hbar \to 0$ interpretation of Liouville supergravity amplitudes with classical black hole physics in the dilaton supergravity model \eqref{superactgen}. As a warmup, we first redo the bosonic case presented in \cite{Mertens:2020hbs} but using our rescaled length variables (as mentioned in the introduction). We start with the fixed-length disk amplitude without any operator insertions. This can be found by letting $\beta_M \to 0$ in \eqref{final2}, using the results of Section \ref{s:planchbos}:
\begin{align}
\left\langle 1\right\rangle_{\ell} &= \int ds \, e^{-\ell \frac{\cosh 2 \pi b s}{2 \pi b^2 \sin \pi b^2}} \sinh 2\pi b s \sinh \frac{2\pi s}{b} \\
&\sim \int_0^\infty dE \, e^{-\ell E}\sinh\left(\frac{1}{b^2} \operatorname{arccosh}\left((2\pi b^2 \sin \pi b^2) E\right)\right).
\end{align}
We now interpret this expression as a thermal partition function with $\ell = \beta = T^{-1}$, the inverse temperature, and with the density of states $\rho(E)$ explicitly visible on the second line. Approximating $\sinh \sim \frac{1}{2}\exp$, we get the saddle-point equation
\begin{equation}
\label{bosET}
\sqrt{E^2 - \frac{1}{(2\pi b^2 \sin \pi b^2)^2}} = \frac{T}{b^2}.
\end{equation}
As a check, in the JT limit $b\to 0$, we set $E = \frac{1}{2\pi b^2 \sin \pi b^2} + E_{\text{JT}}$ and obtain
\begin{equation}
\sqrt{\frac{E_{\text{JT}}}{\pi^2 b^4 }} = \frac{T}{b^2} \quad \implies \quad E_{\text{JT}} = \pi^2 T^2,
\end{equation}
which is the well-known JT gravity black hole first law \cite{Almheiri:2014cka, Cotler:2016fpe, Stanford:2017thb}, upon setting the Schwarzian coupling coefficient to $C = 1/2$ and dropping an arbitrary offset $E_0$. The coefficient $C$ contains information on the rate of divergence of the dilaton field at the boundary \cite{Jensen:2016pah, Maldacena:2016upp, Engelsoy:2016xyb}, and defines the specific model under consideration. We will choose conventions below for generic dilaton gravity models that reproduce the value $C=1/2$ in the JT limit.

Given a bosonic dilaton potential $W(\Phi)$ as in \eqref{secondorder}, up to diffeomorphisms, one can always bring the 2d metric and dilaton to the form \cite{Gegenberg:1994pv, Witten:2020ert}\footnote{One can identify the conformal factor $e^{2\rho} = 4A(r)$ immediately, since the purely radial coordinate transformation $dr_{\text{new}} = \frac{dr}{2A(r)}$ maps the metric to conformal form.}
\begin{equation}
\label{bhgen}
ds^2 = 4A(r) dt^2 + \frac{dr^2}{A(r)}, \qquad \Phi(r) = r,
\end{equation}
where the asymptotic region is $r\to +\infty$ (for which the boundary condition on the dilaton field is fixed as above)\footnote{More generally, one is allowed the asymptotics $\Phi(r) = a r$ as $r\to+\infty$; the value of $a$ defines the model at hand. We have also included an extra factor of $4$ in $g_{tt}$ compared to \cite{Gegenberg:1994pv, Witten:2020ert}, which is a simple rescaling of the time coordinate $t$. This is a possibility that is closely related to the choice of $a$. All of these options in the JT gravity regime ($b\to 0$) correspond to a choice of Schwarzian coupling coefficient $C$. They can be mapped to a choice of prefactor for the boundary Hamiltonian in the Poisson sigma model framework that we develop below.} and where
\begin{equation}
A(r) = \int_{r_h}^{r} dr'\, W(r'),
\end{equation}
with $r=r_h$ being the location of the black hole horizon. This black hole has an energy-temperature relation fully determined by knowledge of the dilaton potential as:
\begin{equation}
\label{ET}
E = \int^{W^{-1}(2\pi T)} W(\Phi)\, d\Phi + E_0.
\end{equation}
For instance, for JT gravity where $W(\Phi)=2\Phi$, we immediately get $E = \pi^2 T^2$ if we set $E_0=0$. Starting with our $E(T)$ relation \eqref{bosET}, we can solve for the dilaton potential in a unique fashion:
\begin{equation}
\label{dilatonpotL}
V(\Phi) = \frac{1}{2} W(\Phi) = \frac{\sinh 2\pi b^2 \Phi}{2\sin \pi b^2}.
\end{equation}
Using this dilaton potential, we write down the Euclidean bulk metric and dilaton field:
\begin{equation}
\label{metric}
ds^2 = \frac{4(\cosh 2\pi b^2 r - \cosh 2\pi b^2 r_h)}{2\pi b^2 \sin \pi b^2} dt^2 + \frac{2\pi b^2 \sin \pi b^2}{\cosh 2\pi b^2 r-\cosh 2\pi b^2 r_h} dr^2, \qquad \Phi(r) = r.
\end{equation}
From this, we can read off the asymptotics of the fields in terms of boundary conditions. The metric component $g_{tt}$ diverges as $r\to \infty$. In the same vein as in aAdS holography, we define lengths $\ell$ as measured by the boundary theory using the $t$-coordinate: $d\ell \equiv dt$.

It is instructive to show agreement between this boundary behavior and the rescaling of length variables as discussed in the introduction. From \eqref{bosredef}, we get the following relation between lengths:
\begin{align}
\ell_{\text{L}} \equiv \int_1^2 e^{b\phi} = \left.e^{-\pi b^2\Phi}\right|_{\partial} \int_1^2 e^{\rho},
\end{align}
where the dilaton field $\Phi$ needs to take a constant value along the boundary in order for the boundary lengths, as measured using the different metrics, to be proportional for any choice of boundary segment. From \eqref{metric}, we get the divergent asymptotics
\begin{equation}
\left. e^{\rho}\right|_{r\to \infty} = \lim_{r\to+\infty}\frac{e^{\pi b^2 r}}{\sqrt{\pi b^2 \sin \pi b^2}}.
\end{equation}
Crucially, this divergence perfectly cancels with the dilaton asymptotics of \eqref{metric}, yielding the finite result for the rescaling that relates the length measured in the Liouville metric to the length measured using the boundary metric $dt$ in \eqref{metric}:
\begin{equation}
\label{bdyresc}
\ell_{\text{L}} = \frac{\ell}{\sqrt{\pi b^2 \sin \pi b^2}}.
\end{equation}
Notice that the relation for the timelike Liouville field $\chi$ in \eqref{bosredef} then implies the boundary condition
\begin{equation}
\left.e^{b\chi}\right|_{\partial} \sim \lim_{r\to+\infty}e^{2\pi b^2 r} \, \to +\infty.
\end{equation}
This Dirichlet boundary condition is precisely the vacuum (or ZZ-) brane boundary condition. This is indeed the boundary condition for $\chi$ that we describe in Appendix \ref{app:bosonic}.\footnote{One can also think of it as an infinite-length boundary condition when interpreting the timelike Liouville CFT in a similar fashion as the spacelike CFT: $\ell_\chi \equiv \int_1^2 e^{b\chi} \, \to +\infty$.}

The above analysis was classical. In quantum Liouville gravity amplitudes, we must use the rescaling between lengths of \eqref{rescaleb}. However, reinstating units of $\hbar$, the Liouville parameter $\kappa$ is actually $\kappa = \sqrt{\mu\hbar/\sin (\pi b^2 \hbar)}$ \cite{Fateev:2000ik}. Using that $\kappa \to \sqrt{\mu/\pi b^2}$ in the semiclassical $\hbar \to 0$ limit and inserting the value of $\mu$ from \eqref{mu}, we see that \eqref{rescaleb} matches with \eqref{bdyresc} in the $\hbar \to 0$ limit. An interesting aspect of this matching is that the rescaling of lengths is itself $\hbar$-dependent, with the correct matching to the classical black hole analysis occurring only in the semiclassical $\hbar \to 0$ limit, as should be the case.

We now move on to $\mathcal{N}=1$ dilaton supergravity. The Liouville supergravity disk amplitude can be found by letting $\beta_M \to 0$, and using the results of Section \ref{s:planch}, we get:
\begin{align}
\label{finaldisk}
\left\langle 1\right\rangle_{\ell} &= \int_0^{+\infty} ds \, e^{-\ell \frac{\sinh^2 \pi b s}{16 \sin^2 \frac{\pi b^2}{2}}}\cosh \pi b s \cosh \frac{\pi s}{b} \nonumber \\
&\sim \int_0^{+\infty} \frac{dE}{\sqrt{E}} \, e^{-\ell E}\cosh\left[ \frac{1}{b^2} \operatorname{arcsinh} \left(\sqrt{16 E \sin^2 \frac{\pi b^2}{2}}\right)\right].
\end{align}
In the $\hbar \to 0$ thermodynamic limit, we need to use the classical $\hbar\to 0$ limit of the Casimir \eqref{class} when going from the first to the second line and to evaluate the integral at large $s$, for which we can approximate $\cosh \sim \frac{1}{2}\exp$. The saddle-point relation for the above integral then yields the semiclassical black hole first law for $E(T)$:\footnote{Compared to the result of \cite{Mertens:2020pfe}, we used the rescaled energy and length variables in the introduction, and additionally corrected some missing factors of 2 compared to that work.}
\begin{equation}
\label{first}
\sqrt{E^2 +\frac{E}{16 \sin^2 \frac{\pi b^2}{2}}} = \frac{T}{2b^2}.
\end{equation}
We will now show that we can reproduce this first law directly from a classical black hole solution in the $\mathcal{N}=1$ dilaton supergravity model with precisely the $\sinh$ dilaton superpotential \eqref{sinhpre}.

For the semiclassical saddle solution, the fermions (dilatino and gravitino) are turned off, and the discussion boils down to that for the bosonic subsector given above. The resulting bosonic potential $V(\Phi)$ is related to the prepotential by
\begin{equation}
V(\Phi) = u(\Phi) u'(\Phi),
\end{equation}
as can be seen by comparing the last relation of \eqref{Poisalg} to the analogous bosonic relation \eqref{bosalg}.

Defining a shifted energy variable $\tilde{E} \equiv E + E_0$ with $E_0 = (32 \sin^2 \frac{\pi b^2}{2})^{-1}$, the first law \eqref{first} can be rewritten as
\begin{equation}
\sqrt{\tilde{E}^2 - E_0^2} = \frac{T}{2b^2},
\end{equation}
which is of the same form as the bosonic black hole first law written in \cite{Mertens:2020hbs}. From this, we can immediately write down the dilaton potential:
\begin{equation}
W(\Phi) = \frac{\pi b^2}{8\sin^2 \frac{\pi b^2}{2}} \sinh (4\pi b^2 \Phi),
\end{equation}
which reproduces \eqref{first} when inserted into \eqref{ET} and upon setting $E_0 = (32 \sin^2 \frac{\pi b^2}{2})^{-1}$. We then have the corresponding prepotential
\begin{equation}
\label{dilatonprepotL}
u(\Phi) = \frac{\sinh (2\pi b^2 \Phi)}{4 \sin \frac{\pi b^2}{2}},
\end{equation}
which indeed matches the superspace proposal and prepotential displayed in \eqref{superact}. Hence, starting with this dilaton potential, we indeed agree with the semiclassical first law derived using purely Liouville supergravity techniques in \cite{Mertens:2020pfe}.

\subsection{The (Graded) Poisson Sigma Model}

We next analyze the graded Poisson sigma model in more detail, and uncover the precise way in which the $q$-deformed algebra is realized. The punchline is that the governing U$_q(\mathfrak{sl}(2,\mathbb{R}))$ or U$_q(\mathfrak{osp}(1|2,\mathbb{R}))$ quantum (super)group appears upon quantizing the target space Poisson structure of the Poisson sigma model.\footnote{The Poisson sigma model language was recently used in \cite{Verlinde:2021kgt} to address universality of wormholes for quantum-mechanical systems that include, in particular, all dilaton (super)gravity models.}

\subsubsection*{The Poisson Sigma Model: Classical Analysis}

Consider the (graded) Poisson sigma model on a half-space:
\begin{equation}
\label{PSM}
S = \int \left( A_i \wedge dX^i - \frac{1}{2} A_i \wedge A_j P^{ji}(X) \right) = \int d^2 x \left(-A_{1i} \partial_0 X^i + A_{0i}( \partial_1 X^i - A_{1j} P^{ji}(X)) \right),
\end{equation}
where $i = 1, \ldots, m$ and $m$ is the dimension of the target space $\mathcal{M}$. $\mathcal{M}$ is equipped with a graded Poisson bracket:
\begin{equation}
\left\{X^i, X^j\right\}_{\PB} = P^{ij}(X), \qquad P^{ij} = - (-)^{\sigma_i\sigma_j} P^{ji}, \qquad \partial^R_\ell P^{[ij|} P^{\ell|k]} = 0,
\end{equation}
the latter relations being (anti)symmetry and the Jacobi identity required for the definition of the bracket operation. $\sigma_i = 0, 1$ denotes the grading of the field $X^i$. The ordering of the different objects in \eqref{PSM} is important.

The action \eqref{PSM} is invariant under the local nonlinear symmetry transformations\footnote{Specifically, $\delta(A_i\wedge dX^i - \frac{1}{2}A_i\wedge A_j P^{ji}(X)) = -d\epsilon_i\wedge dX^i$.}
\begin{align}
\label{}
\delta X^i &= -\epsilon_j P^{ji}, \\
\delta A_i &= -d \epsilon_i + A_j \epsilon_k \partial^R_i P^{kj},
\end{align}
written in terms of the right derivative. This nonlinear symmetry algebra for the particular case of dilaton (super)gravity was first discovered in \cite{Ikeda:1993aj, Ikeda:1993fh} without the reinterpretation in terms of a topological Poisson sigma model. See Appendix \ref{app:nonlinear} for some further remarks on this point of view.

The global part of this symmetry transformation leads to $m$ conserved charges. We derive here the classical charge algebra that they satisfy, in particular accommodating Grassmann-valued variables. A convenient reference for the canonical structure of the bosonic model is \cite{CATTANEO_2001}. The variables $X^i$ and $A_{1i}$ are canonically conjugate:
\begin{equation}
\pi_{X^i}(x) \equiv \frac{\partial^L L}{\partial(\partial_0 X^i)} = - (-)^{\sigma_i} A_{1i},
\end{equation}
where we conventionally take the left derivative for the fermionic variables. This leads to the canonical brackets\footnote{Following \cite{Henneaux:1992ig}, we have $\left\{p, q\right\} = -1$, in this specific order, for both commuting and anticommuting variables.}
\begin{equation}
\label{canbra}
\left\{A_{1i}(x), X^j(y)\right\} = (-)^{\sigma_i}\delta_{i}^{j}\delta(x-y),
\end{equation}
and $A_0$ plays the role of a Lagrange multiplier enforcing the first-class constraints
\begin{equation}
\label{constraint}
\partial_1 X^i  -  A_{1j}  P^{ji}(X) = 0, \qquad i = 1, \ldots, m.
\end{equation}
For a half-space, these constraints can be integrated into the relations
\begin{equation}
x^i \equiv X^i(0) = -\int_{0}^{+\infty} du\, A_{1j}(u) P^{ji}(X(u)).
\end{equation}
This shows that the field $X^i(u)$ has only a single degree of freedom on the half-line $[0, +\infty)$. Together with the conjugate fields $A_{1i}(u)$, this in turn shows that the phase space is finite-dimensional of dimension $2m$. This illustrates that the model is topological, with degrees of freedom that can be thought of as living on the boundary line.

Using the canonical brackets \eqref{canbra}, we can derive the following relation for the boundary variables $x^i$:
\begin{align}
\left\{x^i, x^j\right\} &= \left\{-\int_{0}^{+\infty} du\, A_{1k}(u) P^{ki}(X(u)),X^j(0)\right\} \\
&=  -(-)^{\sigma_i\sigma_k} (-)^{\sigma_k}\int_{0}^{+\infty} du\, P^{ki}(X(u))\left\{ A_{1k}(u), X^j(0)\right\} = P^{ij}(x), \label{poisalg}
\end{align}
which hence satisfy the Poisson algebra of the target space, but now as a canonical phase space algebra. The Noether charges associated with the global nonlinear transformations are given by
\begin{equation}
Q^i \equiv \int dx\, \delta^i X^j \pi_{X^j} = (-)^{\sigma_j}\int_0^{+\infty} du\, P^{ij}(X(u)) A_{1j}(u) = -\int_0^{+\infty} du\, A_{1j}(u) P^{ji}(X(u)) = x^i.
\end{equation}
Being identified with the $x^i$, these charges therefore satisfy the nonlinear Poisson algebra \eqref{poisalg} as well:\footnote{For a \emph{linear} symmetry algebra, the Noether charges always satisfy the same canonical algebra as the underlying Lie algebra. This is no longer generically true for a nonlinear symmetry algebra. However, the above shows that it is true by explicit computation for the Poisson sigma model.} 
\begin{equation}
\label{poisalgQ}
\boxed{\left\{Q^i, Q^j\right\} = P^{ij}(Q).}
\end{equation}

\subsubsection*{The Poisson Sigma Model: Classical Casimirs}

For the above system, the Hamiltonian vanishes, and these charges are trivially conserved quantities. If a nontrivial boundary Hamiltonian can be added, then it must commute with these charges:
\begin{equation}
\frac{d Q^i}{dt} = \left\{H, Q^i\right\} = 0,
\end{equation}
and it is a Casimir of the algebra.

We can write explicit expressions for these Casimirs. First, consider the bosonic dilaton gravity model, for which we have a three-dimensional target space with coordinates $X^i \equiv (E^+, E^-, H)$ and Poisson algebra \eqref{bosalg}. For an arbitrary potential $V(H)$, the rank of the Poisson tensor $P^{ij}$ is two, and there is a single independent Casimir function that can be chosen as \cite{Kl_sch_1996}
\begin{equation}
\label{boscas}
\mathcal{C}(X) = E^+E^- + 2\int^{H} V(y)\, dy.
\end{equation}
It can be explicitly checked to satisfy the relation $\left\{\mathcal{C}, X^i\right\} = \frac{\partial \mathcal{C}}{\partial X^j} P^{ji} = 0$.\footnote{A simple example is that of a linear Poisson structure, where one formally makes contact with the BF framework of JT gravity with Lie algebra $\mathfrak{sl}(2,\mathbb{R})$ and $\mathcal{C}(X) = \Tr X^2$ yields the required boundary Hamiltonian \cite{Mertens:2018fds} in terms of the Cartan-Killing metric. However, it is important to stress that the right-hand side of \eqref{boscas} is commutative at this stage, and one only makes contact with the actual quadratic Casimir of the Lie group after quantizing.}

Of particular interest in this work is the specific potential $V(H) = \frac{\sinh 2 \pi b^2 H}{2\sin \pi b^2}$. For this choice, the Poisson algebra \eqref{bosalg} is almost the same as the quantum algebra \eqref{qsl}, up to some factors of $i$ that we will explain later when quantizing. This choice has the corresponding Casimir:
\begin{equation}
\label{caspoiss}
H_\text{bdy} = \mathcal{C}(X) = E^+E^- + \frac{\cosh 2\pi b^2 H}{2\pi b^2\sin \pi b^2}.
\end{equation}

For $\mathcal{N}=1$ dilaton supergravity, whose $3|2$-dimensional target space has coordinates $X^i \equiv (F^+,F^-,E^+,E^-,H)$, the Casimir function can be written analogously \cite{Strobl_1999, Ertl_2001}:
\begin{equation}
\label{susycas}
\mathcal{C}(X) = E^+E^- + \int^{H} (2uu'+2u'' F^-F^+)\, dy =  E^+E^- + 2u'(H) F^-F^+ + u(H)^2,
\end{equation}
satisfying $\left\{\mathcal{C}, X^i\right\} = \frac{\partial^R \mathcal{C}}{\partial X^j} P^{ji} = 0$.

Specializing to the particular case of $u(H) = \frac{\sinh 2\pi b^2 H}{4\sin \frac{\pi b^2}{2}}$, one finds:
\begin{equation}
\label{caspoissS}
H_\text{bdy} = \mathcal{C}(X) =  E^+ E^- +\frac{\pi b^2\cosh 2\pi b^2 H}{\sin \frac{\pi b^2}{2}} F^- F^+ + \frac{\sinh^2 2\pi b^2 H}{16\sin^2 \frac{\pi b^2}{2}}.
\end{equation}

For ordinary Lie algebras, it is well-known that one can always perform a global rotation to align a given vector in the Lie algebra along the Cartan directions. For example, for SU$(2)$, one conventionally aligns the spin vector along the $z$-direction, hence parametrizing the quadratic Casimir as $J_x^2+J_y^2+J_z^2 = J_z^2 = j^2$, where we introduce the classical spin label $j$. For quantum groups, we can analogously consider only ``turning on'' the $H$-direction.\footnote{This can be motivated, e.g., in the context of the $q$-deformed 2d Yang-Mills (YM) or BF models to find explicit expressions for amplitudes through ``abelianization'' \cite{Blau:1993tv, Aganagic:2004js}. We make some comments on this perspective in the current context in the concluding Section \ref{s:concl}.} This leads to a classical description of the Casimir as:
\begin{alignat}{2}
\label{clascomp}
&\text{\phantom{$\mathcal{N}=1$ supersymmetric:}\llap{bosonic:}} \quad & \mathcal{C}(H) &= 2\int^{H} V(y)\, dy = \frac{\cosh 2\pi b^2 H}{2\pi b^2\sin \pi b^2}, \\
\label{susycomp}
&\text{$\mathcal{N}=1$ supersymmetric:} \quad & \mathcal{C}(H) &= u(H)^2 = \frac{\sinh^2 2\pi b^2 H}{16\sin^2 \frac{\pi b^2}{2}}.
\end{alignat}
Here, we think of $H$ as a $c$-number, analogous to the spin label $j$ in the undeformed $b\to 0$ limit. We will come back to this interpretation further on, when we compare to the quantized formulas.

\subsubsection*{The Poisson Sigma Model: Quantization}

Upon quantizing the model, we replace Poisson brackets with commutators, and the charge algebra \eqref{poisalgQ} becomes a commutator algebra of Hermitian charges $(\hat{Q}^i)^\dagger = \hat{Q}^i$:\footnote{We are following the ``constrain first'' approach to constrained quantum systems, since we have already implemented \eqref{constraint} at the classical level.}
\begin{equation}
\left[\hat{Q}^i, \hat{Q}^j\right] \stackrel{?}{=} i \hbar \hat{P}^{ij}(\hat{Q}).
\end{equation}
However, when the Poisson tensor is nonlinear, ordering ambiguities can appear here. In particular, the above commutator must satisfy the Jacobi identity, which is different than the previous one since now:
\begin{equation}
\left[\hat{P}^{ij}(\hat{Q}), \hat{Q}^k\right] \neq i \hbar \partial_\ell^R\hat{P}^{ij}(\hat{Q}) \hat{P}^{\ell k}(\hat{Q}).
\end{equation}
Instead, one has a more complicated ``ordered'' version of the derivative. Moreover, consistency with hermiticity of the charges requires 
\begin{equation}
\label{hermiticity}
(\hat{P}^{ij}(\hat{Q}))^\dagger = \hat{P}^{ij}(\hat{Q}), \qquad i, j = 1, \ldots, m.
\end{equation}

It is important not to confuse this physical quantization in $\hbar$ with the mathematical ``quantization'' or $q$-deformation of the underlying algebraic structure. Both of these occur independently and concurrently in this section.

For the specific case of bosonic dilaton gravity, the algebra itself does not need to change thanks to the internal commutativity of each entry in the Poisson tensor of \eqref{bosalg}:
\begin{equation}
P^{H\pm}(X) = \pm E^\pm, \qquad P^{+-}(X) = 2V(H).
\end{equation}
Moreover, since $V(\cdot)$ is a real function, \eqref{hermiticity} is also satisfied, and one finds the quantized charge algebra ($\hbar =1$):
\begin{align}
\label{bosalgquant}
\left[\hat{H},\hat{E}^\pm\right] = \pm i \hat{E}^\pm, \qquad \left[\hat{E}^+,\hat{E}^-\right] = 2 i V(\hat{H}).
\end{align}
For the particular case where 
\begin{equation}
\label{sinhpot}
V(\hat{H}) = \frac{\sinh 2 \pi b^2 \hat{H}}{2\sin \pi b^2},
\end{equation}
and upon identifying the Hermitian charges with the antihermitian generators via $\hat{H} = i H$ and $\hat{E}^\pm = i E^\pm$, the resulting algebra becomes precisely \eqref{qsl}:
\begin{align}
\label{uqsl}
[H, E^\pm] = \pm E^\pm, \qquad [E^+, E^-] = \frac{\sin 2 \pi b^2H}{\sin \pi b^2}.
\end{align}
Thus we conclude that:
\begin{quote}
\emph{The conserved charges in the Poisson sigma model description of dilaton gravity with potential \eqref{sinhpot} satisfy an algebra that can be identified with the \emph{U}$_q(\mathfrak{sl}(2,\mathbb{R}))$ algebra.}
\end{quote}

Next, we consider $\mathcal{N}=1$ dilaton supergravity, for which the nonlinear commutator algebra is different from the corresponding classical Poisson algebra \eqref{Poisalg}. In particular, consistency with the noncommutative version of the Jacobi identity requires, in addition to the ``seed'' commutation relations
\begin{gather}
\label{seed}
\left[\hat{H}, \hat{F}^\pm\right] = \pm \frac{i\hbar}{2}\hat{F}^\pm, \qquad \left\{\hat{F}^+, \hat{F}^-\right\} = \frac{i\hbar}{2}u(\hat{H}), \qquad \left\{\hat{F}^\pm, \hat{F}^\pm\right\} = \pm \frac{i\hbar}{2} \hat{E}^\pm,
\end{gather}
also the modified relations:\footnote{Note that $f(\hat{H})\hat{F}^\pm = \hat{F}^\pm f(\hat{H}\pm \frac{i\hbar}{2})$, which follows inductively from the first relation in \eqref{seed}.}
\begin{align}
\left[\hat{E}^+,\hat{F}^-\right] &= 2 \left( u\left(\hat{H}-\frac{i\hbar}{2}\right)-u(\hat{H}) \right) \hat{F}^+, \nonumber \\
\label{seedmodrel}
\left[\hat{E}^-,\hat{F}^+\right] &= 2 \left( u(\hat{H})-u\left(\hat{H}+\frac{i\hbar}{2}\right) \right) \hat{F}^-, \\
\left[\hat{E}^+,\hat{E}^-\right] &= 4\left( u(\hat{H})-u\left(\hat{H}-\frac{i\hbar}{2}\right) \right)u(\hat{H}) \nonumber \\
&\phantom{==} + \frac{8}{i\hbar}\left(u\left(\hat{H}-\frac{i\hbar}{2}\right) - 2u(\hat{H})+u\left(\hat{H}+\frac{i\hbar}{2}\right) \right) \hat{F}^-\hat{F}^+. \nonumber
\end{align}
One checks explicitly that these expressions satisfy the hermiticity property \eqref{hermiticity}, consistent with a set of Hermitian charges $\hat{H}, \hat{F}^\pm, \hat{E}^\pm$.\footnote{This corresponds to defining the adjoint of a product of graded operators as $(AB)^\dagger = (-)^{\sigma_A \sigma_B} B^\dagger A^\dagger$. For odd variables, this definition of conjugation corresponds to the order-preserving convention for Grassmann numbers $(\vartheta_i \vartheta_j)^* = \vartheta_i^* \vartheta_j^*$ that was used in \cite{Fan:2021wsb}.}

For the specific choice where
\begin{equation}
\label{sinhpre}
u(\hat{H}) = \frac{\sinh 2\pi b^2 \hat{H}}{4\sin \frac{\pi b^2}{2}},
\end{equation} 
and upon setting $\hat{H} = i H$, $\hat{F}^\pm = i F^\pm$, and $\hat{E}^\pm = i E^\pm$, this algebra becomes the $q$-deformed algebra \eqref{salgeb} of U$_q(\mathfrak{osp}(1|2,\mathbb{R}))$ but with a sign flip in the anticommutator, as should be the case ($\hbar = 1$):\footnote{This is the usual superalgebra rather than the opposite superalgebra. The former is relevant when understanding discrete or finite-dimensional representations (such as those used in defining the Lagrangian of the BF model), whereas the latter is relevant when understanding the continuous representations whose generators are Grassmann-valued operators.}
\begin{align}
\label{uqosp}
[H,F^\pm] = \pm\frac{1}{2} F^{\pm} , \qquad \left\{F^+,F^-\right\} = \frac{\sin 2\pi b^2 H}{8\sin \frac{\pi b^2}{2}}, \qquad \left\{F^\pm, F^\pm\right\} = \pm \frac{1}{2} E^\pm .
\end{align}
This is our main statement:
\begin{quote}
\emph{The conserved charges in the Poisson sigma model description of $\mathcal{N}=1$ dilaton supergravity with prepotential \eqref{sinhpre} satisfy an algebra that can be identified with the \emph{U}$_q(\mathfrak{osp}(1|2,\mathbb{R}))$ algebra.}
\end{quote}

As a check of \eqref{seed} and \eqref{seedmodrel}, the classical $\hbar \to 0$ limit yields back the right-hand side of the Poisson algebra \eqref{Poisalg}:
\begin{gather}
\left\{F^+, F^-\right\} = \frac{i\hbar}{2}u(H), \quad \left\{F^\pm, F^\pm\right\} = \pm\frac{i\hbar}{2} E^\pm, \nonumber \\
\left[H, F^\pm\right] = \pm\frac{i\hbar}{2}F^\pm, \nonumber \\
\left[E^\pm, F^\mp\right] = -i\hbar u'(H) F^\pm, \vphantom{\frac{1}{2}} \nonumber \\
\left[E^+, E^-\right] = 2i\hbar(u(H)u'(H) + u''(H)F^- F^+). \vphantom{\frac{1}{}}
\end{gather}

\subsubsection*{The Poisson Sigma Model: Quantum Casimirs}

For bosonic dilaton gravity, whereas the right-hand side of the Poisson algebra \eqref{bosalg} remains identical after quantizing, the Casimir function does get modified upon quantization. Instead of \eqref{boscas}, the result is
\begin{equation}
\mathcal{C}(\hat{X}) \, \sim \, \frac{1}{2}\hat{E}^+\hat{E}^- + \frac{1}{2}\hat{E}^-\hat{E}^+ + f(\hat{H}),
\end{equation}
where $f$ is the solution to the linear difference equation\footnote{It is easy to see that the general solution can be written as a particular solution (depending on $V$) plus a periodic (or homogeneous) function with periodicity $i\hbar$. Let us write down the general solution in more detail. Denoting the right-hand side by $s(H)$, the function $f$ can be written in closed form in terms of the $\mathcal{Z}$-transform of $s$ as:
\begin{equation}
f(i\hbar H)= \mathcal{Z}^{-1}\left[\frac{\mathcal{Z}[s](z)}{z-1}\right](i\hbar H).
\end{equation}
The $\mathcal{Z}$-transform is defined as $\mathcal{Z}[s](z) = \sum_{n=-\infty}^{+\infty}s(n)z^{-n}$. The solution \eqref{solcas} is unique up to a choice of $f(0)$ and upon assuming $H \in \mathbb{N}$. To analytically continue to non-integer $H$ and satisfy \eqref{solcas}, one has the ambiguity of adding an arbitrary periodic function. This piece, however, is the homogeneous (source-free) part of the solution. Imagining one could write down a solution for an arbitrary $V(H)$, we can for instance isolate the homogeneous piece by setting $V=0$ and choose to remove it.}
\begin{equation}
\label{solcas}
f(H+i\hbar)-f(H) = i\hbar(V(H+i\hbar)+V(H)).
\end{equation}
In the classical $\hbar \to 0$ limit, this equation reduces to $df/dH = 2V(H)$, which is \eqref{boscas}. For the potential of interest $V(H) = \frac{\sinh 2 \pi b^2 H}{2\sin \pi b^2}$, the result of solving this equation is:
\begin{align}
\hat{H}_\text{bdy} = \mathcal{C}(\hat{X}) &= -\frac{\sin \pi b^2 \hbar}{\pi b^2 \hbar}\left(\frac{1}{2}\hat{E}^+\hat{E}^- + \frac{1}{2}\hat{E}^-\hat{E}^+\right) - \frac{\cos \pi b^2 \hbar}{2 \pi b^2 \sin \pi b^2 }\cosh 2 \pi b^2 \hat{H} \\
&= -\frac{\sin \pi b^2 \hbar}{\pi b^2 \hbar}\hat{E}^- \hat{E}^+ - \frac{\cosh 2\pi b^2(\hat{H}+\frac{ i \hbar}{2})}{2\pi b^2 \sin \pi b^2 },
\end{align}
where we have used a carefully chosen normalization. This is a specifically ordered version of the classical $\hbar \to 0$ result \eqref{caspoiss}, and matches with \eqref{cas} upon using $\hat{H} = iH$ since the quantized symmetry algebra \eqref{uqsl} is precisely the U$_q(\mathfrak{sl}(2,\mathbb{R}))$ algebra.

The last way of writing this expression has all raising operators on the right and lowering operators on the left. When computing its expectation value in a highest-weight state of a finite-dimensional irrep as
\begin{equation}
\left\langle \text{l.w.}\right|\mathcal{C}(\hat{X})\left|\text{h.w.}\right\rangle,
\end{equation}
we hence extract only the last term, where we set $h$ as the $\hat{H}$ eigenvalue of the state. This is the $q$-analogue of the statement that the classical SU$(2)$ Casimir $j^2$ is replaced by $j(j+1)$ when quantizing, which is proven by elementary techniques in a similar fashion. It is useful to compare the classical description of the Casimir \eqref{clascomp} to this quantized description: 
\begin{alignat}{2}
\label{classbos}
&\text{classical ($\hbar \to 0$):} \quad & & {-\frac{\cosh 2\pi b^2 h_{\text{cl}}}{2 \pi b^2 \sin \pi b^2}}, \\
\label{quantbos}
&\text{\phantom{classical ($\hbar \to 0$):}\llap{quantum:}} \quad & & {-\frac{\cosh 2\pi b^2(h+\frac{i\hbar}{2})}{2 \pi b^2 \sin \pi b^2}}.
\end{alignat}
We now compare the quantized result \eqref{quantbos} to the explicit result \eqref{casirrepbos} for the continuous series irreps. We can identify $h$ directly with the representation label $\lambda$ by $h = -\frac{i}{b}\lambda$ where $\lambda = \frac{b\hbar}{2}+ is$. In the semiclassical $\hbar \to 0$ limit, we get $h \, \to \, h_{\text{cl}} = -\frac{i}{b}\lambda_{\text{cl}}$ where $\lambda_{\text{cl}} = is$, in agreement with \eqref{classbos}.

This shift is expected in the undeformed $b\to 0$ limit: for the group SL$(2,\mathbb{R})$, it corresponds to setting either $j=ik$ or $j=-1/2+ik$, where the $-1/2$ shift is a one-loop effect from the perspective of Borel-Weil-Bott coadjoint orbit quantum mechanics as a tool for reproducing group theory. More generally, this is a shift of the weight vector by the Weyl vector, which has been studied extensively in physics language in many works; see, e.g., \cite{Gukov:2008ve, Fan:2018wya}. In the end, when plugging the correct values of $\lambda$ into the above expressions, \eqref{classbos} and \eqref{quantbos} are identical.

For a general $\mathcal{N}=1$ dilaton supergravity model, we can write down an analogous ansatz for a sCasimir operator as
\begin{equation}
\label{ansscas}
\mathcal{Q}(\hat{X}) \,\sim \, \hat{F}^-\hat{F}^+ + f(\hat{H}),
\end{equation}
where demanding that this expression commutes with $\hat{H}$ and anticommutes with $\hat{F}^\pm$ leads to the linear difference equation
\begin{equation}
\label{solscas}
f(H) + f(H - i\hbar/2) = - \frac{i\hbar}{2} u(H).
\end{equation}
Plugging the solution into \eqref{ansscas} and squaring leads to an operator $\mathcal{C} = \mathcal{Q}^2$ that commutes with all generators for a general prepotential $u(H)$.

For the specific case $u(\hat{H}) = \frac{\sinh 2\pi b^2 \hat{H}}{4\sin \frac{\pi b^2}{2}}$, we obtain the suitably scaled sCasimir operator
\begin{equation}
\label{qscas}
\mathcal{Q}(\hat{X}) = \frac{i\sinh \pi b^2 (2\hat{H}+\frac{i\hbar}{2})}{4 \sin \frac{\pi b^2}{2}} - \frac{4}{\hbar} \cos \frac{\pi b^2 \hbar }{2} \hat{F}^- \hat{F}^+,
\end{equation}
leading to the following explicit expression for the Casimir operator $\mathcal{C}$:
\begin{align}
\hat{H}_\text{bdy} = \mathcal{C}(\hat{X}) = -\frac{\sinh^2 \left(\pi b^2 (2\hat{H}+\frac{i\hbar}{2})\right)}{16 \sin^2 \frac{\pi b^2}{2}} &- \frac{\sin (\pi b^2 \hbar) \cosh \left(2\pi b^2 (\hat{H}+\frac{i\hbar}{2})\right)}{\hbar\sin \frac{\pi b^2}{2}}\hat{F}^-\hat{F}^+ \nonumber \\
&- \frac{16}{\hbar^2} \cos^2 \frac{\pi b^2 \hbar }{2}(\hat{F}^-)^2 (\hat{F}^+)^2.
\end{align}
Since the quantized algebra \eqref{uqosp} is precisely that of U$_q(\mathfrak{osp}(1|2,\mathbb{R}))$, this Casimir operator is related to \eqref{qsusycas} by $\hat{H}=iH$ and $\hat{F}^\pm = iF^\pm$ after including the explicit $\hbar$ dependence.

This represents a specifically ordered version of the classical $\hbar \to 0$ expression \eqref{caspoissS}, as can be immediately seen. Notice also that the sCasimir operator $\mathcal{Q}$ in \eqref{qscas} does not have a good $\hbar \to 0$ limit. This is because there is no such object in a Poisson superalgebra.

Taking the expectation value of this operator in a highest-weight state $\left\langle \text{l.w.}\right|\mathcal{C}(\hat{X})\left|\text{h.w.}\right\rangle$, we again distill only the first term, and we find an explicit expression for the Casimir on any particular representation space in terms of the $c$-number eigenvalue $h$ of $\hat{H}$. It is again useful at this point to compare the classical expression \eqref{susycomp} with the resulting quantized one:
\begin{alignat}{2}
\label{class}
&\text{classical ($\hbar \to 0$):} \quad & & {-\frac{\sinh^2 2\pi b^2h_{\text{cl}}}{16 \sin^2 \frac{\pi b^2}{2}}}, \\
\label{quant}
&\text{\phantom{classical ($\hbar \to 0$):}\llap{quantum:}} \quad & & {-\frac{\sinh^2 2\pi b^2(h+\frac{i\hbar}{4})}{16 \sin^2 \frac{\pi b^2}{2}}}.
\end{alignat}
For the continuous series irreps, we can identify the quantum expression \eqref{quant} with the explicit result in \eqref{Casexpl} by setting $h = -\frac{i}{2b}\lambda$ where $\lambda = \frac{b\hbar}{2} + is$. In the semiclassical limit where $h \, \to \, h_{\text{cl}}= -\frac{1}{2b}\lambda_{\text{cl}}$ with $\lambda_{\text{cl}}=is$, we reproduce \eqref{class}. Notice that in terms of $s$, both of these expressions are identical.

This completes our discussion of the quantization of the model.

\section{Discussion and Open Problems}
\label{s:concl}

In this work, we have investigated the quantum-group-theoretic properties that underlie Liouville (super)gravity models. An important role is played by a special representation matrix element known as the Whittaker function. We provided details on how this object leads to Liouville gravity amplitudes. As our main result, we presented an explicit group-theoretic computation of this mixed parabolic matrix element for the $q$-deformation of OSp$(1|2, \mathbb{R})$, which is relevant to $\mathcal{N}=1$ Liouville supergravity. We provided several \emph{a posteriori} checks of this proposal, and explained how it is included in a calculation of boundary operator insertions in Liouville supergravity.

We moreover explained the presence of this quantum group directly from the symmetry algebra at the Lagrangian level when using the Poisson sigma model description of dilaton gravity. We gave several arguments for the equivalence between Liouville (super)gravity and $\sinh$ dilaton (super)gravity.

There is clearly much left to explore. We end here by stating some open problems and speculations, for which we defer the full treatment to future work.

\subsubsection*{Relation to Integrability Techniques}

Representation matrix elements of mixed parabolic type have appeared before in the context of integrability of open Toda chains \cite{Sklyanin:1984sb} and their ``relativistic'' (or $q$-deformed) counterparts \cite{Ruijsenaars:1990}. For these models, simultaneous eigenfunctions of the $N$-Toda Hamiltonians (and their duals) can be found by applying the quantum inverse scattering method (QISM) to obtain eigenfunctions written in the Mellin-Barnes integral representation \cite{Kharchev:1999bh, Kharchev:2000ug, Kharchev:2000yj, Kharchev:2001rs}; see also \cite{Sciarappa:2017hds}. These eigenfunctions coincide with Whittaker functions as constructed purely from conventional representation theory techniques. In our work, we have applied the representation theory framework to find the supersymmetric $q$-deformed Whittaker functions of the simplest supergroup OSp$(1|2,\mathbb{R})$. We moreover showed that they solve a system of finite difference equations \eqref{susyfde2}. It would be interesting to learn whether integrability techniques could be applied to supersymmetric Toda systems to provide an alternative derivation of these Whittaker functions. More broadly, as far as we know, this is---next to the Toda chain systems---only the second time that these particular representation matrix elements have appeared in a direct physical context. It would be interesting to see whether integrability techniques could be applied to more deeply understand these gravitational systems.

As an immediate example, we have been focusing only on the 2-Toda chains ($N=2$), whereas relatively explicit answers for the Whittaker functions have been constructed for general $N$. The eigenfunctions of these $N$-Toda chain models can then be viewed as the required ingredients for computations in higher-spin $\mathfrak{sl}(N)$ JT and Liouville gravity. We leave this as an open avenue for the future.

\subsubsection*{$q$ a Root of Unity}

When $q$ is a root of unity, and specifically when $b^2 = p'/p$ with $p'=2$, it is well-known that the representation theory of quantum groups is more involved in the sense that some highest- (or lowest-) weight irreps become reducible but indecomposable due to the appearance of additional relations of the type $(E^\pm)^p=0$ \cite{Pasquier:1989kd} (see \cite{Slingerland:2001ea} for a nice review). For the modular double \eqref{md}, however, next to it not having either a highest- or lowest-weight irrep (and hence invalidating the presence of the above relations), it is impossible for both $q$ and $\tilde{q}$ to be roots of unity simultaneously. Since the representations of the modular double are defined to be simultaneous representations of both quantum groups, no additional irreps appear, and one is left with only the continuous self-dual representations to figure as the complete set of states in gravitational calculations even when $q$ is a root of unity.

We have indeed seen this in \cite{Mertens:2020hbs, Mertens:2020pfe}: the case that $q$ is an odd root of unity corresponds to the $(2, p)$ minimal string with $p$ odd, for which the structure of the amplitudes \eqref{final2} is similar. In particular, the Whittaker function is still the same as for generic values of $q$. The special features of the representation theory for $q$ a root of unity do play a role, though, for the minimal string: the boundary operator insertions are taken from a discrete set of values for which $S_b(2\beta_M)$ diverges. In detail, $\beta_M = -bj$ where $j= 0, \frac{1}{2}, \ldots, \frac{k}{2}$ and $q = e^{\frac{2\pi i}{p}} = e^{\frac{2\pi i}{k-2}}$. This range of values for boundary operators (the Kac table) is in one-to-one correspondence with the integrable representations of $\widehat{\mathfrak{sl}(2,\mathbb{R})}_k$, or with the type II finite-dimensional representations of U$_q(\mathfrak{sl}(2,\mathbb{R}))$.

Even though the irreps of the modular double do not qualitatively change when $q$ is a root of unity, there are some features worth mentioning. When constructing the Whittaker vectors for the continuous representations in Sections \ref{s:two} and \ref{s:three}, we found that the system of difference equations (e.g., \eqref{fde}) only has a unique solution (up to normalization) when $q$ is \emph{not} a root of unity. The solutions given in those sections are still valid, but there might exist more exotic solutions when $q$ is a root of unity.

\subsubsection*{Gravitational Boundary Conditions}

The importance of the mixed parabolic matrix elements in JT (super)gravity is immediate since they implement the asymptotic AdS boundary conditions of Brown and Henneaux \cite{Brown:1986nw, Henneaux:1999ib}. These asymptotic conditions were originally derived in $(2 + 1)$d, but JT gravity is a direct spherical dimensional reduction and hence inherits the same boundary conditions. Liouville (super)gravity, on the other hand, is different; in the bosonic case, it was argued in \cite{Mertens:2020hbs} and reviewed above to correspond to a dilaton gravity theory with a $\sinh$ dilaton potential. We have argued for a similar statement in the supersymmetric case. The classical solutions of these dilaton gravity models can be interpreted as Yang-Baxter deformations of the JT solutions \cite{Kyono:2017jtc, Kyono:2017pxs}, and in particular, the asymptotics is drastically modified, with a curvature singularity at the holographic boundary. It would be interesting to better understand this asymptotic behavior in the context of holography. It is tempting to speculate that one can understand this in terms of a $q$-deformed boundary CFT (see, e.g., \cite{Shiraishi:1995rp}), but more work is needed. Can we reason along these lines to understand why the same mixed parabolic representation matrix elements appear to play a role here?

\subsubsection*{Relation to $q$-Deformed BF Description?}

Aside from the Poisson sigma model Lagrangian discussed in Section \ref{s:dilgrav}, there is a second description that seems to come closer to reproducing the actual structure of amplitudes: this is $q$-deformed BF theory \cite{Blau:1993tv, Aganagic:2004js}.

Let us try to set up the problem a bit more explicitly. For simplicity, we focus on the bosonic model governed by U$_q(\mathfrak{sl}(2,\mathbb{R}))$. One way to write down a Lagrangian for the $q$-deformed BF model is to make the $B$-field periodic in the undeformed BF model:
\begin{equation}
\label{qBF}
S_{\text{BF}} = \int \Tr(BF), \qquad B \sim B + \frac{2i}{b^2}.
\end{equation}
The calculation of the disk partition function proceeds by abelianization of the $B$-field, after which the periodicity constraint on $B$ causes the Jacobian of this procedure to ``$q$-deform.'' For compact groups, this Jacobian becomes the quantum dimension $\dim_q R$ of the representation $R$ appearing in amplitudes. For noncompact $\mathfrak{sl}(2, \mathbb{R})$, the problem is to find a way to end up with the continuous measure $\rho(s) = \sinh 2\pi b s \sinh \frac{2\pi s}{b}$ in terms of the $s$-label of the continuous series irreps.

Secondly, we add the classical boundary Hamiltonian \eqref{caspoiss}:\footnote{This term will break large gauge invariance, as it should. Gauge transformations that vanish at the boundary are preserved.}
\begin{equation}
\label{qBFo}
S_\partial = \oint H_{\text{bdy}}(B), \qquad H_{\text{bdy}}(B) = B_+ B_- + \frac{\cosh 2\pi b^2 B_H}{2\pi b^2\sin \pi b^2},
\end{equation}
where $(B_+, B_-, B_H)$ are the three $\mathfrak{sl}(2,\mathbb{R})$ components of $B$. The abelianization procedure mentioned above \cite{Blau:1993tv, Aganagic:2004js} effectively sets $B_+ = B_- = 0$ and reduces the calculation to an integral over the Cartan contribution $B_H$. This reduces the above Casimir to its classical description \eqref{clascomp}, and was in fact a motivation for writing that expression in the first place.

The result of this calculation should then be directly matched with Liouville gravity amplitudes \cite{Mertens:2020hbs}. It would be interesting to fill in the details of this argument.

It is important to emphasize that we implement two independent modifications compared to the undeformed case: the periodicity of $B$ (following \cite{Blau:1993tv}), and the change of boundary Hamiltonian from a quadratic function to a hyperbolic-cosine function. More broadly, earlier work has classified which ingredients in 2d YM amplitudes (the measure, the Casimir, the exponential) become $q$-deformed \cite{Szabo:2013vva}, and it would be interesting to understand how and why Liouville (super)gravity (or $\sinh$ dilaton (super)gravity) requires these specific deformations of the JT amplitudes.

A direct Lagrangian rewriting of the $q$-deformed BF model \eqref{qBF}--\eqref{qBFo} as the specific dilaton gravity model \eqref{actgen} or in terms of its Poisson sigma model description would shed significant light on these results.

\subsubsection*{3d Gravity}

It has been known for a long time that calculations in pure 3d gravity, when described in Chern-Simons language, are likewise governed by $q$-deformed SL$(2, \mathbb{R})$ ingredients (see, e.g., \cite{Jackson:2014nla, McGough:2013gka}). However, there are some differences. In particular, when computing the solid torus amplitude (which is the 3d analogue of the 2d disk diagram), one finds that the Casimir contribution is \emph{not} $q$-deformed but the measure $\rho(s)$ is.\footnote{This calculation will be reported elsewhere.}

The fact that the Casimir operator is not $q$-deformed in 3d gravity amplitudes can be appreciated rather quickly by writing out the Chern-Simons action on a solid torus $\mathcal{M}$ in $(t, r, \phi)$-coordinates as:
\begin{align}
S_{\text{CS}} &= \frac{k}{4\pi}\int_\mathcal{M} d^3 x\, \epsilon^{\mu\nu\rho}\Tr\left(A_\mu \partial_\nu A_\rho + \frac{2}{3} A_\mu A_\nu A_\rho \right) \\
&= \frac{k}{2\pi} \int_\mathcal{M} \Tr\left(A_\phi (\partial_t A_r - \partial_r A_t) +A_r \partial_\phi A_t  + A_\phi[A_t,A_r] \right) +  \frac{k}{4\pi}  \oint_{\partial \mathcal{M}} \Tr A_t^2, \nonumber
\end{align}
imposing the boundary condition $A_t=\left.A_\phi\right|_{\partial \mathcal{M}}$. This action dimensionally reduces to the 2d BF action upon identifying $B \equiv A_\phi$ and setting $\partial_\phi =0$. In particular, one can already see the appearance of the boundary Hamiltonian in the form of the quadratic Casimir, and \emph{not} in terms of a $q$-deformed version of it. We leave further investigation to future work.

\subsubsection*{Arbitrary Dilaton Gravity Models}

\begin{figure}[!htb]
\centering
\includegraphics[width=0.15\textwidth]{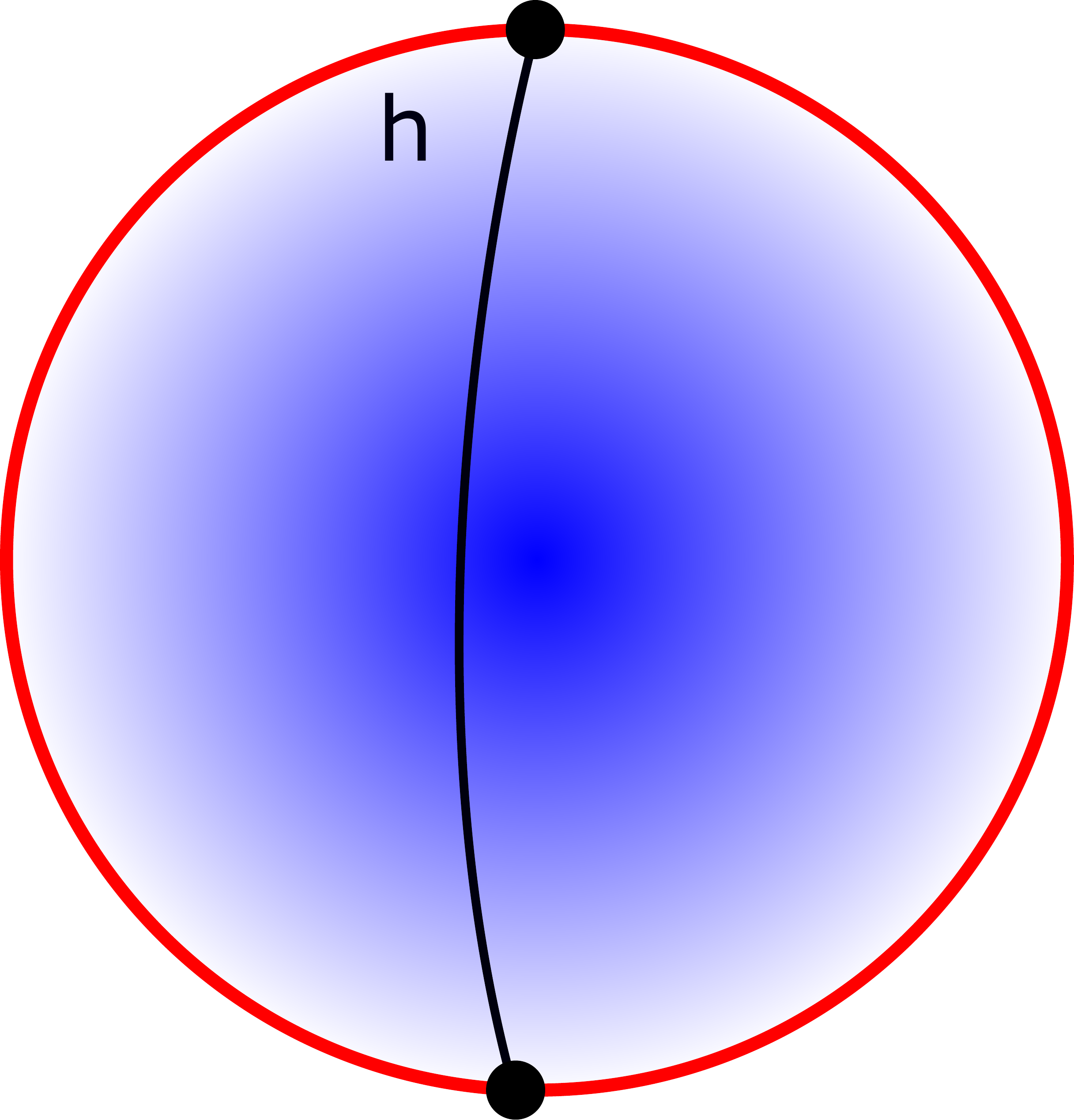}
\caption{Deformed JT gravity disk amplitude with two boundary operators, interpretable as undeformed JT gravity with a gas of defects localized in the interior of the disk (blue blob).}
\label{vertexcoupling}
\end{figure}

Recent work \cite{Witten:2020ert, Maxfield:2020ale, Witten:2020wvy} has analyzed deformations of JT gravity. These deformations correspond to modified dilaton potentials of the type
\begin{equation}
\label{deform}
W(\Phi) = 2\Phi + \sum_i \epsilon_i e^{-\alpha_i \Phi}, \qquad \pi < \alpha_i < 2\pi.
\end{equation}
Such potentials preserve the JT asymptotics as $\Phi \to \infty$, which, owing to the coordinate choice of \eqref{bhgen} where $\Phi = r$, matches with the asymptotic AdS$_2$ region $r\to +\infty$. Amplitudes in such models can be found by series-expanding the corrections and interpreting them as a gas of defects (of the type studied in \cite{Mertens:2019tcm}) within JT gravity. The result is a modified density of states that incorporates this defect gas. It is not difficult to convince oneself, again by series-expanding the deformation, that a similar procedure is possible when including boundary operators (Figure \ref{vertexcoupling}).

The result is that in any amplitude with multiple boundary operators, one only replaces the density factors by the deformed ones:
\begin{equation}
\rho_{\text{JT}}(k) = k \sinh(2\pi k) \to \rho_{\text{def}}(k),
\end{equation}
where $\rho_{\text{def}}(k)$ is given in \cite{Maxfield:2020ale, Witten:2020wvy}. The vertex functions (or $3j$-symbols) and propagation factors in the amplitude are the same as for undeformed JT gravity. This observation was also made in \cite{Iliesiu:2021ari}, and argued to hold even when including higher-genus corrections.

Liouville (super)gravity provides an exception to these statements, since the $3j$-symbols that we require there \eqref{idwhitsb} are not those appearing in JT gravity amplitudes \eqref{3jcl}. The interpretation is that the dilaton (pre)potential of Liouville (super)gravity does not fall into the class of JT deformations of \eqref{deform}, due to the asymptotics of the $\sinh$ function.

This set of observations has an intuitive bulk interpretation. The vertex functions themselves are drawn as three-vertices at the holographic boundary (Figure \ref{vertexcoupling}). Since the gas of defects for the deformations \eqref{deform} does not reach the actual boundary, these local three-vertices do not feel the deformation. However, if the gas of defects does reach the boundary, as it does when viewing Liouville (super)gravity as a deformation of JT (super)gravity, then the resulting vertex functions are different. This leads to a division of dilaton gravity models into different classes, where all entries within a given class have the same dilaton asymptotics and hence the same set of vertex functions, but different densities of states.\footnote{It would be interesting to collect evidence for this classification, e.g., by looking at the set of deformations that preserve the $\sinh$ asymptotics of Liouville gravity \cite{Turiaci:2020fjj}.} This is especially intriguing when combined with the Poisson sigma model description of Section \ref{s:dilgrav}, in which generic dilaton potentials lead to nonlinear symmetry algebras and apparently similar structures for the amplitudes, even though there is no particularly useful known group structure underlying the generic case. It would be interesting to understand this situation better.

\subsubsection*{From Gauge Theory to Gravity}

Locally, lower-dimensional gravity is described by a gauge theory. Globally, however, there are mismatches that need to be properly appreciated before making detailed comparisons. Let us interpret our results from this perspective.

The analysis of Sections \ref{s:two} and \ref{s:three} required explicit knowledge of the modular double of the quantum group U$_q(\mathfrak{osp}(1|2,\mathbb{R}))$, and not just the quantum group itself. In Section \ref{s:dilgrav}, we saw no indication of this modified structure. This observation parallels what happens in the undeformed case \cite{Fan:2021wsb}, where manipulations at the level of the Lagrangian and its symmetry group are insensitive to global algebraic information. This information is contained in the precise path integration cycle for the gauge field $A_\mu$ in the Poisson sigma model (PSM) description or the $q$-deformed BF description mentioned above. The restriction is enforced by demanding that the gauge field $A_\mu$ correspond to smooth geometries (no punctures or cusps) \cite{Verlinde:1989ua, Witten:2007kt}, which can be done very explicitly in 2d. See, e.g., Section 4.1 of \cite{Fan:2021wsb} for an intuitive argument. In the undeformed case, the more precise algebraic structure that implements this smoothness constraint is the positive subsemigroup, which can in turn be found as the $b\to 0$ limit of the modular double of the quantum group U$_q(\mathfrak{sl}(2, \mathbb{R}))$. This was the main motivation for pursuing the subsemigroup approach in \cite{Blommaert:2018oro, Blommaert:2018iqz, Fan:2021wsb}.\footnote{We remark that going to the modular double might look like an expansion instead of a reduction of the model, but it has been shown \cite{Ponsot:1999uf, Ponsot:2000mt, Bytsko:2002br, Bytsko:2006ut, Ip} that the nontrivial irreducible representations of the modular double consist only of the continuous representations studied in Sections \ref{s:two} and \ref{s:three}, and are hence in fact a subset of those of the quantum group that one started with.}

Next to this modification of the algebraic structure, a second modification is necessary to make contact with gravity: one needs to take into account the overcounting induced by large diffeomorphisms. For JT gravity, this is done in the context of hyperbolic geometry, and is baked into the definition of Weil-Petersson volumes \cite{Saad:2019lba, Dijkgraaf:2018vnm}. For Liouville gravity, in the particular case where the matter sector is the $(2,p)$ minimal model with $p$ an odd integer, an analogous statement was made in \cite{Mertens:2020hbs} in terms of a $q$-deformation of these Weil-Petersson volumes. This modification is required to make contact with the description of the theory in terms of Hermitian matrix models.

Finally, a third step is to sum over topologies by hand. This is not natural from the gauge theory perspective, but it is easy to accommodate at least at the perturbative level in the genus expansion.

Our discussion implies that all three steps for going from gauge theory to gravity (restricting the algebraic structure, modding out by large diffeomorphisms, and summing over topologies) are also necessary in the specific case of Liouville gravity. From the dilaton gravity perspective, both JT and Liouville gravity simply correspond to specific choices of dilaton potential. Therefore, it is natural to conjecture more generally that going from $q$-deformed BF or PSM amplitudes to dilaton gravity amplitudes requires the same three-step modification.\footnote{As preliminary evidence, performing the second and third steps is natural whenever one needs to match onto a matrix model description. And indeed, at least for the dilaton gravity models of the class \eqref{deform}, it was proven in \cite{Maxfield:2020ale, Witten:2020wvy} that these models are dual to matrix integrals.}

We summarize the multi-step process to go from gauge theory to gravity in Table \ref{gaugegravity}.

\begin{table}[!htb]
\centering
\begin{equation*}
\begin{array}{c||c|c}
 \textit{Start with gauge theory}& \text{SL}(2,\mathbb{R}) \text{ BF} &  \text{U}_q(\mathfrak{sl}(2,\mathbb{R})) \, q\text{-BF or PSM} \\
\hline \hline
\begin{array}{c} \textbf{Step 1} \\ \text{Restrict to smooth geometries}\end{array} & \begin{array}{c} \text{positive subsemigroup} \\ \text{SL}^+(2,\mathbb{R}) \end{array}& \begin{array}{c} \text{modular double } \\ \text{U}_q(\mathfrak{sl}(2,\mathbb{R})) \otimes \text{U}_{\tilde{q}}(\mathfrak{sl}(2,\mathbb{R})) \end{array}\\
\hline
\begin{array}{c}  \textbf{Step 2} \\ \text{Remove large diffeos} \end{array} & \text{Weil-Petersson volumes} & \text{$q$-Weil-Petersson volumes} \\
\hline 
\begin{array}{c} \textbf{Step 3} \\ \text{Sum over topologies}\end{array}  & \text{ad hoc} & \text{ad hoc} \\
\hline
\textit{End with gravity} & \text{JT gravity} & \text{Liouville gravity}
\end{array}
\end{equation*}
\caption{Passage from gauge theory to gravity as a multi-step process.}
\label{gaugegravity}
\end{table}

\section*{Acknowledgements}

We thank D.\ Grumiller for an early discussion on 2d dilaton supergravity. The work of YF was supported by the National Science Foundation under Grant No.\ PHY-1914679.  TM gratefully acknowledges financial support from Research Foundation Flanders (FWO Vlaanderen).

\appendix

\section{Liouville (Super)gravity: Setup and Fixed-Length Amplitudes}
\label{app:bosonic}

We recall here the definition of 2d Liouville gravity and supergravity, as well as the computation of fixed-length amplitudes therein.

The non-critical string is defined by coupling a 2d CFT described by the action $S_M[\chi;g]$ to the gravitational fields as
\begin{equation}
Z = \sum_\text{topologies} \int \frac{\mathcal{D}g\, \mathcal{D}\chi}{\operatorname{Vol}(\operatorname{Diff})} e^{ - S_M[\chi; g] - \mu_0 \int_\Sigma d^2 z \sqrt{g}},
\end{equation}
where a bare cosmological constant term has been added. It is well-known that upon going to conformal gauge $g_{\mu\nu} = e^{2b\phi} \hat{g}_{\mu\nu}$, where $\hat{g}$ is a reference metric, and taking into account the conformal anomaly, the action reduces to a sum of three 2d CFTs: $S_L + S_M + S_\text{gh}$, with vanishing conformal anomaly $c_L + c_M + c_\text{gh} =0$. The three pieces are as follows:
\begin{itemize}
\item
The Liouville action describing the conformal factor of the 2d geometry:
\begin{equation}
\label{liouact}
S_L = \frac{1}{4\pi} \int_{\Sigma} d^2 x\left[ (\hat{\nabla} \phi)^2 + Q \hat{R} \phi + 4 \pi \mu e^{2 b \phi} \right] + \frac{1}{2\pi} \oint_{\partial \Sigma} dx\left[ Q \hat{K} \phi + 2\pi \mu_B e^{b \phi} \right],
\end{equation}
where $Q=b+1/b$ and $c_L = 1 + 6 Q^2 > 25$. For our purposes, we added a boundary term with boundary cosmological constant $\mu_B$, allowing for Neumann-like boundary conditions on $\phi$ describing a piece of FZZT-brane on which the worldsheet can end. 
\item
The matter 2d CFT with $c_M = 1- 6 \mathfrak{q}^2$ where $\mathfrak{q}=1/b-b$. For the purposes of this paper, it is convenient to parametrize it as a timelike Liouville CFT:
\begin{equation}
S_M[\chi] = \frac{1}{4\pi} \int_{\Sigma} d^2 x\left[ -(\hat{\nabla} \chi)^2 - \mathfrak{q} \hat{R} \chi + 4 \pi \mu_M e^{2 b \chi} \right]  - \frac{1}{2\pi} \oint_{\partial \Sigma} dx\, \mathfrak{q} \hat{K} \chi\, ,
\end{equation}
where we take Dirichlet boundary conditions on the field $\chi \to \infty$ on any boundary. This corresponds to the vacuum brane boundary condition.
\item
The $bc$ ghost CFT $S_\text{gh}$ with $c_\text{gh} = -26$.
\end{itemize}
We are interested in boundary vertex operators. Within the Liouville parametrization above, primary CFT operators are constructed as:
\begin{alignat}{2}
\text{Liouville :}& \mbox{ } e^{\beta \phi} \qquad & \Delta_\beta &= \beta(Q-\beta), \\
\text{Matter :}& \mbox{ } e^{\beta_M \chi} \qquad & \Delta_{\beta_M} &= \beta_M(\mathfrak{q}+\beta_M), \label{bdylioupar}
\end{alignat}
such that we get the open string tachyon vertex operators by gravitationally dressing the matter part as
\begin{equation}
\mathcal{B}_{\beta_M} \sim \oint_{\partial\Sigma} dx\, e^{\beta_M \chi}e^{\beta \phi} \simeq ce^{\beta_M \chi}e^{\beta \phi}
\end{equation}
with the restriction that $\Delta_{\beta_M} + \Delta_\beta = 1$, which leads to $\beta=b-\beta_M$.\footnote{The solution $\beta = 1/b + \beta_M$ is related by a Liouville reflection $\beta \to Q - \beta$.} Amplitudes with insertions of these vertex operators on different geometries can then be computed, in principle, using string theory techniques.

From the 2d gravity perspective, we can obtain amplitudes of fixed boundary length by Fourier transforming any amplitude as
\begin{equation}
{-i}\int_{i\mathbb{R}} d\mu_B\, e^{\mu_B \ell_{\text{L}}} \cdots,
\end{equation}
since by \eqref{liouact}, we bring down a factor of $\delta (\ell_{\text{L}} - \int e^{b\phi})$ in the path integral, where $\int e^{b\phi}$ is precisely the boundary length as measured by the 2d metric $g_{\mu\nu}$ that we started with.

The extension to Liouville supergravity proceeds along similar lines. We only point out some of the differences here.

Starting with any 2d matter SCFT and coupling to 2d supergravity, we can reach the combined action $S_L + S_M + S_\text{gh}$ in terms of the $\mathcal{N}=1$ super-Liouville CFT, the matter SCFT that we started with, and the $bc$ and $\beta \gamma$ ghost systems. The total central charge again vanishes, $c_L + c_M + c_\text{gh} =0$, where $c_\text{gh}=-15$ in this case.

Similarly as in the bosonic case, we can construct worldsheet diffeomorphism-invariant boundary operator insertions as
\begin{equation}
\mathcal{B} = \bigl( c e^{-\varphi}\bigr) e^{\frac{\beta}{2}\phi} e^{\frac{\beta_M}{2} \chi},
\end{equation}
where now $\Delta_\beta = \frac{1}{2}\beta(Q-\beta)$ and $\Delta_{\beta_M} = \frac{1}{2}\beta_M(\mathfrak{q}+\beta_M)$, restricted according to $\Delta_\beta + \Delta_{\beta_M} = 1/2$. The solution is again $\beta = b - \beta_M$. The factor of $c e^{-\varphi}$ is the ghost piece of the vertex operator. However, the open string tachyon vertex operators of interest here can be written as follows:
\begin{equation}
\mathcal{B}_{\beta_M} = \bigl(c e^{-\varphi}\bigr) \left[e^{\frac{\beta}{2} \phi} e^{\frac{\beta_M}{2} \chi} + (\text{superpartner})\right], 
\end{equation}
where we added the worldsheet superpartner of the operator in a particular linear combination. It was observed in \cite{Mertens:2020pfe} that it is this combined boundary operator whose amplitudes behave well in the fixed-length basis.

For the super-Liouville part, one can analogously define FZZT boundary conditions and, from there, transform to the fixed-length basis in the original metric:
\begin{equation}
{-i}\int_{\mathcal{C}} d\mu_B\, e^{\mu_B^2\ell_{\text{L}}} \cdots, \qquad \mathcal{C}={\mu_B^2 - \mu :-i\infty \to+i\infty},
\end{equation}
where the integration is performed along the half-hyperbola $\mathcal{C}$ in the $\mu_B$-plane. This again brings down a factor of $\delta (\ell_{\text{L}} - \int e^{b\phi})$ in the functional integral.

\section{Special Functions and Identities}
\label{app:spec}

We collect and define here the double sine functions and Barnes identities that we need in the main text.

\subsection{Double Sine Functions \texorpdfstring{$S_b(x)$}{Sb(x)}}
\label{app:def}

The Barnes double gamma function $\Gamma_2(z|\omega_1,\omega_2)$ is defined by the series expression
\begin{equation}
\log \Gamma_2(z|\omega_1,\omega_2) \equiv \left.\left(\frac{d}{dt}\sum_{n_1,n_2=0}^{+\infty}\frac{1}{(z+\omega_1n_1+\omega_2n_2)^{t}} \right) \right|_{t=0}.
\end{equation}
The ``$b$-deformed'' gamma function $\Gamma_b(x)$ is conventionally defined so that $\Gamma_b(Q/2)=1$:
\begin{equation}
\Gamma_b(x) \equiv \frac{\Gamma_2(x | b, b^{-1})}{\Gamma_2 (Q/2|b,b^{-1})}.
\end{equation}
It satisfies the shift properties
\begin{align}
\Gamma_b(x+b) = \frac{\sqrt{2\pi}b^{bx-1/2}}{\Gamma(bx)}\Gamma_b(x), \qquad \Gamma_b(x+1/b) = \frac{\sqrt{2\pi}b^{-x/b+1/2}}{\Gamma(x/b)}\Gamma_b(x). 
\end{align}
The double sine function $S_b(x)$ is then constructed as
\begin{equation}
S_b(x)\equiv \frac{\Gamma_b(x)}{\Gamma_b(Q-x)},
\end{equation}
which satisfies the defining functional relations
\begin{equation}
S_b(Q-x) = 1/S_b(x), \qquad S_b(x+b) = 2 \sin \pi  b x \, S_b(x), \qquad S_b\left(x+\frac{1}{b}\right) = 2 \sin \frac{\pi x}{b} \, S_b(x).
\end{equation}
The double sine functions satisfy the following $q$-deformed generalization of the first Barnes lemma:
\begin{align}
\label{qbarnebos}
\int_{-\infty}^{+\infty} d\tau\, e^{\pi \tau (\alpha+\beta+\gamma+\delta)}&S_b(\alpha+i\tau)S_b(\beta+i\tau)S_b(\gamma-i\tau)S_b(\delta-i\tau) \\
&= e^{\pi i (\alpha\beta-\gamma\delta)} \frac{S_b(\alpha+\gamma)S_b(\alpha+\delta)S_b(\beta+\gamma)S_b(\beta+\delta)}{S_b(\alpha+\beta+\gamma+\delta)}. \nonumber
\end{align}
In Section \ref{s:three}, we require the combinations
\begin{align}
S_{\NS}(x) &= S_b\left(\frac{x}{2}\right) S_b\left(\frac{x}{2} + \frac{Q}{2}\right), \qquad S_{\R}(x) = S_b\left(\frac{x}{2} + \frac{b}{2}\right) S_b\left(\frac{x}{2} + \frac{1}{2b}\right),
\end{align}
which satisfy
\begin{equation}
S_{\NS}(Q-x) = 1/S_{\NS}(x), \qquad S_{\R}(Q-x) = 1/S_{\R}(x),
\end{equation}
as well as the crucial functional shift relations
\begin{alignat}{2}
S_{\NS}(x+b) &= 2 \cos \left(\frac{\pi b x}{2} \right) S_{\R}(x), \qquad & S_{\NS}\left(x+\frac{1}{b}\right) &= 2 \cos \left(\frac{\pi x}{2b} \right) S_{\R}(x), \\
S_{\R}(x+b) &= 2 \sin \left(\frac{\pi b x}{2} \right) S_{\NS}(x), \qquad & S_{\R}\left(x+\frac{1}{b}\right) &= 2 \sin \left(\frac{\pi x}{2b} \right) S_{\NS}(x). \nonumber
\end{alignat}

\subsection{\texorpdfstring{$q$}{q}-Deformed Supersymmetric Barnes Identity}
\label{app:qbarnes}

For external indices $\rho_{A,B,C} = 0, 1$, as well as $S_{0}(x) = S_{\rm R}(x)$ and $S_1(x) = S_{\rm NS}(x)$, we have the $q$-deformed supersymmetric Barnes identity \cite{Hadasz:2013bwa}:
\begin{align}
\sum_{\sigma=0,1} \int_{-\infty}^{+\infty} & d\tau\, e^{-\frac{\pi \tau}{2}(\alpha+\beta+\gamma+\delta)}S_{\rho_A+\sigma}(\alpha+i\tau)S_{\rho_B+\sigma}(\beta+i\tau)S_{\rho_C+\sigma}(\gamma-i\tau)S_{1+\sigma}(\delta-i\tau) \nonumber\\
&= 2 e^{-\frac{i\pi}{2}(\alpha\beta-\gamma\delta)}\frac{S_{\rho_A+\rho_C+1}(\alpha+\gamma)S_{\rho_A}(\alpha+\delta)S_{\rho_B+\rho_C+1}(\beta+\gamma)S_{\rho_B}(\beta+\delta)}{S_{\rho_A+\rho_B+\rho_C}(\alpha+\beta+\gamma+\delta)}.
\end{align}
Addition of indices takes place modulo 2. It is convenient for later reference to unpack this identity into
\begin{align}
\int_{-\infty}^{+\infty} & d\tau\, e^{-\frac{\pi \tau}{2}(\alpha+\beta+\gamma+\delta)}\times \left[S_{\R}(\alpha+i\tau)S_{\R}(\beta+i\tau)S_{\NS}(\gamma-i\tau)S_{\NS}(\delta-i\tau) \right. \nonumber \\
&\hspace{5cm} \left. {} + S_{\NS}(\alpha+i\tau)S_{\NS}(\beta+i\tau)S_{\R}(\gamma-i\tau)S_{\R}(\delta-i\tau)\right] \nonumber \\
&=2 e^{-\frac{i\pi}{2}(\alpha\beta-\gamma\delta)}\frac{S_{\R}(\alpha+\gamma)S_{\R}(\alpha+\delta)S_{\R}(\beta+\gamma)S_{\R}(\beta+\delta)}{S_{\NS}(\alpha+\beta+\gamma+\delta)}
\end{align}
and
\begin{align}
\int_{-\infty}^{+\infty} & d\tau\, e^{-\frac{\pi \tau}{2}(\alpha+\beta+\gamma+\delta)}\times \left[S_{\NS}(\alpha+i\tau)S_{\NS}(\beta+i\tau)S_{\NS}(\gamma-i\tau)S_{\NS}(\delta-i\tau) \right. \nonumber \\
&\hspace{5cm} \left. {} + S_{\R}(\alpha+i\tau)S_{\R}(\beta+i\tau)S_{\R}(\gamma-i\tau)S_{\R}(\delta-i\tau)\right]\nonumber\\
&=2 e^{-\frac{i\pi}{2}(\alpha\beta-\gamma\delta)}\frac{S_{\NS}(\alpha+\gamma)S_{\NS}(\alpha+\delta)S_{\NS}(\beta+\gamma)S_{\NS}(\beta+\delta)}{S_{\NS}(\alpha+\beta+\gamma+\delta)}
\end{align}
and
\begin{align}
\int_{-\infty}^{+\infty} & d\tau\, e^{-\frac{\pi \tau}{2}(\alpha+\beta+\gamma+\delta)} \times\left[S_{\R}(\alpha+i\tau)S_{\NS}(\beta+i\tau)S_{\R}(\gamma-i\tau)S_{\NS}(\delta-i\tau) \right. \nonumber \\
&\hspace{5cm} \left. {} + S_{\NS}(\alpha+i\tau)S_{\R}(\beta+i\tau)S_{\NS}(\gamma-i\tau)S_{\R}(\delta-i\tau)\right]\nonumber\\
&=2 e^{-\frac{i\pi}{2}(\alpha\beta-\gamma\delta)}\frac{S_{\NS}(\alpha+\gamma)S_{\R}(\alpha+\delta)S_{\R}(\beta+\gamma)S_{\NS}(\beta+\delta)}{S_{\NS}(\alpha+\beta+\gamma+\delta)}. \label{qbarnes}
\end{align}
It is this last identity that we need in Section \ref{s:three}.

\section{Alternative Realization of the \texorpdfstring{U$_q(\mathfrak{osp}(1|2, \mathbb{R}))$}{Uq(osp(1|2, R))} Continuous Series}
\label{alternative}

An alternative realization of the algebra
\begin{align}
\label{salgeba}
KF^\pm = q^{\pm\frac{1}{2}}F^{\pm}K , \qquad \left\{F^+,F^-\right\} = -\frac{K^2-K^{-2}}{8(q^{1/2}-q^{-1/2})},
\end{align}
mirroring the bosonic realization \eqref{KLT} studied in \cite{Kharchev:2001rs}, is found by setting
\begin{align}
K &= e^{-\pi i b^2 j }T_{ib/2} \left(\begin{array}{c|c}
1 &0 \\
\hline
0 &q^{1/2} 
\end{array}\right), \\
F^+ &= \left(  \begin{array}{c|c}
0 &-\frac{1}{2} e^{2\pi b  t}\frac{q^{1/2} e^{2 i \pi b^2 j} + q^{3/2} e^{-2i\pi b^2j} T_{ib}}{q^{1/2}+q^{-1/2}} \\
\hline
\frac{1}{4}\frac{e^{2 i \pi b^2j} - e^{-2i\pi b^2 j}T_{ib}}{q^{1/2}-q^{-1/2}} & 0 
\end{array}\right), \\
F^- &=  \left(  \begin{array}{c|c}
0 & \frac{1}{2}\frac{q^{1/2} +  q^{-1/2} T_{-ib}}{q^{1/2}+q^{-1/2}} \\
\hline
 -\frac{1}{4}e^{-2\pi b t}\frac{-1 + T_{-ib}}{q^{1/2}-q^{-1/2}} & 0 
\end{array}\right),
\end{align}
where $\alpha = 1/2b - 2bj$ as in the quantum algebra \eqref{KLT} in the main text.

Expanding these operators in the $b\to 0$ limit, we obtain the differential operators
\begin{align}
\hat{H} &= \left(\begin{array}{c|c}
x\partial_x -j &0 \\
\hline
0 &x\partial_x  -j +1/2
\end{array}\right) = x\partial_x + \frac{1}{2} \vartheta \partial_\vartheta -j, \\
\hat{F}^+ &= \left(  \begin{array}{c|c}
0 &  - \frac{1}{2}x \\
\hline
- \frac{1}{2}x \partial_x + j & 0 
\end{array}\right) = - \frac{1}{2}x \partial_\vartheta - \frac{1}{2}x \vartheta \partial_x + j \vartheta, \\
\hat{F}^- &= \left(  \begin{array}{c|c}
0 & \frac{1}{2} \\
\hline
\frac{1}{2}\partial_x & 0 
\end{array}\right) = \frac{1}{2}\left(\partial_\vartheta + \vartheta \partial_x\right), \\
\hat{E}^+ &= \left(\begin{array}{c|c}
-x^2\partial_x + 2jx  &0 \\
\hline
0 & -x^2\partial_x  + 2jx - x
\end{array}\right) = -x^2 \partial_x - x \vartheta \partial_\vartheta + 2jx, \\
\hat{E}^- &= \left(\begin{array}{c|c}
\partial_x &0 \\
\hline
0 &\partial_x
\end{array}\right) = \partial_x, 
\end{align}
which constitute the infinitesimal version of the group action of OSp$^+(1|2,\mathbb{R})$ on $L^2(\mathbb{R}^{+1|1})$ \cite{Fan:2021wsb}:
\begin{align}
(g \cdot f) (x,\vartheta) &= (bx+d+\delta\vartheta)^{2j}f\left(\frac{ax+c+\beta\vartheta}{bx+d+\delta\vartheta}, -\frac{\alpha x + \gamma -e \vartheta}{bx+d + \delta \vartheta} \right).
\end{align}
This is the Borel-Weil realization of the principal series representations of OSp$^+(1|2,\mathbb{R})$, defined on $L^2$ functions on the super half-line $\mathbb{R}^{+1|1} \equiv \left\{(x,\vartheta) \,|\, x>0\right\}$.

This construction mirrors the results \eqref{BW} in the bosonic case. It would be interesting to develop the story from the perspective of this carrier space.

\section{\texorpdfstring{$q$}{q}-Deformed BesselI Function}
\label{app:besseli}

\begin{figure}[!htb]
\centering
\includegraphics[width=0.7\textwidth]{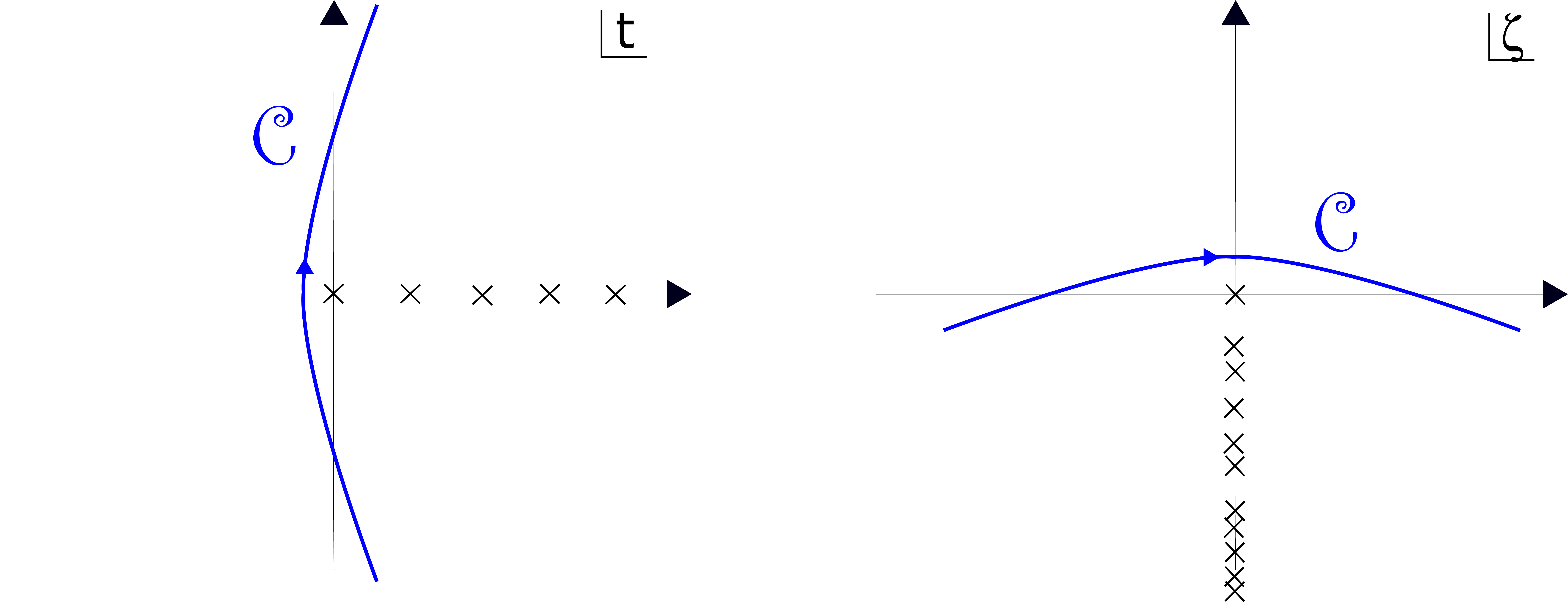}
\caption{Left: contour $\mathcal{C}$ used to define $I_\alpha(x)$. Right: contour $\mathcal{C}$ used to define the $q$-deformed version $\mathcal{I}^{\epsilon}_{\alpha}(x)$.}
\label{contourI}
\end{figure}

The classical modified Bessel function of the first kind can be written as
\begin{align}
\label{besseli}
I_\alpha(x) = \frac{1}{2\pi i}\int_{\mathcal{C}} dt\, \frac{\Gamma(-t)}{\Gamma(\alpha+t+1)}i^{2t}\left(\frac{x}{2}\right)^{2t+\alpha} = \sum_{n=0}^{+\infty} \frac{1}{n! \, \Gamma(n+\alpha+1)}\left(\frac{x}{2}\right)^{2n+\alpha},
\end{align}
where the second equality results from picking up the residues from the poles of the $\Gamma(-t)$ in the right half-plane. The initial contour $\mathcal{C}$ runs from infinity at $-\pi/2 < \arg(t) < 0$ to infinity at $0 < \arg(t) < \pi/2$, encircling the origin, as in Figure \ref{contourI} (left).

Since $I_\alpha(x)$ is real for $\alpha \in \mathbb{R}$, we can write it equivalently as
\begin{align}
I_\alpha(x) = \frac{1}{2\pi i}\int_{\mathcal{C}} dt\, \frac{\Gamma(-t)}{\Gamma(\alpha+t+1)}\cos \pi t \left(\frac{x}{2}\right)^{2t+\alpha}.
\end{align}
In this form, one readily proves the equality\footnote{Namely, by using the Euler reflection formula, and a shift $t \to t+\alpha$. After combining both integrals, one can move the contour to follow $i\mathbb{R}$, to the left of the origin.}
\begin{equation}
\label{relKI}
K_\alpha(x) = \frac{\pi}{2 \sin \pi \alpha}\left( I_{-\alpha}(x) - I_{\alpha}(x) \right).
\end{equation}
We define the $q$-deformed BesselI function by the integral
\begin{equation}
\label{qbesseli}
\boxed{\mathcal{I}^{\epsilon}_{\alpha}(x) \equiv b e^{-\alpha \pi x} \int_{\mathcal{C}} d\zeta\, g_\mu^{i\zeta} g_\nu^{i\zeta+\alpha}  \frac{S_b(-i\zeta)}{S_b(i\zeta+\alpha +b)}\cosh \frac{\pi \zeta}{b}\, e^{-\pi i \epsilon(\zeta^2 - \alpha i \zeta)}e^{-2\pi i  \zeta x},}
\end{equation}
with the contour $\mathcal{C}$ of Figure \ref{contourI} (right).

Using the elegant residue formula
\begin{equation}
\operatorname{Res}\left.S_b\right|_{x=-mb-n/b} = \frac{S_b(Q)}{2\pi}\frac{(-)^{n+m+nm}}{S_b(mb+n/b+Q)},
\end{equation}
the integral \eqref{qbesseli} can be evaluated by contour deformation into the lower half-plane, e.g.:
\begin{align}
\label{taylor}
&\mathcal{I}^{\epsilon=0}_{\alpha}(x) = b S_b(Q) \sum_{n,m=0}^{+\infty} \frac{(-)^{n+nm}g_\mu^{mb + n/b}g_\nu^{mb + n/b+\alpha}\cos(\frac{\pi n}{b^2})}{S_b(bm+n/b+Q)S_b(bm+n/b+\alpha + b)}e^{-(2mb+2n/b+\alpha) \pi x} \\
&+ b S_b(Q) \sum_{n,m=0}^{+\infty} \frac{(-)^{n+nm}g_\mu^{mb + n/b-\alpha+1/b}g_\nu^{mb + n/b+1/b}\cos(\frac{\pi }{b^2}(n+1)-\frac{\pi \alpha}{b})}{S_b(bm+n/b+Q)S_b(\frac{2}{b} + bm+n/b-\alpha + b)}e^{-(2mb+2n/b+2/b-\alpha) \pi x}. \nonumber 
\end{align}
The classical $b\to 0$ limit can also be calculated directly, and indeed yields the usual BesselI function ($\phi = \pi b x$):
\begin{equation}
\lim_{b\to 0} \mathcal{I}^{\epsilon}_{\alpha}(x) = I_\frac{\alpha}{b}\bigl(2 \sqrt{\mu \nu} e^{-\phi}\bigr).
\end{equation}
One can see this either from the Mellin-Barnes integral representation of \eqref{qbesseli} or from the Taylor series expansion \eqref{taylor},\footnote{We need to set $n=0$ in the latter evaluation, since the poles with $n\neq 0$ from the $S_b$ function are shifted to infinity in the classical limit. The poles resulting in the second line of \eqref{taylor} (arising from the zeros of $S_b$ in the denominator of \eqref{qbesseli}) also shift to infinity.} yielding back the classical formulas \eqref{besseli}.

The function \eqref{qbesseli} is an eigenfunction of the finite difference equation
\begin{align}
\label{difeq}
\left( T_{ib} + T_{-ib} + g_\mu^b g_\nu^b  q^{\epsilon} e^{-2\pi b x} T_{-\epsilon i b }\right) f(x) = 2 \cos \pi b \alpha \, f(x),
\end{align}
where we left the parameter $\epsilon$ arbitrary (corresponding to different quantizations of the same underlying classical problem). We are most interested in the cases $\epsilon =\pm1$ and $\epsilon =0$.\footnote{The relevant Whittaker function for Liouville gravity turns out to have $\epsilon=\pm 1$, whereas the $\epsilon=0$ case is the simplest toy example.}

The second eigenfunction of this finite difference equation is
\begin{equation}
\mathcal{K}^{\epsilon}_{\alpha,M}(x) = b\pi e^{-\alpha \pi x} \int_{-\infty}^{+\infty} d\zeta\, g_\mu^{i\zeta} g_\nu^{i\zeta + \alpha} S_b(-i\zeta) S_b(-\alpha -i \zeta ) e^{-\pi i \epsilon (\zeta^2 + 2s \zeta)} e^{-2\pi i \zeta x},
\end{equation}
which is the usual $q$-deformed BesselK. Writing the $q$-deformed BesselI as \eqref{qbesseli}, we have the $q$-deformed relation
\begin{equation}
\mathcal{K}^{\epsilon}_{\alpha}(x) = \frac{\pi}{2 \sin \frac{\pi \alpha}{b}} \left(\mathcal{I}^{\epsilon}_{-\alpha}(x)- \mathcal{I}^{\epsilon}_{\alpha}(x)  \right),
\end{equation}
generalizing the classical version \eqref{relKI}. Satisfying this relation can be viewed as a requirement for any candidate $q$-deformation of the BesselI function.

The limit $g_\mu^b g_\nu^b \to 0$ allows for a truncation of the series \eqref{taylor} to the $n=m=0$ term of the first line, and gives an exponential function:
\begin{equation}
\lim_{ g_\mu^b g_\nu^b\to 0} \mathcal{I}_{\alpha}(x) = \frac{b g_\nu^{\alpha}}{S_b(\alpha+b)}e^{-\alpha \pi x}.
\end{equation}
We do not care about the prefactor since this can be absorbed into a redefinition of the operator of interest.

As $b\to 0$, the difference equation \eqref{difeq} limits to ($\phi = \pi b x$):
\begin{equation}
\left(-\partial_\phi^2 +  \mu\nu e^{2 \phi} \right) f(\phi) = - \frac{\alpha^2}{b^2} f(\phi),
\end{equation}
with independent solutions
\begin{equation}
I_\frac{\alpha}{b}\bigl(2 \sqrt{\mu \nu} e^{-\phi}\bigr), \qquad K_\frac{\alpha}{b}\bigl(2 \sqrt{\mu \nu} e^{-\phi}\bigr).
\end{equation}

\subsection{Extension to \texorpdfstring{U$_q(\mathfrak{osp}(1|2, \mathbb{R}))$}{Uq(osp(1|2, R))}}

Let us try to generalize this discussion to the case of U$_q(\mathfrak{osp}(1|2,\mathbb{R}))$. We look for a second solution to the system of difference equations \eqref{susyfde2}, aside from \eqref{qmelb}. We propose the function
\begin{align}
\label{sqbesseli}
\psi^{\epsilon, \pm}_{\alpha,g_\mu g_\nu}(x) \equiv b e^{-\alpha \pi x} \int_{\mathcal{C}} d\zeta\, & g_\mu^{i\zeta} g_\nu^{i\zeta +\alpha} e^{-\pi i \frac{\epsilon}{2} (\zeta^2 - \alpha i \zeta)}e^{-\pi i  \zeta x} \\
&\times\left(\frac{S_{\NS}(-i\zeta)}{S_{\NS}(i\zeta+\alpha +b)}\cosh \frac{\pi \zeta}{2b} \pm \frac{S_{\R}(-i\zeta)}{S_{\R}(i\zeta+\alpha +b)}\sinh \frac{\pi \zeta}{2b} \right), \nonumber
\end{align}
which satisfies the same pair of difference equations as before:
\begin{align}
\left( T_{ib} - T_{-ib} - i g_\mu^b g_\nu^b e^{-\pi b x} q^{\epsilon/2} T_{-i\epsilon b} \right) \psi^{\epsilon,+}_{\alpha,g_\mu g_\nu}(x) &= -2i \sin \pi b \alpha \, \psi^{\epsilon,-}_{\alpha,g_\mu g_\nu}(x), \\
\left( T_{ib} - T_{-ib} + i g_\mu^b g_\nu^b e^{-\pi b x} q^{\epsilon/2} T_{-i\epsilon b} \right) \psi^{\epsilon,-}_{\alpha,g_\mu g_\nu}(x) &= -2i \sin \pi b \alpha \, \psi^{\epsilon,+}_{\alpha,g_\mu g_\nu}(x). 
\end{align}
A quick way to see this is to rewrite \eqref{sqbesseli} as
\begin{align}
\psi^{\epsilon,\pm}_{\alpha}(x) = 2b e^{-\alpha \pi x} \int_{\mathcal{C}} d\zeta\, & g_\mu^{i\zeta} g_\nu^{i\zeta +\alpha} e^{-\pi i \frac{\epsilon}{2} (\zeta^2 - \alpha i \zeta)}e^{-\pi i  \zeta x} \\
&\times \bigg( S_{\NS}(-i\zeta)S_{\R}(-i\zeta-\alpha)\cos\left(\frac{\pi}{2b}(-i\zeta-2\alpha)\right)\cosh \frac{\pi \zeta}{2b} \nonumber \\
&\phantom{====} \pm S_{\R}(-i\zeta) S_{\NS}(-i\zeta-\alpha) \sin\left(\frac{\pi}{2b}(-i\zeta-2\alpha)\right)\sinh \frac{\pi \zeta}{2b} \bigg). \nonumber
\end{align}
This expression is almost the same as that for the Whittaker function \eqref{qmelb}, up to setting $\alpha = is$ and the presence of the hyperbolic functions $\cosh$ and $\sinh$. The latter map into one another under $\zeta \to \zeta + ib$. Hence this is a solution to the finite difference equations as long as the same is true of the Whittaker function \eqref{qmelb}, which we already know.

In the $g_\mu^b g_\nu^b \to 0$ limit of \eqref{sqbesseli}, the pole at $\zeta=0$ dominates the first term (just like in the bosonic case) and the pole at $\zeta=-ib$ dominates the second term. This leads schematically to the function
\begin{equation}
\# e^{-\pi \alpha x} + \# e^{-\pi (\alpha +b) x},
\end{equation}
which will also be seen in the classical limit \eqref{sclas}.

In the classical limit, where we let $\phi = \pi b x$, $g_\mu = (4\pi b^2 \mu)^{\frac{1}{2b}}$, $g_\nu = (4\pi b^2 \nu)^{\frac{1}{2b}}$, we get
\begin{equation}
\label{sclas}
\psi^{\epsilon,\pm}_{\alpha}\left(x\right)\to \sqrt{\frac{e^{-\phi}}{2}}\left( I_{-1/2+\frac{\alpha}{b}}\left(2\sqrt{\mu\nu} e^{-\phi}\right) \pm I_{1/2+\frac{\alpha}{b}}\left(2\sqrt{\mu\nu} e^{-\phi}\right) \right),
\end{equation}
which, for $z= 2\sqrt{\mu\nu}e^{-\phi}$, is the set of functions solving the second-order ODE
\begin{equation}
\left(-z^2\partial_z^2 - z\partial_z \pm z+ z^2 \right) \psi(z) = -\alpha^2 \psi(z).
\end{equation}

\section{Whittaker Function in the \texorpdfstring{$\eta = -1$}{eta = -1} Sector}
\label{app:NS}

We can emulate our construction of amplitudes in the $\eta=-1$ sector by considering the following ad hoc ``Whittaker function'' given by the expression
\begin{align}
\tilde{\psi}^{\epsilon,\pm}_{s,g_\mu g_\nu}(x) = e^{-\pi i  s x} \int_{-\infty}^{+\infty} d\zeta\, & g_\mu^{i\zeta} g_\nu^{i\zeta + 2is} e^{-\pi i \frac{\epsilon}{2} (\zeta^2 + 2s \zeta)} e^{-\pi i \zeta x} \\
&\times \left[S_{\NS}(-i\zeta) S_{\NS}(-2i s -i \zeta ) \pm S_{\R}(-i\zeta) S_{\R}(-2i s -i \zeta )\right]. \nonumber
\end{align}
The only difference between this expression and \eqref{qmelb} is the combination of $S_{\NS}$ and $S_{\R}$ appearing on the second line.

These functions satisfy the following system of difference equations:\footnote{For convenience, we note that
\begin{align}
2\cosh (\pi b (\zeta +s)) &= 4 \sinh \frac{\pi b}{2}(\zeta +2s) \sinh \frac{\pi b \zeta}{2} +2 \cosh \pi b s \\
&= 4 \cosh \frac{\pi b}{2}(\zeta +2s) \cosh \frac{\pi b \zeta}{2} - 2 \cosh \pi b s.
\end{align}}
\begin{align}
\left(T_{ib} + T_{-ib} \right) \tilde{\psi}^{\epsilon,+}_{s,g_\mu g_\nu}(x) - g_\mu^b g_\nu^b e^{-\pi b x} q^{\epsilon/2} T_{-i\epsilon b} \tilde{\psi}^{\epsilon,-}_{s,g_\mu g_\nu}(x ) &= 2 \cosh \pi b s \, \tilde{\psi}^{\epsilon,-}_{s,g_\mu g_\nu}(x), \\
\left(T_{ib} + T_{-ib} \right) \tilde{\psi}^{\epsilon,-}_{s,g_\mu g_\nu}(x) + g_\mu^b g_\nu^b e^{-\pi b x} q^{\epsilon/2} T_{-i\epsilon b} \tilde{\psi}^{\epsilon,+}_{s,g_\mu g_\nu}(x) &= 2 \cosh \pi b s \, \tilde{\psi}^{\epsilon,+}_{s,g_\mu g_\nu}(x).
\end{align}
It would be interesting to understand whether there exists a proper group-theoretic origin of these functions. If so, it would need to involve $q$-deformation in an essential way, since the underlying classical superalgebra $\mathfrak{osp}(1|2,\mathbb{R})$ does not contain this additional freedom.

These ``Whittaker functions'' satisfy the orthogonality property
\begin{align}
\int_{-\infty}^{+\infty} dx \, \tilde{\psi}^{\epsilon,\pm}_{s_1,g_\mu g_\nu} (x) \tilde{\psi}^{\epsilon,\pm}_{s_2,g_\mu g_\nu} (x)^* = \frac{4\delta(s_1-s_2)}{\sinh \frac{\pi s_2}{b} \sinh \pi b s_2},
\end{align}
from which one finds a Plancherel measure $\rho(s) \sim \sinh \frac{\pi s}{b} \sinh \pi b s$ quite similar to that in the bosonic U$_q(\mathfrak{sl}(2,\mathbb{R}))$ scenario.

\section{Nonlinear Lie (Super)algebras}
\label{app:nonlinear}

An alternative to the language of (graded) Poisson sigma models is that of nonlinear (super)gauge theory.  Namely, a theory of 2D dilaton (super)gravity can be viewed as a gauge theory based on a nonlinear Lie (super)algebra \cite{Ikeda:1993aj, Ikeda:1993fh, Ikeda:1993dr}.  We briefly review this terminology here.

It is important to note that, from the Poisson algebra point of view, all of the nonlinear Lie algebras below describe ``classical'' Poisson brackets, \emph{before} quantization of the Poisson structure.  Therefore, in all equations that present an abstract bracket on the left-hand side, the multiplication operation on the right-hand side is (graded-)commutative.

\subsubsection*{Nonlinear Gauge Theory}

A nonlinear Lie algebra with basis $\{T^i\}$ is specified by a generalized Lie bracket
\begin{equation}
[T^i, T^j] = P^{ij}(T),
\end{equation}
where $P^{ij} = -P^{ji}$ is a polynomial.  Multiplication takes place in the polynomial ring of the $\{T^i\}$, and is hence commutative.  The generalized Jacobi identity reads
\begin{equation}
\partial_\ell P^{[ij}P^{k]\ell} = 0.
\label{jacobiid}
\end{equation}
An ordinary Lie algebra with structure constants $f^{ijk}$ is recovered upon setting $P^{ij}(T) = f^{ijk}T_k$.

The minimal field content of a nonlinear gauge theory consists of an adjoint scalar $X^i$ and a gauge field $A_i$.  A nonlinear gauge transformation with parameter $\epsilon$ takes the form
\begin{align}
\delta X^i &= -P^{ji}(X)\epsilon_j, \nonumber \\
\delta A_i &= -d\epsilon_i - \partial_i P^{jk}(X)A_j\epsilon_k. \label{nonlineargauge}
\end{align}
In two dimensions, there exists an invariant action of the form $S = \int \Omega$ where
\begin{align}
\Omega &\equiv X^i dA_i + \frac{1}{2}P^{jk}(X)A_j\wedge A_k, \label{invariantaction} \\
\delta\Omega &= d((P^{jk}(X) - X^i\partial_i P^{jk}(X))A_j\epsilon_k). \nonumber
\end{align}
The equations of motion following from \eqref{invariantaction} are $F_i = 0$ and $DX^i = 0$, where we define the field strength and covariant derivative by
\begin{align}
F_i &= dA_i + \frac{1}{2}\partial_i P^{jk}(X)A_j\wedge A_k, \\
DX^i &= dX^i + P^{ij}(X)A_j.
\end{align}
Note that
\begin{align}
\delta F_i &= -\partial_i P^{jk}(X)F_j\epsilon_k - DX^\ell\partial_\ell\partial_i P^{jk}(X)\wedge A_j\epsilon_k, \\
\delta(DX^i) &= -DX^k\partial_k P^{ji}(X)\epsilon_j.
\end{align}
In particular, the field strength $F_i$ transforms inhomogeneously, where the inhomogeneous terms vanish in the case of a linear gauge theory.  By \eqref{jacobiid}, the commutator algebra of the transformations \eqref{nonlineargauge} closes on shell with respect to the action \eqref{invariantaction}.

Liouville gravity with dilaton potential \eqref{dilatonpotL} is a nonlinear gauge theory with gauge algebra U$_q(\mathfrak{sl}(2, \mathbb{R}))$ ($q = e^{\pi ib^2}$).  The connection between sinh dilaton gravity and nonlinear $q$-gauge theory was already observed in \cite{Kyono:2017pxs}.  In the limit $b\to 0$, we obtain JT gravity with $V(\Phi) = \Phi$, whose first-order action is that of an $\mathfrak{sl}(2, \mathbb{R})$ BF theory.

\subsubsection*{Nonlinear Supergauge Theory}

A nonlinear Lie superalgebra with basis $\{\mathcal{T}^i\}$ takes the form
\begin{equation}
[\mathcal{T}^i, \mathcal{T}^j] = P^{ij}(\mathcal{T}),
\end{equation}
where the bracket is $\mathbb{Z}_2$-graded and $P^{ij} = -(-)^{\sigma_i\sigma_j}P^{ji}$.  Multiplication takes place in the $\mathbb{Z}_2$-graded polynomial ring of the $\{\mathcal{T}^i\}$.  ``Classically,'' the graded Jacobi identity takes the form
\begin{equation}
\sum_{\operatorname{cyc}(ijk)} (-)^{\sigma_i\sigma_k}P^{i\ell}\partial_\ell P^{jk} = 0, \qquad \sum_{\operatorname{cyc}(ijk)} (-)^{\sigma_i\sigma_k}\partial_\ell^R P^{ij}P^{\ell k} = 0,
\label{classicaljacobishort}
\end{equation}
written in terms of left or right derivatives, respectively.\footnote{These relations correspond to two equivalent ways of writing the ``quantum'' graded Jacobi identity:
\begin{align}
0 &= (-)^{\sigma_i\sigma_k}[\mathcal{T}_i, [\mathcal{T}_j, \mathcal{T}_k]] + (-)^{\sigma_j\sigma_i}[\mathcal{T}_j, [\mathcal{T}_k, \mathcal{T}_i]] + (-)^{\sigma_k\sigma_j}[\mathcal{T}_k, [\mathcal{T}_i, \mathcal{T}_j]] \\
&= (-)^{\sigma_j\sigma_i}[[\mathcal{T}_j, \mathcal{T}_k], \mathcal{T}_i] + (-)^{\sigma_k\sigma_j}[[\mathcal{T}_k, \mathcal{T}_i], \mathcal{T}_j] + (-)^{\sigma_i\sigma_k}[[\mathcal{T}_i, \mathcal{T}_j], \mathcal{T}_k].
\end{align}}  When we take derivatives to act from the left (right), we write variations on the left (right) so as to obtain the correct signs when anticommuting fermionic quantities.

Our nonlinear gauge transformations are now
\begin{align}
\delta X^i &= -\epsilon_j P^{ji}(X), \nonumber \\
\delta A_i &= -d\epsilon_i + A_j\epsilon_k\partial_i^R P^{kj}(X).
\end{align}
We have the invariant action $S = \int \Omega$ where
\begin{align}
\Omega &\equiv dA_i X^i - \frac{1}{2}A_i\wedge A_j P^{ji}(X), \label{bfaction} \\
\delta\Omega &= d(A_j\epsilon_k(\partial_i^R P^{kj}(X)X^i - P^{kj}(X))). \nonumber
\end{align}
Note that the ``BF-type'' action \eqref{bfaction} differs by a total derivative from the graded Poisson sigma model action \eqref{PSM} used in the main text:
\begin{equation}
\Omega - d(A_i X^i) = A_i\wedge dX^i - \frac{1}{2}A_i\wedge A_j P^{ji}(X).
\end{equation}
If all fields are bosonic, then the preceding formulas reduce to the bosonic ones.

$\mathcal{N} = 1$ Liouville supergravity with dilaton prepotential \eqref{dilatonprepotL} is a nonlinear supergauge theory with gauge superalgebra U$_q(\mathfrak{osp}(1|2, \mathbb{R}))$ ($q = e^{\pi ib^2}$).  In the limit $b\to 0$, we obtain $\mathcal{N} = 1$ JT supergravity with $u(\Phi) = \Phi$, whose first-order action is that of an $\mathfrak{osp}(1|2, \mathbb{R})$ BF theory.

\mciteSetMidEndSepPunct{}{\ifmciteBstWouldAddEndPunct.\else\fi}{\relax}
\bibliographystyle{utphys}
{\small \bibliography{references2022}{}}

\end{document}